\documentclass[aps,singlecolumn,prd,preprintnumbers,superscriptaddress,nofootinbib]{revtex4-2}
\usepackage{hyperref}
\hypersetup{colorlinks=true,linkcolor=purple,anchorcolor=blue,citecolor=blue, filecolor=blue,urlcolor=blue,bookmarksnumbered=true,
	pdfview=FitB
}
\usepackage{ragged2e}
\usepackage[justification=centering]{caption}
\usepackage{subcaption}
\usepackage{color}
\usepackage{xcolor}
\usepackage{amsmath}
\usepackage{ulem}
\usepackage{csquotes}
\colorlet{purple1}{blue!70!red}
\colorlet{darkred}{red!50!black}
\usepackage{graphicx}
\usepackage{psfrag}
\usepackage{color}
\usepackage{comment}
\usepackage{amssymb}
\usepackage{amsmath}
\usepackage{epstopdf}
\usepackage{epsf}

\usepackage{natbib}

\newcommand{\nslash}{\kern 0.2 em n\kern -0.50em /}
\newcommand{\kslash}{\kern 0.2 em k\kern -0.45em /}
\newcommand{\lslash}{\kern 0.2 em l\kern -0.50em /}
\newcommand{\pslash}{\kern 0.2 em p\kern -0.50em /}
\newcommand{\Sslash}{\kern 0.2 em S\kern -0.50em /}
\newcommand{\Pslash}{\kern 0.2 em P\kern -0.50em /}
\newcommand{\Dslash}{\kern 0.2 em D\kern -0.65em /\kern 0.15em}

\newcommand{\be}{\begin{eqnarray}}
\newcommand{\ee}{\end{eqnarray}}

\newcommand{\bfk}{\mathbf{k}_{\perp}}

\newcommand{\bfb}{{\bf b}_{\perp}}

\begin{document}
	
	\title{Gluon generalized transverse momentum dependent parton distributions and Wigner functions of the Proton}
\author{Dipankar~Chakrabarti}
	\email{dipankar@iitk.ac.in} 
	\affiliation{Department of Physics, Indian Institute of Technology Kanpur, Kanpur-208016, India}

 \author{Bheemsehan~Gurjar}
	\email{gbheem@iitk.ac.in} 
	\affiliation{Department of Physics, Indian Institute of Technology Kanpur, Kanpur-208016, India}
	
 \author{Asmita~Mukherjee}
	\email{asmita@phy.iitb.ac.in} 
	\affiliation{Department of Physics, Indian Institute of Technology Bombay, Powai, Mumbai 400076, India}

    \author{Kauship~Saha}
	\email{kauships24@iitk.ac.in} 
	\affiliation{Department of Physics, Indian Institute of Technology Kanpur, Kanpur-208016, India}
\date{\today}
\begin{abstract}
{{We present a calculation of the leading-twist (twist-2) generalized transverse momentum distributions (GTMDs) for unpolarized and longitudinally polarized gluons in the proton, using a recently developed light-front gluon spectator model inspired by the soft-wall AdS/QCD prediction. We calculate the gluon GTMDs for non-zero skewness, using overlaps in terms of light-front wave functions. We also compute the gluon Wigner distributions}} for various polarization configurations of the proton target, including unpolarized, longitudinally polarized, and transversely polarized states. We present these Wigner distributions and the associated spin densities in both the transverse momentum plane and the transverse impact parameter plane. Furthermore, we evaluate the gluon orbital angular momentum and spin-orbit correlations, and extract the generalized parton distribution and transverse momentum-dependent distribution limits of the GTMDs. 
\end{abstract}
\maketitle
\section{Introduction}
Nucleons (the protons and neutrons) are the fundamental constituents of visible matter. They are composed of quarks and gluons, which are bound together by the strong interaction governed by quantum chromodynamics (QCD). The internal structure of these nucleons has been probed through several key experiments over the past decades~\cite{AbdulKhalek:2021gbh}. In particular, deep inelastic scattering (DIS) experiments have offered one-dimensional information into parton distribution functions (PDFs), {{ which describe how the longitudinal momentum is distributed among the partons at different momentum scales~\cite{Collins:1981uw,Martin:1998sq}. These are so-called collinear PDFs, where the information about the transverse motion of the partons is missing. More recently, single spin asymmetries observed in experiments as well as deeply virtual Compton scattering and exclusive meson production showed that the transverse motion of the partons as well as their spatial distribution inside the nucleon are important.}} The spatial distribution of the partons inside hadrons is encoded in the generalized parton distributions (GPDs), and the transverse motion is encoded in the transverse momentum dependent parton distributions (TMDs). TMDs can be probed experimentally through semi-inclusive processes such as semi-inclusive deep inelastic scattering (SIDIS) and the Drell–Yan process~\cite{Mulders:1995dh,Barone:2001sp,Bacchetta:2006tn,Arnold:2008kf,Anselmino:2020vlp}, while GPDs are accessible via hard exclusive processes like deeply virtual Compton scattering (DVCS) and deeply virtual meson production (DVMP)~\cite{{Ji:1996nm,Goeke:2001tz,Belitsky:2005qn,Diehl:2003ny}}. Both TMDs and GPDs provide a three-dimensional description of nucleons, offering a more detailed and comprehensive multi-dimensional picture than traditional one-dimensional PDFs (see the recent reviews~\cite{Lorce:2025aqp,boussarie2023tmd} and references therein).

Meanwhile, the concept of phase-space distributions, known as Wigner distributions, was first introduced by Wigner in 1932~\cite{Wigner:1932eb}. They serve as the quantum-mechanical counterparts of classical phase-space distributions, offering a unified description of the momentum and spatial distributions of partons, and were later incorporated into QCD by Ji~\cite{Ji:2003ak}. In general, for the non-relativistic systems, the Wigner distributions were approximated as six-dimensional phase-space distributions that depend on three position and three momentum variables~\cite{Belitsky:2003nz}. {A more intuitive picture is obtained } within the light-front formalism, where it is often more practical to work with five-dimensional Wigner distributions, expressed as functions of the longitudinal momentum fraction $x$, two transverse momentum components $\mathbf{k}_{\perp}$, and two transverse position coordinates $\mathbf{b}_{\perp}$~\cite{Lorce:2015sqe}. {{The Wigner distributions are not directly measurable in experiments, 
however, they are related to the generalized transverse momentum distributions (GTMDs) at zero skewness~\cite{Lorce:2013pza,Meissner:2009ww,Meissner:2008ay}. GTMDs are derived by integrating the generalized parton correlation functions (GPCFs) over the light-cone energy component ($k^{-}$)~\cite{Meissner:2009ww}. GPCFs are fully unintegrated, off-diagonal two-parton correlators that depend on the four-momentum of the parton $k$ and the four-momentum transfer $\Delta=p^{\prime}-p$ between the initial and final hadron states~\cite{Collins:2007ph}.  They encode the most comprehensive information about the partonic structure of nucleons. GTMDs are functions of the longitudinal momentum fraction $x$, transverse momentum $\mathbf{k}_{\perp}$, longitudinal momentum transfer fraction $\xi$, and transverse momentum transfer $\mathbf{\Delta}_{\perp}$. The Wigner distributions are obtained by Fourier transforming the GTMDs over $\mathbf{\Delta}_{\perp}$ at zero $\xi$ (skewness). Significant progress has been made in recent years in identifying scattering processes that are sensitive to GTMDs. In particular, the quark GTMDs typically appear at subleading twist, where standard collinear factorization is not applicable and can be accessed through the exclusive double Drell–Yan process~\cite{Bhattacharya:2017bvs}, and exclusive $\pi^{0}$ production process in electron-proton collision~\cite{Bhattacharya:2023hbq,Bhattacharya:2023yvo}. On the other hand, diffractive vector meson production~\cite{Martin:1999wb}, Higgs production at Tevatron and LHC~\cite{Khoze:2000cy} and dijet production in deep-inelastic lepton–nucleon and lepton–nucleus scattering~\cite{Bhattacharya:2022vvo,Bhattacharya:2024sck, Hatta:2016aoc,Ji:2016jgn,Boer:2021upt}, ultraperipheral proton–nucleus collisions~\cite{Hagiwara:2017fye,Linek:2023kga,Boer:2023mip}, and virtual photon–nucleus quasielastic scattering processes are promising avenues for probing gluon GTMDs at high energies (small $x$)~\cite{Zhou:2016rnt,Agrawal:2023mzm}. On the other hand, Wigner distributions}}  have been extensively investigated in various QCD-inspired models to gain insights into the multidimensional structure of nucleons;  quark Wigner distributions have been investigated in many recent publications, for example ~\cite{Lorce:2011ni,Lorce:2011kd,Mukherjee:2014nya,More:2017zqq,Liu:2015eqa,Chakrabarti:2016yuw,Chakrabarti:2017teq,Gutsche:2016gcd,Ojha:2022fls,Yang:2025neu,Luan:2024nwc,Chakrabarti:2019wjx,Maji:2022tog}, while studies focusing on gluons remain relatively limited~\cite{Mukherjee:2015aja,More:2017zqp,Tan:2023vvi,Jana:2023btd}. The Wigner distributions are often referred to as the ``mother distributions'' of TMDs and GPDs~\cite{Meissner:2009ww}. The GPDs in position space, also known as impact parameter dependent parton distributions (IPDs)~\cite{Burkardt:2000za,Burkardt:2002hr}, are obtained by integrating the Wigner distributions over the transverse momentum $\mathbf{k}_{\perp}$. These IPDs depend on the longitudinal momentum fraction $x$ and the transverse position (impact parameter) $\mathbf{b}_{\perp}$. While TMDs are extracted by integrating the Wigner distributions over the transverse position $\mathbf{b}_{\perp}$ and are functions of longitudinal momentum fraction $x$ and the transverse momentum $\mathbf{k}_{\perp}$. The uncertainty principle doesn't allow the Wigner distributions to become positive definite over the entire phase-space, and it has no probabilistic interpretation~\cite{Lorce:2011kd}.

 Depending on the various spin-orbit correlations between the nucleon and the gluon, there are 16 gluon GTMDs at leading-twist~\cite{Meissner:2009ww,Lorce:2013pza}. Among them, the GTMDs $F^{g}_{1,4}$ and $G^{g}_{1,1}$ encode crucial information about the nucleon spin structure through the gluon canonical orbital angular momentum (OAM)~\cite{Ji:2012sj,Hatta:2011ku,Leader:2013jra} and the spin-orbit correlations of partons~\cite{Lorce:2014mxa,Chakrabarti:2024hwx}, respectively. Additionally, the GTMDs $F^{g}_{1,1}$ and $G^{g}_{1,4}$ are related to the unpolarized and helicity-dependent parton distributions, respectively~\cite{Maji:2022tog}. Moreover, the unpolarized ($H$ and $E$) and the helicity-dependent ($\tilde{H}$ and $\tilde{E}$) gluon GPDs are related to the unpolarized and longitudinally polarized gluon GTMDs~\cite{Chakrabarti:2024hwx}. 

The skewness variable $\xi$, which represents the fraction of longitudinal momentum transfer in a physical process, plays a vital role in extracting experimental information. Most previous studies on GTMDs have been focused on the zero-skewness case ($\xi = 0$), which corresponds to purely transverse momentum transfer~\cite{Lorce:2011kd,Chakrabarti:2019wjx,Chakrabarti:2016yuw,Chakrabarti:2017teq}. However, developing a deeper understanding of GTMDs at nonzero skewness is essential~\cite{Maji:2022tog,Tan:2024dmz,Ojha:2022fls,Jana:2023btd}. In this work, we investigate the non-skewed leading-twist unpolarized and longitudinally polarized gluon GTMDs of the proton within the Dokshitzer–Gribov–Lipatov–Altarelli–Parisi (DGLAP) region ($1>x>\xi$), using the light-front gluon spectator model~\cite{Chakrabarti:2023djs}. The light-front wave functions (LFWFs) in this model are adopted from the two-particle effective wave functions derived in the soft-wall AdS/QCD framework~\cite{Brodsky:2014yha}. This model has previously been successfully applied to describe several properties of the proton, such as PDFs, GPDs, TMDs, gravitational form factors (GFFs), and mechanical properties~\cite{Chakrabarti:2024hwx,Sain:2025kup,Chakrabarti:2023djs}. We also demonstrate that, in specific kinematical regimes, the GTMDs reduce to GPDs and TMDs through appropriate projections~\cite{Meissner:2009ww,Lorce:2011dv}. Using these GTMDs and the projected GPDs, we further compute the canonical and kinetic OAM and the spin–orbit correlation factor. Finally, we obtain the Wigner distributions for unpolarized, longitudinally polarized, and linearly polarized gluons inside an unpolarized, longitudinally polarized, and transversely polarized proton by performing a two-dimensional Fourier transform of the GTMDs at zero skewness~\cite{Lorce:2011kd}. 

The rest of the paper is organized as follows: In Section~\ref{sec:spectator model}, we briefly discuss the proton LFWFs in the light-front gluon spectator model motivated by the soft-wall AdS/QCD framework. Section~\ref{sec:GTMDs and Wigners defination} presents the bilinear decomposition of the GTMD correlators and their parameterizations with gluon GTMDs and corresponding Wigner distributions. In Section~\ref{sec:results and discussion}, we present and discuss the numerical results. Finally, Section~\ref{sec:conclusion} summarize  our findings and concludes the paper.
\section{Light-Front gluon spectator model}\label{sec:spectator model}
The gluon spectator models developed in recent years all share the core idea of describing the nucleon as an effective two-body system, where the active parton is a gluon and the remainder of the nucleon is treated as a single spectator, but they differ in construction and physical emphasis. For instance, the Lu–Ma spectator model~\cite{Lu:2016vqu} employs a light-cone wave function with a spin-1/2 spectator to study the gluon Sivers function, and later extended to gluon GPDs, GTMDs, and Wigner functions, emphasizing orbital angular momentum and spin–orbit correlations. The model uses non-Gaussian scalar LFWFs inspired by electron wave functions, and incorporates nonperturbative color interactions at the gluon–nucleon–spectator vertex through the Brodsky–Huang–Lepage prescription~\cite{Brodsky:1981jv}. Another spectator model by Bacchetta \textit{et al.}~\cite{Bacchetta:2020vty,Bacchetta:2024fci} instead introduces a spectral mass distribution for the spectator, together with a flexible effective nucleon-gluon-spectator vertex, allowing a systematic treatment of both T-even and T-odd gluon TMDs. Motivated by the works discussed above, in Ref.~\cite{Chakrabarti:2023djs}, we introduce a similar type of gluon spectator model in which the transverse momentum dependence of the wave functions follows a Gaussian form inspired by AdS/QCD. Within this framework, initially we studied the T-even gluon TMDs and was later extended to investigate gluon GPDs~\cite{Chakrabarti:2024hwx}, as well as gluonic GFFs and the mechanical properties of the proton~\cite{Sain:2025kup}. The AdS/QCD wavefunctions were previously used to study the quark distribution functions in our quark–diquark model, where the nucleon was treated as a bound state of three valence quarks represented by the Fock state of an active quark and a spectator diquark, i.e., $|q(qq)\rangle$~\cite{Mondal:2015uha,Maji:2016yqo}. In the present gluon spectator model, we consider the minimal Fock state of the proton as a system consisting of an active gluon and a composite spectator made of the three valence quarks, i.e., $|g(qqq)\rangle$. Due to the complexity of the full four-particle system, we approximate the proton target state as an effective two-particle system, composed of an active spin-1 gluon $g$ and a spin-1/2 spectator particle $X$ of mass $M_{X}$. In a reference frame where the proton carries no transverse momentum, i.e., ${P} = (P^{+}, \frac{M^2}{P^{+}}, \mathbf{0}_{\perp})$, the gluon is taken as the active parton with longitudinal momentum fraction $x$ and relative transverse momentum $\bfk$. The two-particle Fock-state expansion of the proton with helicity, $J_{z} = \pm \tfrac{1}{2}$~\cite{Brodsky:2000ii} can then be written as
\begin{align}\label{eq:proton state}\nonumber
		|P;\uparrow(\downarrow)\rangle
		= \int \frac{\mathrm{d}^2 {\bf k}_\perp \mathrm{d} x}{16 \pi^3 \sqrt{x(1-x)}}&\times \Bigg[\psi_{+1+\frac{1}{2}}^{\uparrow(\downarrow)}\left(x, {\bf k}_\perp\right)\left|+1,+\frac{1}{2} ; x P^{+}, {\bf k}_\perp\right\rangle
		+\psi_{+1-\frac{1}{2}}^{\uparrow(\downarrow)}\left(x, {\bf k}_\perp \right)\left|+1,-\frac{1}{2} ; x P^{+}, {\bf k}_\perp \right\rangle\nonumber\\ 
		&~~+\psi_{-1+\frac{1}{2}}^{\uparrow(\downarrow)}\left(x, {\bf k}_\perp \right)\left|-1,+\frac{1}{2} ; x P^{+}, {\bf k}_\perp \right\rangle+\psi_{-1-\frac{1}{2}}^{\uparrow(\downarrow)}\left(x, {\bf k}_\perp\right)\left|-1,-\frac{1}{2} ; x P^{+}, {\bf k}_\perp\right\rangle\Bigg].
	\end{align}	
    here, $\psi^{\uparrow(\downarrow)}_{\lambda_{g}\lambda_{X}}$ represents the LFWFs with proton helicities $\uparrow(\downarrow)$ corresponds to $J_{z}=\pm 1/2$, which encodes the probability amplitude for the two-particle state, $|\lambda_{g}, \lambda_{X}; xP^{+}, \mathbf{k}_{\perp}\rangle$ with, $\lambda_{g} = \pm1$ corresponds to the gluon helicity, while $\lambda_{X} = \pm1/2$ denotes the helicity of the spectator. The two-particle LFWFs of the proton of helicity $J_{z} = +1/2$~\cite{Chakrabarti:2023djs} are given by,
\begin{eqnarray} \label{LFWFsuparrow}   
		\psi_{+1+\frac{1}{2}}^{\uparrow}\left(x,{\bf k}_\perp\right)&=&-\sqrt{2}\frac{(-k^{(1)}_{\perp}+ik^{(2)}_{\perp})}{x(1-x)}\varphi(x,{\bf k}_\perp^2), \\ \nonumber
		\psi_{+1-\frac{1}{2}}^{\uparrow}\left(x, {\bf k}_\perp\right)&=&-\sqrt{2}\bigg( M-\frac{M_{X}}{(1-x)} \bigg) \varphi(x,{\bf k}_\perp^2), \\ \nonumber
		\psi_{-1+\frac{1}{2}}^{\uparrow}\left(x, {\bf k}_\perp\right)&=&-\sqrt{2}\frac{(k^{(1)}_{\perp}+ik^{(2)}_{\perp})}{x}\varphi(x,{\bf k}_\perp^2), \\
		\psi_{-1-\frac{1}{2}}^{\uparrow}\left(x, {\bf k}_\perp\right)&=&0\,.
\end{eqnarray}
Similarly, the two-particle LFWFs  of a proton with $J_{z}=-1/2$ have  the form,
	\begin{eqnarray} \label{LFWFsdownarrow}   \nonumber
		\psi_{+1+\frac{1}{2}}^{\downarrow}\left(x, {\bf k}_\perp\right)&=& 0, \\ \nonumber
		\psi_{+1-\frac{1}{2}}^{\downarrow}\left(x,{\bf k}_\perp\right)&=&-\sqrt{2}\frac{(-k^{(1)}_{\perp}+ik^{(2)}_{\perp})}{x}\varphi(x,{\bf k}_\perp^2), \\ \nonumber
		\psi_{-1+\frac{1}{2}}^{\downarrow}\left(x, {\bf k}_\perp\right)&=&-\sqrt{2}\bigg( M-\frac{M_{X}}{(1-x)} \bigg) \varphi(x,{\bf k}_\perp^2),  \\
		\psi_{-1-\frac{1}{2}}^{\downarrow}\left(x, {\bf k}_\perp \right)&=& -\sqrt{2}\frac{(k^{(1)}_{\perp}+ik^{(2)}_{\perp})}{x(1-x)}\varphi(x,{\bf k}_\perp^2)\,.
	\end{eqnarray}
    where $M$ and $M_{X}$ denote the proton and spectator masses, respectively. The scalar function $\varphi(x,\bfk^2)$ is a modified form of the soft-wall AdS/QCD wave function~\cite{Brodsky:2014yha,Gutsche:2013zia}, incorporating additional parameters $a$ and $b$ that regulate the asymptotic behavior of the gluon PDFs, and is expressed as~\cite{Chakrabarti:2023djs}, 
\begin{align}\label{AdSphi}
\varphi(x,{\bf k}_\perp^2)=&N_{g}\frac{4\pi}{\kappa}\sqrt{\frac{\log[1/(1-x)]}{x}}x^{b}(1-x)^{a}\exp{\Big[-\frac{\log[1/(1-x)]}{2\kappa^{2}x^2}{\bf k}_\perp^{2}\Big]}\,.
\end{align}
where $\kappa$ is an emergent mass scale parameter that describes the transverse dynamics of gluons within a hadron. It is determined, along with the normalization constant $N_{g}$ and the parameters $a$ and $b$, by fitting to the gluon unpolarized PDFs using NNPDF3.0 data set at the scale $Q_{0} = 2~\text{GeV}$~\cite{Chakrabarti:2023djs,Sain:2025kup}. To ensure proton stability, the spectator mass is set to $M_{X} = 0.985^{+0.044}_{-0.045}$ GeV, which is larger than the proton mass, while the gluon mass is taken to be zero. 
\section{Gluon GTMDs and Wigner distributions in spectator model}\label{sec:GTMDs and Wigners defination}
In this section, we present the bilinear decomposition of the fully unintegrated gluon-gluon correlator for proton and the detailed calculations of leading-twist gluon GTMDs {{ with non-zero skewness and the corresponding Wigner distributions, and discuss the calculation using light-front spectator model.  The gluon GTMDs encode a wealth of information about the role of quark-gluon interaction on the internal hadronic structure and can be obtained from the fully unintegrated two-parton correlator,}} also known as the GPCFs, after performing the light-front energy ($k^{-}$) integration at equal light-front time $z^{+}=0$~\cite{Meissner:2009ww,Lorce:2013pza}.

\subsection{Gluon GTMDs}\label{subsec:GTMDs defination}
The off-forward gluon-gluon GTMD correlator at equal light-front time is defined as~\cite{Lorce:2013pza},
\begin{align}\label{eq:GTMDs correlator}
W^{g[ij]}_{\lambda^{\prime\prime}\lambda^{\prime}}(x,\xi,\mathbf{\Delta}_{\perp},\mathbf{k}_\perp)=&\int \frac{{\rm d}z^-{\rm d}^2\bm{z}_\perp}{(2\pi)^3 xP^+}e^{ik\cdot z}  \langle P^{\prime\prime},\lambda^{\prime\prime}|\Gamma^{ij}F^{+i}_a(-z/2)\mathcal{W}_{ab}(-z/2,z/2) F^{+j}_b(z/2) | P^{\prime},\lambda^{\prime} \rangle|_{z^+=0}\,.
\end{align}
where $| P^{\prime},\lambda^{\prime} \rangle$ and $| P^{\prime\prime},\lambda^{\prime\prime} \rangle$ denote the initial and final nucleon states with momenta $P^{\prime}$, $P^{\prime\prime}$, and helicities $\lambda^{\prime},~\lambda^{\prime\prime}$, respectively. The $\Gamma^{ij}$ denotes the leading-twist generic gluon operator, i.e., $\Gamma^{ij}=\delta^{ij}_{\perp},~-i\epsilon_{\perp}^{ij}$ corresponding to unpolarized and longitudinally polarized gluons, respectively.   $F_{a(b)}^{\mu\nu}$ represents the gluon field strength tensor, while $\mathcal{W}_{ab}$ denotes the Wilson line (or gauge link) that connects the two gluon field operators, ensuring the gauge invariance of the bilocal gluon correlator. Within the light-cone gauge, $A^{+}=0$, the Wilson line reduces to unity. The GTMD correlators can be parametrized in terms of gluon GTMDs, which can be further projected onto gluon GPDs and TMDs~\cite{Meissner:2009ww,Lorce:2013pza}. A detailed parametrization of the quark-quark and gluon-gluon GTMD correlators in terms of GTMDs can be found in Ref.~\cite{Meissner:2009ww} for quarks and Ref.~\cite{Lorce:2013pza} for gluons, respectively. As one can see in Eq.~\eqref{eq:GTMDs correlator}, the GTMDs are the six-dimensional distribution functions with $x=k^{+}/P^{+}$ being the longitudinal momentum fraction carried by the gluon, {{here}}, $\xi=-\Delta^{+}/2P^{+}$ denote the fraction of longitudinal momentum transfer with respect to the nucleon, $\mathbf{k}_{\perp}$ and $\mathbf{\Delta}_{\perp}$ are the two-dimensional transverse components of gluon momentum, and transverse momentum transfer to the hadron, respectively. Within the light-cone conventions $x^{\pm}=(x^{0}\pm x^{3})$, the kinematic variables are defines as follows~\cite{Maji:2022tog},
\begin{align}
  P=&\Big(P^{+},\frac{M^2+\mathbf{\Delta}_{\perp}^{2}/4}{(1-\xi^2)P^{+}},\mathbf{0}_{\perp}\Big)\,, \quad\quad\quad\quad
  k=(xP^{+},k^{-},\mathbf{k}_{\perp})\,, \quad\quad\quad\quad
    \Delta=\Big(-2\xi P^{+},\frac{t+\mathbf{\Delta}_{\perp}^{2}}{-2\xi P^{+}},\mathbf{\Delta}_{\perp}\Big)\,.
\end{align}
In a symmetric frame, the average four momentum of proton is denoted by $P=(P^{\prime\prime}+P^{\prime})/2$ and the four momentum transfer $\Delta=(P^{\prime\prime}-P^{\prime})$. Meanwhile, the initial and final state proton momenta are given as,
\begin{align}
    P^{\prime}=&\Big((1+\xi)P^{+},\frac{M^2+\mathbf{\Delta}_{\perp}^{2}/4}{(1+\xi)P^{+}},-\mathbf{\Delta}_{\perp}/2\Big)\,,\quad\quad
     P^{\prime\prime}=\Big((1-\xi)P^{+},\frac{M^2+\mathbf{\Delta}_{\perp}^{2}/4}{(1-\xi)P^{+}},~\mathbf{\Delta}_{\perp}/2\Big)\,.
\end{align}  
The detailed bilinear decomposition of gluon-gluon correlator in terms of leading-twist gluon GTMDs is presented in the Appendix~\ref{AppendixA}. Meanwhile, the two-particle LFWF overlap representation of gluon-gluon correlator for unpolarized (with $\Gamma^{ij}=\delta^{ij}_{\perp}$) and longitudinally polarized (with $\Gamma^{ij}=-i\epsilon^{ij}_{\perp}$) gluons is given as follows~\cite{More:2017zqp},
\begin{align}\label{eq:correlator1}
    W_{\lambda^{\prime\prime}\lambda^{\prime}}^{g}(x,\xi,\mathbf{k}_{\perp},\mathbf{\Delta}_{\perp})=&\frac{1}{16\pi^{3}}\sum_{\lambda_{g},\lambda_{X}}\psi^{\lambda^{\prime\prime\ast}}_{\lambda_{g}\lambda_{X}}(x^{\prime\prime},\mathbf{k}_{\perp}^{\prime\prime})\psi^{\lambda^{\prime}}_{\lambda_{g}\lambda_{X}}(x^{\prime},\mathbf{k}_{\perp}^{\prime})(\varepsilon^{1}_{\lambda_{g}}\varepsilon^{1\ast}_{\lambda_{g}}+\varepsilon^{2}_{\lambda_{g}}\varepsilon^{2\ast}_{\lambda_{g}})\,,\\
    \widetilde{W}_{\lambda^{\prime\prime}\lambda^{\prime}}^{g}(x,\xi,\mathbf{k}_{\perp},\mathbf{\Delta}_{\perp})=&-\frac{i}{16\pi^{3}}\sum_{\lambda_{g},\lambda_{X}}\psi^{\lambda^{\prime\prime\ast}}_{\lambda_{g}\lambda_{X}}(x^{\prime\prime},\mathbf{k}_{\perp}^{\prime\prime})\psi^{\lambda^{\prime}}_{\lambda_{g}\lambda_{X}}(x^{\prime},\mathbf{k}_{\perp}^{\prime})(\varepsilon^{1}_{\lambda_{g}}\varepsilon^{2\ast}_{\lambda_{g}}-\varepsilon^{2}_{\lambda_{g}}\varepsilon^{1\ast}_{\lambda_{g}})\,.
    \label{eq:correlator2}
\end{align}
where, $\varepsilon_{\lambda_{g}}^{\mu}$ is the gluon polarization vector with, $\varepsilon_{\pm}^{\perp}=\frac{1}{\sqrt{2}}(1,\pm i)$, $\mathbf{k}_{\perp}^{\prime}$ and $\mathbf{k}_{\perp}^{\prime\prime}$ are the transverse momenta of struck gluon in the initial and final nucleon states and are defined as follows,
\begin{align}
    \mathbf{k}_{\perp}^{\prime}=&\mathbf{k}_{\perp}+(1-x^{\prime})\frac{\mathbf{\Delta}_{\perp}}{2}\,,\quad\quad\text{with}\quad\quad x^{\prime}=\frac{x-\xi}{1-\xi}\,,\nonumber\\
       \mathbf{k}_{\perp}^{\prime\prime}=&\mathbf{k}_{\perp}-(1-x^{\prime\prime})\frac{\mathbf{\Delta}_{\perp}}{2}\,,\quad\quad\text{with}\quad\quad x^{\prime\prime}=\frac{x+\xi}{1+\xi}\,.
\end{align}
respectively. By substituting the model results for the unpolarized gluon GTMD correlators from Eqs.~\eqref{eq:unpolcorrelatorupup}–\eqref{eq:unpolcorrelatordownup} into the parametrization of the gluon GTMDs given in Eqs.~\eqref{eq:F11}–\eqref{eq:GTMDsparametrization13}, we obtain explicit analytical expressions for the four $F$-type leading-twist gluon GTMDs corresponding to unpolarized gluons, as follows:
\begin{align}
F^{g}_{1,1}(x,\xi,\mathbf{k}_{\perp},\mathbf{\Delta}_{\perp})=&\frac{2N_{g}^{2}}{\pi\kappa^2}\sqrt{1-\xi^2}\Big[\mathcal{F}_{a}(x^{\prime},x^{\prime\prime})\Big(\mathbf{k}_{\perp}^{2}-(1-x^{\prime})(1-x^{\prime\prime})\frac{\mathbf{\Delta}_{\perp}^{2}}{4}\Big)+\mathcal{F}_{b}(x^{\prime},x^{\prime\prime})+\frac{1}{2}(\mathbf{k}_{\perp}.\mathbf{\Delta}_{\perp})\nonumber\\
   &\times\Big(\mathcal{F}_{c}(x^{\prime},x^{\prime\prime})-\mathcal{F}_{d}(x^{\prime},x^{\prime\prime})\Big)\Big]\exp\Big[-\mathcal{A}(x^{\prime},x^{\prime\prime})\mathbf{k}_{\perp}^{2}- \mathcal{B}(x^{\prime},x^{\prime\prime})\frac{\mathbf{\Delta}_{\perp}^{2}}{4}- \mathcal{C} (x^{\prime},x^{\prime\prime})(\mathbf{k}_{\perp}.\mathbf{\Delta}_{\perp})\Big]\,,\label{eq:F11}\\  \nonumber
   F^{g}_{1,2}(x,\xi,\mathbf{k}_{\perp},\mathbf{\Delta}_{\perp})=&\frac{2N_{g}^{2}}{\pi\kappa^2}\frac{M}{\sqrt{1-\xi^2}}\mathcal{F}_{e}(x^{\prime},x^{\prime\prime})\exp\Big[-\mathcal{A}(x^{\prime},x^{\prime\prime})\mathbf{k}_{\perp}^{2}- \mathcal{B}(x^{\prime},x^{\prime\prime})\frac{\mathbf{\Delta}_{\perp}^{2}}{4}- \mathcal{C} (x^{\prime},x^{\prime\prime})(\mathbf{k}_{\perp}.\mathbf{\Delta}_{\perp})\Big]\label{eq:F12}\\ 
    & -\frac{\Delta_{\perp}^2}{2M^2}\frac{\xi}{(1-\xi^2)}F^{g}_{1,4}\,,\\ \nonumber
    F^{g}_{1,3}(x,\xi,\mathbf{k}_{\perp},\mathbf{\Delta}_{\perp})=&\frac{1}{2(1-\xi^2)}F^{g}_{1,1}+\frac{1}{2M^2}\frac{\xi}{(1-\xi^2)}(\mathbf{k}_{\perp}.\mathbf{\Delta}_{\perp})F^{g}_{1,4}-\frac{N_{g}^{2}}{\pi\kappa^2}\frac{M}{\sqrt{1-\xi^2}}\mathcal{F}_{f}(x^{\prime},x^{\prime\prime})\label{eq:F13}\\ 
    &\times\exp\Big[-\mathcal{A}(x^{\prime},x^{\prime\prime})\mathbf{k}_{\perp}^{2}- \mathcal{B}(x^{\prime},x^{\prime\prime})\frac{\mathbf{\Delta}_{\perp}^{2}}{4}- \mathcal{C} (x^{\prime},x^{\prime\prime})(\mathbf{k}_{\perp}.\mathbf{\Delta}_{\perp})\Big]\,,\\
    F^{g}_{1,4}(x,\xi,\mathbf{k}_{\perp},\mathbf{\Delta}_{\perp})=&\frac{N_{g}^{2}}{\pi\kappa^2}{M^{2}}\sqrt{1-\xi^2}\Big(\mathcal{F}_{c}(x^{\prime},x^{\prime\prime})+\mathcal{F}_{d}(x^{\prime},x^{\prime\prime})\Big)
    \nonumber\\
    &\times \exp\Big[-\mathcal{A}(x^{\prime},x^{\prime\prime})\mathbf{k}_{\perp}^{2}- \mathcal{B}(x^{\prime},x^{\prime\prime})\frac{\mathbf{\Delta}_{\perp}^{2}}{4}- \mathcal{C} (x^{\prime},x^{\prime\prime})(\mathbf{k}_{\perp}.\mathbf{\Delta}_{\perp})\Big]\,.
    \label{eq:F14}
\end{align}
Similarly, by substituting the model results for the longitudinally polarized gluon GTMD correlators from Eqs.~\eqref{eq:longpolcorrelatorupup}–\eqref{eq:longpolcorrelatordownup} into the parametrization of the longitudinally polarized gluon GTMDs given in Eqs.~\eqref{eq:G14}–\eqref{eq:parametrizationG13}, we obtain explicit analytical expressions for the four $G$-type gluon GTMDs corresponding to longitudinally polarized gluons, as follows:
\begin{align}
    G^{g}_{1,1}(x,\xi,\mathbf{k}_{\perp},\mathbf{\Delta}_{\perp})=&-\frac{N_{g}^{2}}{\pi\kappa^2}{M^{2}}\sqrt{1-\xi^2}\Big(\mathcal{F}_{g}(x^{\prime},x^{\prime\prime})+\mathcal{F}_{h}(x^{\prime},x^{\prime\prime})\Big)
    \nonumber\\
    &\times\exp\Big[-\mathcal{A}(x^{\prime},x^{\prime\prime})\mathbf{k}_{\perp}^{2}- \mathcal{B}(x^{\prime},x^{\prime\prime})\frac{\mathbf{\Delta}_{\perp}^{2}}{4}- \mathcal{C} (x^{\prime},x^{\prime\prime})(\mathbf{k}_{\perp}.\mathbf{\Delta}_{\perp})\Big]\,,\\   
    G^{g}_{1,2}(x,\xi,\mathbf{k}_{\perp},\mathbf{\Delta}_{\perp})=&
     \frac{\Delta_{\perp}^{2}}{2M^{2}(1-\xi^2)}G^{g}_{1,1}
     -\frac{2N_{g}^{2}}{\pi\kappa^2}\frac{M}{\sqrt{1-\xi^2}}\mathcal{F}_{k}(x^{\prime},x^{\prime\prime})\nonumber\\
     &\times\exp\Big[-\mathcal{A}(x^{\prime},x^{\prime\prime})\mathbf{k}_{\perp}^{2}- \mathcal{B}(x^{\prime},x^{\prime\prime})\frac{\mathbf{\Delta}_{\perp}^{2}}{4}- \mathcal{C} (x^{\prime},x^{\prime\prime})(\mathbf{k}_{\perp}.\mathbf{\Delta}_{\perp})\Big]\,,\\
     G^{g}_{1,3}(x,\xi,\mathbf{k}_{\perp},\mathbf{\Delta}_{\perp})=&\frac{N_{g}^{2}}{\pi\kappa^2}\frac{M}{\sqrt{1-\xi^2}}\mathcal{F}_{j}(x^{\prime},x^{\prime\prime})\exp\Big[-\mathcal{A}(x^{\prime},x^{\prime\prime})\mathbf{k}_{\perp}^{2}- \mathcal{B}(x^{\prime},x^{\prime\prime})\frac{\mathbf{\Delta}_{\perp}^{2}}{4}- \mathcal{C} (x^{\prime},x^{\prime\prime})(\mathbf{k}_{\perp}.\mathbf{\Delta}_{\perp})\Big]\label{eq:G13}\nonumber\\
 &  +\frac{\xi}{2(1-\xi^2)}G^{g}_{1,4}-\frac{(\mathbf{k}_{\perp}.\mathbf{\Delta}_{\perp})}{2M^{2}(1-\xi^2)}G^{g}_{1,1}\,,\\
 G^{g}_{1,4}(x,\xi,\mathbf{k}_{\perp},\mathbf{\Delta}_{\perp})=&\frac{2N_{g}^{2}}{\pi\kappa^2}\sqrt{1-\xi^2}\Bigg[\mathcal{F}_{i}(x^{\prime},x^{\prime\prime})\Big(\mathbf{k}_{\perp}^{2}-(1-x^{\prime})(1-x^{\prime\prime})\frac{\mathbf{\Delta}_{\perp}^{2}}{4}\Big)+\mathcal{F}_{b}(x^{\prime},x^{\prime\prime})+\frac{1}{2}(\mathbf{k}_{\perp}.\mathbf{\Delta}_{\perp})\nonumber\\
   &\times\Big(\mathcal{F}_{g}(x^{\prime},x^{\prime\prime})-\mathcal{F}_{h}(x^{\prime},x^{\prime\prime})\Big)\Bigg]\exp\Big[-\mathcal{A}(x^{\prime},x^{\prime\prime})\mathbf{k}_{\perp}^{2}- \mathcal{B}(x^{\prime},x^{\prime\prime})\frac{\mathbf{\Delta}_{\perp}^{2}}{4}- \mathcal{C} (x^{\prime},x^{\prime\prime})(\mathbf{k}_{\perp}.\mathbf{\Delta}_{\perp})\Big]\,.
\end{align}
where, the functions $\mathcal{F}_{a,b...,k}~,\mathcal{A},~\mathcal{B}$ and $\mathcal{C}$ are parametrized in terms of gluon longitudinal momentum fractions in the incoming proton state ($x^{\prime}$) and outgoing proton state ($x^{\prime\prime}$) as given in Eqs.~\eqref{eq:AppendixFaxx}-\eqref{eq:AppendixCxx}. Within the forward limit, $\Delta=0$ and at zero skewness variable, $\xi=0$, those gluon GTMDs reduces to the leading-twist gluon TMDs which are reported in our previous work~\cite{Chakrabarti:2023djs}. Whereas, on the other hand, the transverse momentum integrated gluon GTMDs provides the gluon GPDs as presented in our earlier work~\cite{Chakrabarti:2024hwx}. Specially, the $F$-type gluon GTMDs $F_{1,1}^{g},~F_{1,2}^{g}$ and $F_{1,3}^{g}$ provides the unpolarized gluon GPDs $H^{g}$ and $E^{g}$, while the $G$-type gluon GTMDs $G_{1,1}^{g},~G_{1,2}^{g}$ and $G_{1,4}^{g}$ gives the polarized gluon GPDs $\widetilde{H}^{g}$ and $\widetilde{E}^{g}$, respectively as follows~\cite{Maji:2022tog},
\begin{align}\label{eq:GPDHg}
    H^{g}(x,\xi,\mathbf{\Delta}_{\perp})=&\int{\rm d}^{2}\mathbf{k}_{\perp}\Big[F_{1,1}^{g}+2\xi^{2}\Big(\frac{\mathbf{k}_{\perp}\cdot\mathbf{\Delta}_{\perp}}{\mathbf{\Delta}_{\perp}^{2}}F_{1,2}^{g}+F_{1,3}^{g}\Big)\Big]\,,\\
    E^{g}(x,\xi,\mathbf{\Delta}_{\perp})=&\int{\rm d}^{2}\mathbf{k}_{\perp}\Big[-F_{1,1}^{g}+2(1-\xi^{2})\Big(\frac{\mathbf{k}_{\perp}\cdot\mathbf{\Delta}_{\perp}}{\mathbf{\Delta}_{\perp}^{2}}F_{1,2}^{g}+F_{1,3}^{g}\Big)\Big]\,,\label{eq:GPDEg}\\
    \widetilde{H}^{g}(x,\xi,\mathbf{\Delta}_{\perp})=&\int{\rm d}^{2}\mathbf{k}_{\perp}\Big[2\xi\Big(\frac{\mathbf{k}_{\perp}\cdot\mathbf{\Delta}_{\perp}}{\mathbf{\Delta}_{\perp}^{2}}G_{1,2}^{g}+G_{1,3}^{g}\Big)+G_{1,4}^{g}\Big]\,,\label{eq:GPDHtildeg}\\
    \widetilde{E}^{g}(x,\xi,\mathbf{\Delta}_{\perp})=&\int{\rm d}^{2}\mathbf{k}_{\perp}\Big[\frac{2(1-\xi^{2})}{\xi}\Big(\frac{\mathbf{k}_{\perp}\cdot\mathbf{\Delta}_{\perp}}{\mathbf{\Delta}_{\perp}^{2}}G_{1,2}^{g}+G_{1,3}^{g}\Big)-G_{1,4}^{g}\Big]\,.
\end{align}
The above obtained gluon GPDs are consistent with the previous results in our earlier work~\cite{Chakrabarti:2024hwx}. For nonzero skewness, i.e., $\xi \neq 0$, those GTMDs contain a $\mathbf{k}_{\perp} \cdot \mathbf{\Delta}_{\perp}$ term, which breaks the axial symmetry of the distributions in the transverse momentum plane for fixed values of transverse momentum transfer. As shown in Eqs.~\eqref{eq:F12} and~\eqref{eq:F13} the $F$-type nonzero skewed gluon GTMDs, $F_{1,2}^{g}$ and $F_{1,3}^{g}$, contain additional contributions from $F_{1,4}^{g}$, while the $G$-type non-skewed gluon GTMD, $G_{1,3}^{g}$ includes $G_{1,4}^{g}$ (see Eq.~\eqref{eq:G13}). These additional contributions vanish in the zero-skewness limit. We present the numerical results and corresponding discussion on GTMDs in section~\ref{sec:results and discussion}. In the zero skewness limit, the GTMDs $F_{1,2}^{g}$ and $G_{1,3}^{g}$ vanishes due to the following relations among the proton LFWFs: $\psi^{\uparrow\ast}_{+1-\frac{1}{2}}=\psi^{\downarrow}_{-1+\frac{1}{2}}\,,~\psi^{\uparrow\ast}_{-1+\frac{1}{2}}=-\psi^{\downarrow}_{+1-\frac{1}{2}}\,,~\text{and}~\psi^{\uparrow\ast}_{+1+\frac{1}{2}}=\psi^{\downarrow}_{-1-\frac{1}{2}}$. 
These relations lead to the vanishing of $\mathcal{F}_{e}(x,x)$ and $\mathcal{F}_{j}(x,x)$ as shown in Eqs.~\eqref{eq:Fe} and \eqref{eq:Fj}.

According to Ji's spin sum rule~\cite{Ji:1996nm}, the nucleon spin structure can be investigated through the Mellin moments of zero-skewed GPDs in the forward limit, as follows:
\begin{align}\label{eq:Jg}
    J_{z}^{g}=\frac{1}{2}\int{\rm d}x\Big{\{}x[H^{g}(x,0,0)+E^{g}(x,0,0)]\Big{\}}\,,
\end{align}
By inserting the zero-skewed unpolarized and helicity flip gluon GPDs from Eqs.~\eqref{eq:GPDHg} and \eqref{eq:GPDEg} into Eq.~\eqref{eq:Jg}, we obtain the  gluon angular momentum as, $J_{z}^{g} = 0.205\pm0.013$. We find that our result is consistent with several recent determinations of the total gluon angular momentum at a scale of $2$ GeV. These include the lattice QCD calculation by the Extended Twisted Mass (ETM) Collaboration, which reports $J_{z}^{g} = 0.187(46)(10)$ in the $\overline{\text{MS}}$ scheme~\cite{Alexandrou:2020sml}; the basis light-front quantization (BLFQ) result of $J_{z}^{g} = 0.22(02)$~\cite{Nair:2025sfr}; the Dyson-Schwinger equation (DSE) approach yielding $J_{z}^{g} \sim 0.20$~\cite{Tandy:2025tea}; lattice QCD predictions based on dipole and $z$-expansion fits to the proton GFFs with $J_{z}^{g} = 0.25(13)$ and $J_{z}^{g} = 0.234(27)$, respectively~\cite{Hackett:2023rif}; and the result from the $\chi$QCD Collaboration, $J_{z}^{g} = 0.23(11)(22)$~\cite{Wang:2021vqy}. The kinetic OAM can also be obtained in terms of zero-skewed GPDs in the forward limit as follows,
\begin{align}\label{eq:kinetic_OAM}
    L_{z}^{g}=\int{\rm d}x\Big{\{}\frac{1}{2}x[H^{g}(x,0,0)+E^{g}(x,0,0)]-\widetilde{H}^{g}(x,0,0)\Big{\}}
\end{align}
Again, by employing the zero-skewed GPDs from Eqs.~\eqref{eq:GPDHg}, \eqref{eq:GPDEg}, and \eqref{eq:GPDHtildeg} into Eq.~\eqref{eq:kinetic_OAM}, we obtain the gluon kinetic orbital angular momentum as, $L_{z}^{g} = -0.22$, which is larger than the value $L_{z}^{g} = -0.123$ reported in Ref.~\cite{Tan:2023kbl}.
\subsection{Gluon Wigner Distributions}\label{subsec:Wigners defination defination}
The gluon Wigner distribution correlation functions can be obtained by performing the two-dimensional Fourier transform of the gluon-gluon GTMD correlator (see Eq.~\ref{eq:GTMDs correlator}) at zero skewness as~\cite{Lorce:2011kd,Meissner:2009ww},
\begin{align}
    \mathcal{W}^{g}_{\lambda^{\prime\prime}\lambda^{\prime}}(x,\mathbf{k}_{\perp},\mathbf{b}_{\perp})=\int\frac{{\rm d}^{2}\mathbf{\Delta}_{\perp}}{(2\pi)^{2}}e^{-i\mathbf{\Delta}_{\perp}\cdot\mathbf{b}_{\perp}}\int \frac{{\rm d}z^-{\rm d}^2\bm{z}_\perp}{(2\pi)^3 xP^+ }e^{ik\cdot z}  \langle P^{\prime\prime},\lambda^{\prime\prime}|\Gamma^{ij}F^{+i}_a(-z/2)\mathcal{W}_{ab}(-z/2,z/2) F^{+j}_b(z/2) | P^{\prime},\lambda^{\prime} \rangle|_{z^+=0}\,.
\end{align}
where $\mathbf{b}_{\perp}$ is the Fourier conjugate of the transverse momentum transfer $\mathbf{\Delta}_{\perp}$ and is known as the impact parameter space variable. Depending on all possible polarization states of the gluon ($\Lambda_{g}$) and the proton ($\Lambda_{p}$), the analytical expressions for the gluon Wigner distributions $\mathcal{W}_{\Lambda_{p}\Lambda_{g}}$ are given as follows:
\begin{itemize}
    \item The Wigner distributions for the unpolarized, longitudinally polarized, and linearly polarized ($L$ and $R$) gluons inside an unpolarized proton
\end{itemize}
\begin{align}\nonumber
    \mathcal{W}_{UU}(x,\mathbf{k}_{\perp},\mathbf{b}_{\perp})=&\frac{1}{2}\Big[\mathcal{W}_{\uparrow\uparrow}^{g}(x,\mathbf{k}_{\perp},\mathbf{b}_{\perp})+\mathcal{W}_{\downarrow\downarrow}^{g}(x,\mathbf{k}_{\perp},\mathbf{b}_{\perp})\Big]\\ 
    =&\int\frac{{\rm d}^{2}\mathbf{\Delta}_{\perp}}{(2\pi)^{2}}e^{-i\mathbf{\Delta}_{\perp}\cdot\mathbf{b}_{\perp}}\frac{2N_{g}^{2}}{\pi\kappa^2}\Big[\mathcal{F}_{a}(x,x)\Big(\mathbf{k}_{\perp}^{2}-(1-x)^{2}\frac{\mathbf{\Delta}_{\perp}^{2}}{4}\Big)+\mathcal{F}_{b}(x,x)\Big]\nonumber\\
    &\times\exp\Big[-2\tilde{a}(x)\Big(\mathbf{k}_{\perp}^{2}+(1-x)^{2}\frac{\mathbf{\Delta}_{\perp}^{2}}{4}\Big)\Big]\,,\\
    \mathcal{W}_{UL}(x,\mathbf{k}_{\perp},\mathbf{b}_{\perp})=&\frac{1}{2}\Big[\widetilde{\mathcal{W}}_{\uparrow\uparrow}^{g}(x,\mathbf{k}_{\perp},\mathbf{b}_{\perp})+\widetilde{\mathcal{W}}_{\downarrow\downarrow}^{g}(x,\mathbf{k}_{\perp},\mathbf{b}_{\perp})\Big]\nonumber\\
   =&\int\frac{{\rm d}^{2}\mathbf{\Delta}_{\perp}}{(2\pi)^{2}}e^{-i\mathbf{\Delta}_{\perp}\cdot\mathbf{b}_{\perp}}\frac{N_{g}^{2}}{\pi\kappa^2}\Big[\mathcal{F}_{g}(x,x)+\mathcal{F}_{h}(x,x)\Big]i\varepsilon^{ij}_{\perp}\mathbf{k}_{\perp}^{i}\mathbf{\Delta}_{\perp}^{j}\exp\Big[-2\tilde{a}(x)\Big(\mathbf{k}_{\perp}^{2}+(1-x)^{2}\frac{\mathbf{\Delta}_{\perp}^{2}}{4}\Big)\Big]\,,\\
   \mathcal{W}_{U\mathcal{T}}^{(R)}(x,\mathbf{k}_{\perp},\mathbf{b}_{\perp})=&\frac{1}{2}\Big[\mathcal{W}_{\uparrow\uparrow}^{R}(x,\mathbf{k}_{\perp},\mathbf{b}_{\perp})+\mathcal{W}_{\downarrow\downarrow}^{R}(x,\mathbf{k}_{\perp},\mathbf{b}_{\perp})\Big]\nonumber\\
    =&\int\frac{{\rm d}^{2}\mathbf{\Delta}_{\perp}}{(2\pi)^{2}}e^{-i\mathbf{\Delta}_{\perp}\cdot\mathbf{b}_{\perp}}\frac{2N_{g}^{2}}{\pi\kappa^2}\Big{\{}\Big[\mathcal{F}_{a}(x,x)\Big(\mathbf{k}_{\perp}^{2}-(1-x)^{2}\frac{\mathbf{\Delta}_{\perp}^{2}}{4}\Big)+\mathcal{F}_{b}(x,x)\Big]-\frac{1}{2}\Big(\mathcal{F}_{g}(x,x)+\mathcal{F}_{h}(x,x)\Big)\nonumber\\
     &\times i\varepsilon^{ij}_{\perp}\mathbf{k}_{\perp}^{i}\mathbf{\Delta}_{\perp}^{j}\Big{\}} \exp\Big[-2\tilde{a}(x)\Big(\mathbf{k}_{\perp}^{2}+(1-x)^{2}\frac{\mathbf{\Delta}_{\perp}^{2}}{4}\Big)\Big]\,,\\
    \mathcal{W}_{U\mathcal{T}}^{(L)}(x,\mathbf{k}_{\perp},\mathbf{b}_{\perp})=&\frac{1}{2}\Big[\mathcal{W}_{\uparrow\uparrow}^{L}(x,\mathbf{k}_{\perp},\mathbf{b}_{\perp})+\mathcal{W}_{\downarrow\downarrow}^{L}(x,\mathbf{k}_{\perp},\mathbf{b}_{\perp})\Big]\nonumber\\
    =&\int\frac{{\rm d}^{2}\mathbf{\Delta}_{\perp}}{(2\pi)^{2}}e^{-i\mathbf{\Delta}_{\perp}\cdot\mathbf{b}_{\perp}}\frac{2N_{g}^{2}}{\pi\kappa^2}\Big{\{}\Big[\mathcal{F}_{a}(x,x)\Big(\mathbf{k}_{\perp}^{2}-(1-x)^{2}\frac{\mathbf{\Delta}_{\perp}^{2}}{4}\Big)+\mathcal{F}_{b}(x,x)\Big]+\frac{1}{2}\Big(\mathcal{F}_{g}(x,x)+\mathcal{F}_{h}(x,x)\Big)\nonumber\\
     &\times i\varepsilon^{ij}_{\perp}\mathbf{k}_{\perp}^{i}\mathbf{\Delta}_{\perp}^{j}\Big{\}} \exp\Big[-2\tilde{a}(x)\Big(\mathbf{k}_{\perp}^{2}+(1-x)^{2}\frac{\mathbf{\Delta}_{\perp}^{2}}{4}\Big)\Big]\,.
\end{align}
where $\mathcal{W}^{R}_{\lambda^{\prime\prime}\lambda^{\prime}}$ and $\mathcal{W}^{L}_{\lambda^{\prime\prime}\lambda^{\prime}}$ are the Fourier transforms of corresponding right-handed and left-handed linearly polarized gluon-gluon GTMD correlators, which are given as follows,
\begin{align}
 \mathcal{W}_{\lambda^{\prime\prime}\lambda^{\prime}}^{R}(x,\mathbf{k}_{\perp},\mathbf{b}_{\perp})=&\frac{1}{16\pi^{3}}\sum_{\lambda_{g},\lambda_{X}}\int\frac{{\rm d}^{2}\mathbf{\Delta}_{\perp}}{(2\pi)^{2}}e^{-i\mathbf{\Delta}_{\perp}\cdot\mathbf{b}_{\perp}}\psi^{\lambda^{\prime\prime\ast}}_{\lambda_{g}\lambda_{X}}(x^{\prime\prime},\mathbf{k}_{\perp}^{\prime\prime})\psi^{\lambda^{\prime}}_{\lambda_{g}\lambda_{X}}(x^{\prime},\mathbf{k}_{\perp}^{\prime})\varepsilon^{R}_{\lambda_{g}}\varepsilon^{R\ast}_{\lambda_{g}}\,,\\
    \mathcal{W}_{\lambda^{\prime\prime}\lambda^{\prime}}^{L}(x,\mathbf{k}_{\perp},\mathbf{b}_{\perp})=&\frac{1}{16\pi^{3}}\sum_{\lambda_{g},\lambda_{X}}\int\frac{{\rm d}^{2}\mathbf{\Delta}_{\perp}}{(2\pi)^{2}}e^{-i\mathbf{\Delta}_{\perp}\cdot\mathbf{b}_{\perp}}\psi^{\lambda^{\prime\prime\ast}}_{\lambda_{g}\lambda_{X}}(x^{\prime\prime},\mathbf{k}_{\perp}^{\prime\prime})\psi^{\lambda^{\prime}}_{\lambda_{g}\lambda_{X}}(x^{\prime},\mathbf{k}_{\perp}^{\prime})\varepsilon^{L}_{\lambda_{g}}\varepsilon^{L\ast}_{\lambda_{g}}\,.
    \label{eq:correlator4}
\end{align} 
with the right and left-handed gluon polarization vectors, $\varepsilon^{R(L)}_{\lambda_{g}}=\varepsilon^{1}_{\lambda_{g}}\pm i\varepsilon^{2}_{\lambda_{g}}$, respectively.
\begin{itemize}
    \item The Wigner distributions for the unpolarized, longitudinally polarized, and linearly polarized ($L$ and $R$) gluons inside a longitudinally  polarized proton
\end{itemize}
\begin{align}
    \mathcal{W}_{LU}(x,\mathbf{k}_{\perp},\mathbf{b}_{\perp})=&\frac{1}{2}\Big[\mathcal{W}_{\uparrow\uparrow}^{g}(x,\mathbf{k}_{\perp},\mathbf{b}_{\perp})-\mathcal{W}_{\downarrow\downarrow}^{g}(x,\mathbf{k}_{\perp},\mathbf{b}_{\perp})\Big]\nonumber\\
    =&\int\frac{{\rm d}^{2}\mathbf{\Delta}_{\perp}}{(2\pi)^{2}}e^{-i\vec{\Delta}_{\perp}\cdot\vec{b}_{\perp}}\frac{N_{g}^{2}}{\pi\kappa^2}\Big(\mathcal{F}_{c}(x,x)+\mathcal{F}_{d}(x,x)\Big)i\varepsilon^{ij}_{\perp}\mathbf{k}_{\perp}^{i}\mathbf{\Delta}_{\perp}^{j}\exp\Big[-2\tilde{a}(x)\Big(\mathbf{k}_{\perp}^{2}+(1-x)^{2}\frac{\mathbf{\Delta}_{\perp}^{2}}{4}\Big)\Big]\,,\\
    \mathcal{W}_{LL}(x,\mathbf{k}_{\perp},\mathbf{b}_{\perp})=&\frac{1}{2}\Big[\widetilde{\mathcal{W}}_{\uparrow\uparrow}^{g}(x,\mathbf{k}_{\perp},\mathbf{b}_{\perp})-\widetilde{\mathcal{W}}_{\downarrow\downarrow}^{g}(x,\mathbf{k}_{\perp},\mathbf{b}_{\perp})\Big]\nonumber \\
  =&\int\frac{{\rm d}^{2}\mathbf{\Delta}_{\perp}}{(2\pi)^{2}}e^{-i\vec{\Delta}_{\perp}\cdot\vec{b}_{\perp}}\frac{2N_{g}^{2}}{\pi\kappa^2}\Big[\mathcal{F}_{i}(x,x)\Big(\mathbf{k}_{\perp}^{2}-(1-x)^{2}\frac{\mathbf{\Delta}_{\perp}^{2}}{4}\Big)+\mathcal{F}_{b}(x,x)\Big]\nonumber\\
  &\times \exp\Big[-2\tilde{a}(x)\Big(\mathbf{k}_{\perp}^{2}+(1-x)^{2}\frac{\mathbf{\Delta}_{\perp}^{2}}{4}\Big)\Big]\,,\\
  \mathcal{W}_{L\mathcal{T}}^{(R)}(x,\mathbf{k}_{\perp},\mathbf{b}_{\perp})=&\frac{1}{2}\Big[\mathcal{W}_{\uparrow\uparrow}^{R}(x,\mathbf{k}_{\perp},\mathbf{b}_{\perp})-\mathcal{W}_{\downarrow\downarrow}^{R}(x,\mathbf{k}_{\perp},\mathbf{b}_{\perp})\Big]\,,\nonumber\\
    =&-\int\frac{{\rm d}^{2}\mathbf{\Delta}_{\perp}}{(2\pi)^{2}}e^{-i\mathbf{\Delta}_{\perp}\cdot\mathbf{b}_{\perp}}\frac{2N_{g}^{2}}{\pi\kappa^2}\Big{\{}\Big[\mathcal{F}_{i}(x,x)\Big(\mathbf{k}_{\perp}^{2}-(1-x)^{2}\frac{\mathbf{\Delta}_{\perp}^{2}}{4}\Big)+\mathcal{F}_{b}(x,x)\Big]-\frac{1}{2}\Big(\mathcal{F}_{c}(x,x)+\mathcal{F}_{d}(x,x)\Big)\nonumber\\
     &\times i\varepsilon^{ij}_{\perp}\mathbf{k}_{\perp}^{i}\mathbf{\Delta}_{\perp}^{j}\Big{\}} \exp\Big[-2\tilde{a}(x)\Big(\mathbf{k}_{\perp}^{2}+(1-x)^{2}\frac{\mathbf{\Delta}_{\perp}^{2}}{4}\Big)\Big]\,,
     \end{align}
  \begin{align}
      \mathcal{W}_{L\mathcal{T}}^{(L)}(x,\mathbf{k}_{\perp},\mathbf{b}_{\perp})=&\frac{1}{2}\Big[\mathcal{W}_{\uparrow\uparrow}^{L}(x,\mathbf{k}_{\perp},\mathbf{b}_{\perp})-\mathcal{W}_{\downarrow\downarrow}^{L}(x,\mathbf{k}_{\perp},\mathbf{b}_{\perp})\Big]\nonumber\\
    =&\int\frac{{\rm d}^{2}\mathbf{\Delta}_{\perp}}{(2\pi)^{2}}e^{-i\mathbf{\Delta}_{\perp}\cdot\mathbf{b}_{\perp}}\frac{2N_{g}^{2}}{\pi\kappa^2}\Big{\{}\Big[\mathcal{F}_{i}(x,x)\Big(\mathbf{k}_{\perp}^{2}-(1-x)^{2}\frac{\mathbf{\Delta}_{\perp}^{2}}{4}\Big)+\mathcal{F}_{b}(x,x)\Big]+\frac{1}{2}\Big(\mathcal{F}_{c}(x,x)+\mathcal{F}_{d}(x,x)\Big)\nonumber\\
     & \times i\varepsilon^{ij}_{\perp}\mathbf{k}_{\perp}^{i}\mathbf{\Delta}_{\perp}^{j}\Big{\}} \exp\Big[-2\tilde{a}(x)\Big(\mathbf{k}_{\perp}^{2}+(1-x)^{2}\frac{\mathbf{\Delta}_{\perp}^{2}}{4}\Big)\Big]\,.
\end{align}
\begin{itemize}
    \item The Wigner distributions for the unpolarized, longitudinally polarized, and linearly polarized ($L$ and $R$) gluons inside a transversely polarized proton
\end{itemize}
\begin{align}
    \mathcal{W}^{x}_{TU}(x,\mathbf{k}_{\perp},\mathbf{b}_{\perp})=&\frac{1}{2}\Big[\mathcal{W}^{g}(x,\mathbf{k}_{\perp},\mathbf{b}_{\perp},\hat{e}_{x})-\mathcal{W}^{g}(x,\mathbf{k}_{\perp},\mathbf{b}_{\perp},-\hat{e}_{x})\Big]\nonumber\\
    =&\int\frac{{\rm d}^{2}\mathbf{\Delta}_{\perp}}{(2\pi)^{2}}e^{-i\mathbf{\Delta}_{\perp}\cdot\mathbf{b}_{\perp}}\frac{N_{g}^{2}}{\pi\kappa^2}\Big(-i\Delta_{\perp}^{(2)}\mathcal{F}_{f}(x,x)\Big)\exp\Big[-2\tilde{a}(x)\Big(\mathbf{k}_{\perp}^{2}+(1-x)^{2}\frac{\mathbf{\Delta}_{\perp}^{2}}{4}\Big)\Big]\,,\\  
    \mathcal{W}^{x}_{TL}(x,\mathbf{k}_{\perp},\mathbf{b}_{\perp})=&\frac{1}{2}\Big[\widetilde{\mathcal{W}}^{g}(x,\mathbf{k}_{\perp},\mathbf{b}_{\perp},\hat{e}_{x})-\widetilde{\mathcal{W}}^{g}(x,\mathbf{k}_{\perp},\mathbf{b}_{\perp},-\hat{e}_{x})\Big]\nonumber\\
    =&\int\frac{{\rm d}^{2}\mathbf{\Delta}_{\perp}}{(2\pi)^{2}}e^{-i\mathbf{\Delta}_{\perp}\cdot\mathbf{b}_{\perp}}\frac{N_{g}^{2}}{\pi\kappa^2}\Big(-2 k_{\perp}^{(1)}\mathcal{F}_{k}(x,x)\Big)\exp\Big[-2\tilde{a}(x)\Big(\mathbf{k}_{\perp}^{2}+(1-x)^{2}\frac{\mathbf{\Delta}_{\perp}^{2}}{4}\Big)\Big]\,,\\
    \mathcal{W}^{(R)x}_{T\mathcal{T}}(x,\mathbf{k}_{\perp},\mathbf{b}_{\perp})=&\frac{1}{2}\Big[\mathcal{W}^{R}(x,\mathbf{k}_{\perp},\mathbf{b}_{\perp},\hat{e}_{x})-\mathcal{W}^{R}(x,\mathbf{k}_{\perp},\mathbf{b}_{\perp},-\hat{e}_{x})\Big]\nonumber\\
    =&\int\frac{{\rm d}^{2}\mathbf{\Delta}_{\perp}}{(2\pi)^{2}}e^{-i\mathbf{\Delta}_{\perp}\cdot\mathbf{b}_{\perp}}\frac{N_{g}^{2}}{\pi\kappa^2}\Big(2 k_{\perp}^{(1)}\mathcal{F}_{k}(x,x)-i\Delta_{\perp}^{(2)}\mathcal{F}_{f}(x,x)\Big)\exp\Big[-2\tilde{a}(x)\Big(\mathbf{k}_{\perp}^{2}+(1-x)^{2}\frac{\mathbf{\Delta}_{\perp}^{2}}{4}\Big)\Big]\,,\\
    \mathcal{W}^{(L)x}_{T\mathcal{T}}(x,\mathbf{k}_{\perp},\mathbf{b}_{\perp})=&\frac{1}{2}\Big[\mathcal{W}^{L}(x,\mathbf{k}_{\perp},\mathbf{b}_{\perp},\hat{e}_{x})-\mathcal{W}^{L}(x,\mathbf{k}_{\perp},\mathbf{b}_{\perp},-\hat{e}_{x})\Big]\nonumber\\
    =&-\int\frac{{\rm d}^{2}\mathbf{\Delta}_{\perp}}{(2\pi)^{2}}e^{-i\mathbf{\Delta}_{\perp}\cdot\mathbf{b}_{\perp}}\frac{N_{g}^{2}}{\pi\kappa^2}\Big(2 k_{\perp}^{(1)}\mathcal{F}_{k}(x,x)+i\Delta_{\perp}^{(2)}\mathcal{F}_{f}(x,x)\Big)\exp\Big[-2\tilde{a}(x)\Big(\mathbf{k}_{\perp}^{2}+(1-x)^{2}\frac{\mathbf{\Delta}_{\perp}^{2}}{4}\Big)\Big]\,.
\end{align} 
where $\mathcal{W}^{g}_{\uparrow\uparrow}$ and $\mathcal{W}^{g}_{\downarrow\downarrow}$ are the Wigner distributions corresponding target polarizations along $\hat{\mathbf{e}}_{z}$ and $-\hat{\mathbf{e}}_{z}$, respectively. While $\hat{\mathbf{e}}_{x}$ and $-\hat{\mathbf{e}}_{x}$ denotes the transverse polarization of the target state along the $x$-axis and these can be expressed as a linear combination of helicity states, $|\pm\hat{\mathbf{e}}_{x}\rangle=\frac{1}{\sqrt{2}}(|\frac{1}{2}\rangle\pm\frac{1}{2}\rangle)$. The first index of the Wigner distributions, $\mathcal{W}_{\Lambda_{p}\Lambda_{g}}$, is corresponding to the target polarization ($\Lambda_{p}$) and the second index is for the gluon polarization ($\Lambda_{g}$). In Ref.~\cite{More:2017zqq}, six of the above gluon Wigner distributions for linearly polarized gluons ($L$ and $R$) were rewritten as a linear combination of unpolarized and longitudinally polarized gluons; {{these relations are }}  also valid in this model  and given as follows,
\begin{align}
   \mathcal{W}_{U\mathcal{T}}^{R}(x,\mathbf{k}_{\perp},\mathbf{b}_{\perp})=& \mathcal{W}_{UU}(x,\mathbf{k}_{\perp},\mathbf{b}_{\perp})-W_{UL}(x,\mathbf{k}_{\perp},\mathbf{b}_{\perp})\,,\\
   \mathcal{W}_{U\mathcal{T}}^{L}(x,\mathbf{k}_{\perp},\mathbf{b}_{\perp})=& \mathcal{W}_{UU}(x,\mathbf{k}_{\perp},\mathbf{b}_{\perp})+W_{UL}(x,\mathbf{k}_{\perp},\mathbf{b}_{\perp})\,,\\
   \mathcal{W}_{L\mathcal{T}}^{R}(x,\mathbf{k}_{\perp},\mathbf{b}_{\perp})=& -\mathcal{W}_{LL}(x,\mathbf{k}_{\perp},\mathbf{b}_{\perp})+W_{LU}(x,\mathbf{k}_{\perp},\mathbf{b}_{\perp})\,,\\
   \mathcal{W}_{L\mathcal{T}}^{L}(x,\mathbf{k}_{\perp},\mathbf{b}_{\perp})=& \mathcal{W}_{LL}(x,\mathbf{k}_{\perp},\mathbf{b}_{\perp})+W_{LU}(x,\mathbf{k}_{\perp},\mathbf{b}_{\perp})\,,\\
   \mathcal{W}_{T\mathcal{T}}^{(R)x}(x,\mathbf{k}_{\perp},\mathbf{b}_{\perp})=& -\mathcal{W}_{TL}^{x}(x,\mathbf{k}_{\perp},\mathbf{b}_{\perp})+W_{TU}^{x}(x,\mathbf{k}_{\perp},\mathbf{b}_{\perp})\,,\\
   \mathcal{W}_{T\mathcal{T}}^{(L)x}(x,\mathbf{k}_{\perp},\mathbf{b}_{\perp})=& \mathcal{W}_{TL}^{x}(x,\mathbf{k}_{\perp},\mathbf{b}_{\perp})+W_{TU}^{x}(x,\mathbf{k}_{\perp},\mathbf{b}_{\perp})\,.
\end{align}
There are some additional model-independent relations between the Wigner distributions and the Fourier transformed GTMDs~\cite{Lorce:2011kd}. {{We have verified that these are satisfied in our model. These relations are given below :}}
\begin{align}
   \mathcal{W}_{UU}(x,\mathbf{k}_{\perp},\mathbf{b}_{\perp})=& \mathcal{F}_{1,1}^{g}(x,\mathbf{k}_{\perp},\mathbf{b}_{\perp})\,,\\
   \mathcal{W}_{LU}(x,\mathbf{k}_{\perp},\mathbf{b}_{\perp})=& -\frac{1}{M^2}\varepsilon_{\perp}^{ij}k_{\perp}^{i}\frac{\partial}{\partial b_{\perp}^{j}}\mathcal{F}_{1,4}^{g}(x,\mathbf{k}_{\perp},\mathbf{b}_{\perp})\,,\\
   \mathcal{W}_{UL}(x,\mathbf{k}_{\perp},\mathbf{b}_{\perp})=& \frac{1}{M^2}\varepsilon_{\perp}^{ij}k_{\perp}^{i}\frac{\partial}{\partial b_{\perp}^{j}}\mathcal{G}_{1,1}^{g}(x,\mathbf{k}_{\perp},\mathbf{b}_{\perp})\,,\label{eq:para_W_{LU}}\\
   \mathcal{W}_{LL}(x,\mathbf{k}_{\perp},\mathbf{b}_{\perp})=& \mathcal{G}_{1,4}^{g}(x,\mathbf{k}_{\perp},\mathbf{b}_{\perp})\,.
\end{align}
where $\mathcal{F}_{1,1}^{g},~\mathcal{F}_{1,4}^{g},~\mathcal{G}_{1,1}^{g}$ and $\mathcal{G}_{1,4}^{g}$ represents the Fourier transform of the corresponding GTMDs ${F}_{1,1}^{g},~{F}_{1,4}^{g},~{G}_{1,1}^{g}$ and ${G}_{1,4}^{g}$.

{{We define the spin-spin correlation function of a longitudinally polarized gluon with helicity $\Lambda_{g}$ inside a longitudinally polarized proton with helicity $\Lambda_{p}$ as $\rho_{\Lambda_{p}\Lambda_{g}}^{g}(x,\bfk,\bfb)$. It is obtained by weighting the corresponding Wigner distributions with the gluon and proton helicities and is expressed as follows~\cite{Lorce:2011kd}:
\begin{align}\label{eq:spin-spin correlation}
    \rho_{\Lambda_{p}\Lambda_{g}}(x,\bfk,\bfb)=\frac{1}{2}\Big[\mathcal{W}_{UU}(x,\bfk,\bfb)+\Lambda_{g}\mathcal{W}_{UL}(x,\bfk,\bfb)+\Lambda_{p}\mathcal{W}_{LU}(x,\bfk,\bfb)+\Lambda_{g}\Lambda_{p}\mathcal{W}_{LL}(x,\bfk,\bfb)\Big]\,.
\end{align}
where $\mathcal{W}_{UU}$ denotes the Wigner distribution of unpolarized gluons inside an unpolarized proton, while $\mathcal{W}_{UL}$, $\mathcal{W}_{LU}$, and $\mathcal{W}_{LL}$ correspond to the Wigner distributions involving longitudinal polarization of the gluon, the proton, or both, respectively. The gluon and proton helicities are corresponding to $\{\Lambda_{g},\Lambda_{p}\}=\uparrow,\downarrow$ (where $\uparrow$ and $\downarrow$ are corresponds to $+1$ and $-1$ for longitudinal polarizations respectively). }}

\section{Numerical Results and Discussion}\label{sec:results and discussion}
In this section, we present the numerical results for gluon GTMDs and Wigner distributions {{calculated in the spectator model for the gluon discussed above. The GTMDs are calculated analytically using overlaps of LFWFs in the DGLAP region
($ x > \xi $). The analytic model results are rather lengthy and are given in the appendix. The numerical values of the model parameters are obtained by}}  fitting the unpolarized gluon PDF from NNPDF dataset at the initial model scale of $Q_{0}=2~\text{GeV}$. These are  given by  $a=3.880 \pm 0.223$, $b=-0.530 \pm 0.007$ and $\kappa=2.62  \pm 0.135$~\cite{Chakrabarti:2023djs,Sain:2025kup}.

\subsection[A]{Unpolarized gluon GTMDs}

\begin{figure}
    \centering
    \includegraphics[width=0.245\linewidth]{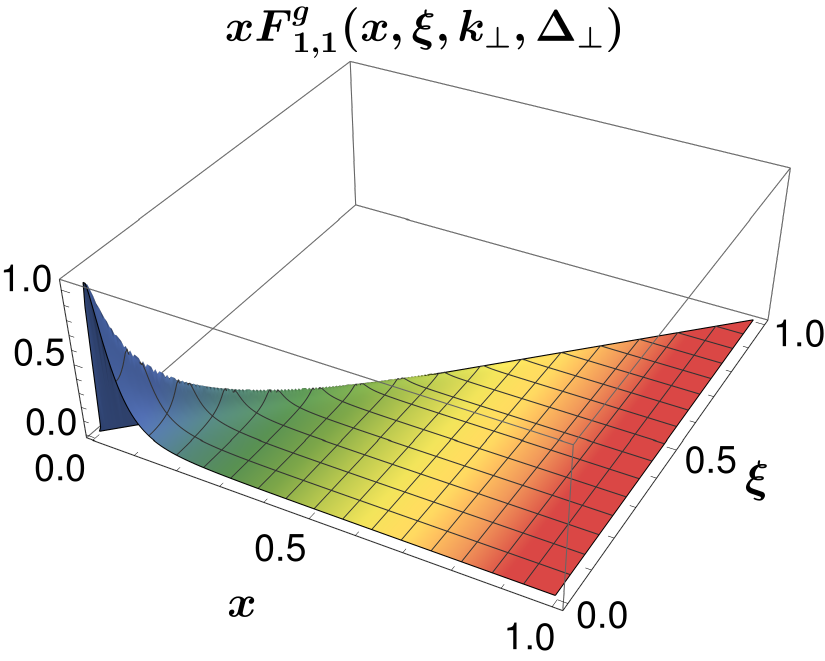}
    \includegraphics[width=0.245\linewidth]{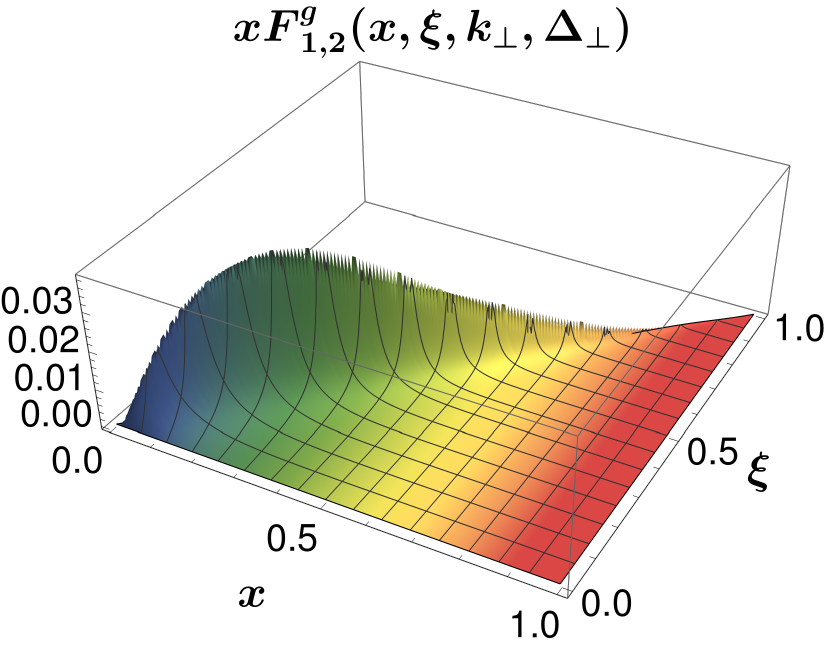}
    \includegraphics[width=0.245\linewidth]{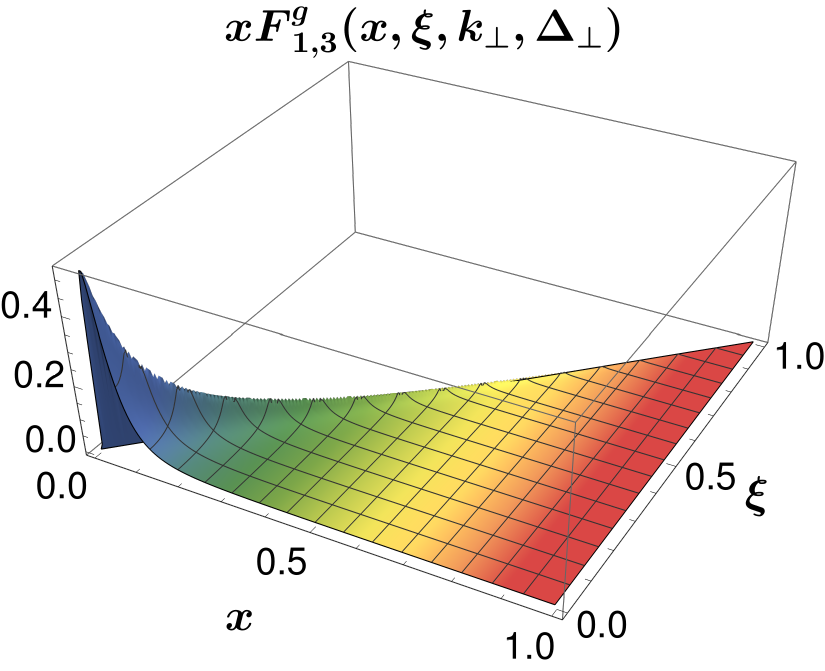}
    \includegraphics[width=0.245\linewidth]{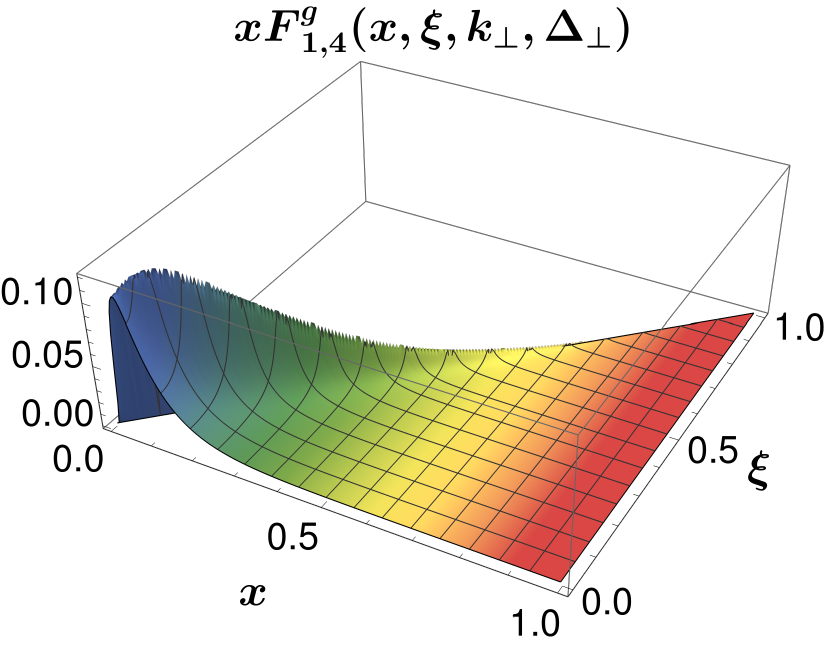}\\
    \includegraphics[width=0.245\linewidth]{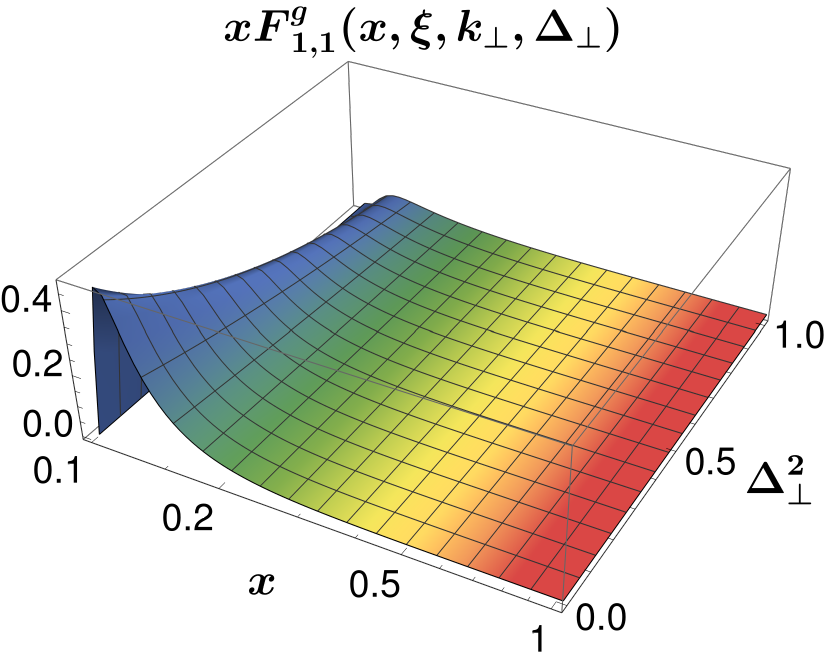}
    \includegraphics[width=0.245\linewidth]{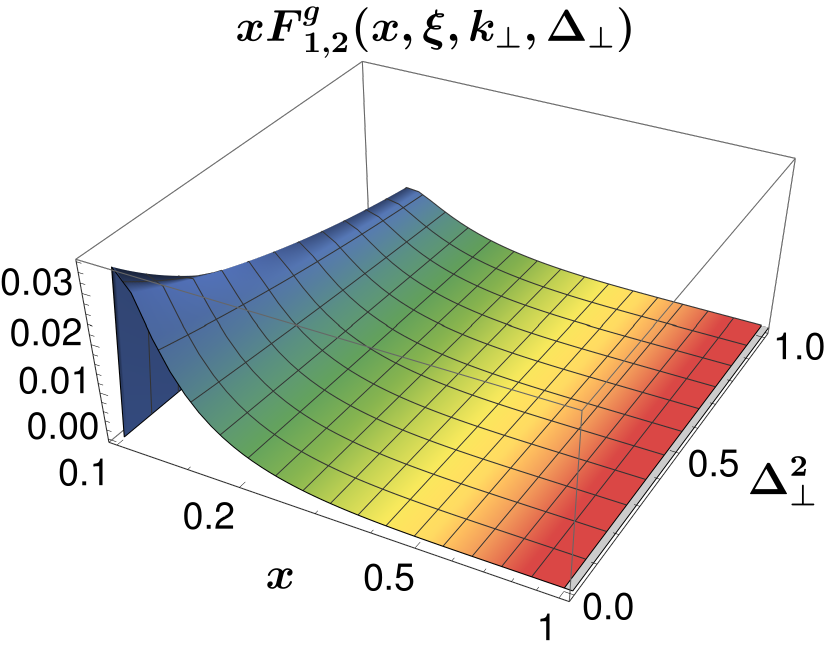}
    \includegraphics[width=0.245\linewidth]{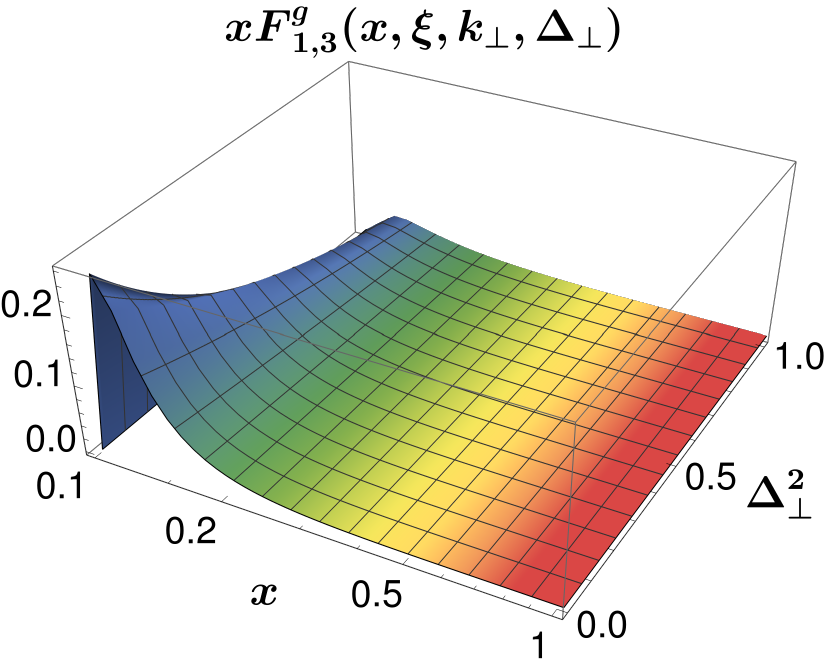}
    \includegraphics[width=0.245\linewidth]{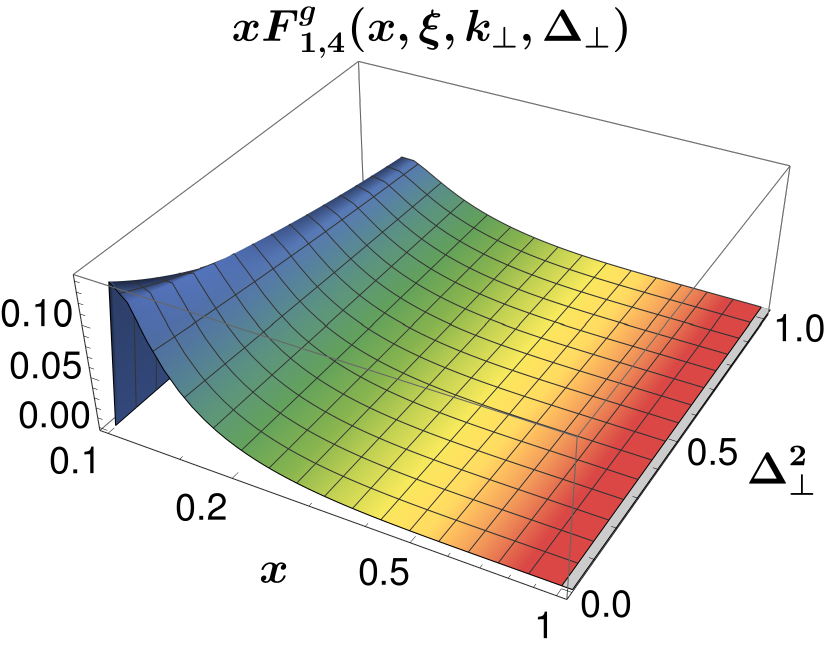}
    \caption{
    \justifying
    Upper panel: Three-dimensional distributions of the four $F$-type gluon GTMDs: $F_{1,1}^{g},~F_{1,2}^{g},~F_{1,3}^{g}$ and $F_{1,4}^{g}$ as functions of the gluon longitudinal momentum fraction $x$ and the skewness parameter $\xi$, evaluated at fixed transverse momentum $k_{\perp}^{2}=0.3~\mathrm{GeV}^2$, transverse momentum transfer $\Delta_{\perp}^{2}=0.2~\mathrm{GeV}^2$, and $k_{\perp}\cdot\Delta_{\perp}=0$.
Lower panel: Same as the upper panel but plotted as functions of $x$ and $\Delta_{\perp}^{2}$, with fixed $k_{\perp}^{2}=0.3~\mathrm{GeV}^2$ and $\xi=0.1$.}
    \label{fig:GTMDs_x_xi}
\end{figure}

In Figure~\ref{fig:GTMDs_x_xi} we present the model results for the gluon GTMDs within the DGLAP region, $x>\xi$. Specially, the upper panel depicts the four $F$-type gluon GTMDs, $F_{1,1}^{g},~F_{1,2}^{g},~F_{1,3}^{g}$ and $F_{1,4}^{g}$ as a function of longitudinal momentum fraction $x$, and the longitudinal momentum transfer fraction, $\xi$ at fixed values of transverse momentum, $k_{\perp}^{2}=0.3~\text{GeV}^2$ as well as transverse momentum transfer, $\Delta_{\perp}^{2}=0.2~\text{GeV}^2$ at $\bfk$ being perpendicular to $\mathbf{\Delta}_{\perp}$, i.e., $\bfk\cdot\mathbf{\Delta}_{\perp}=0$. The GTMDs have their peaks around small $x$; the peaks shift toward higher values of $x$ {and the magnitude decreases} as the momentum transfer increases in the longitudinal direction. Similarly, the lower panel shows the variation of above four $F$-type GTMDs with $x$ and $\Delta_{\perp}^{2}$ for fixed values of skewness variable $\xi=0.1$ and $k_{\perp}^{2}=0.3~\text{GeV}^2$. We found that the overall shape of all plots shows the same profile and peaks around small $x$ and the magnitude decrease on increasing the transverse momentum transfer $\Delta^{2}_{\perp}$. Similarly, on increasing the longitudinal momentum fraction $x$, the magnitude of distributions decreases at a larger rate than $\Delta^{2}_{\perp}$. We found that all four unpolarized gluon GTMDs are positive in $x-\xi$ and $x-\Delta^{2}_{\perp}$ planes, respectively. In Ref.~\cite{Tan:2024dmz}, the authors computed the gluon GTMDs at nonzero skewness within the DGLAP region. They found that $F_{1,1}^{g}$ and $F_{1,4}^{g}$ exhibit positive distributions,  $F_{1,2}^{g}$ remains negative, while $F_{1,3}^{g}$ is positive in the small-$x$ region but turns negative at larger-$x$. Within the light-cone gauge, $A^{+}=0$, in the forward limit at zero skewness the GTMD $F_{1,1}^{g}$ is related to the unpolarized gluon TMD $f_{1}^{g}(x,\bfk^2)$, while within the GPD limit (on doing the $\bfk$ integration at zero skewness), it project out to the unpolarized gluon GPD $H^{g}(x,t)$. On the other hand the GTMD $F_{1,4}^{g}$ is related to the gluon canonical OAM in the forward limit, $\Delta_{\perp}=0$ at zero skewness variable $\xi=0$ and given as follows~\cite{Lorce:2011kd},
\begin{align}\label{eq:canonical_OAM}
    \ell_{z}^g=-\int {\rm d}x{\rm d}^{2}\bfk\frac{\bfk^2}{M^2}F^{g}_{1,4}(x,\xi=0,\bfk, \mathbf{\Delta}_{\perp}=0)\,.
\end{align}
\begin{figure}
    \centering
    \includegraphics[width=0.245\linewidth]{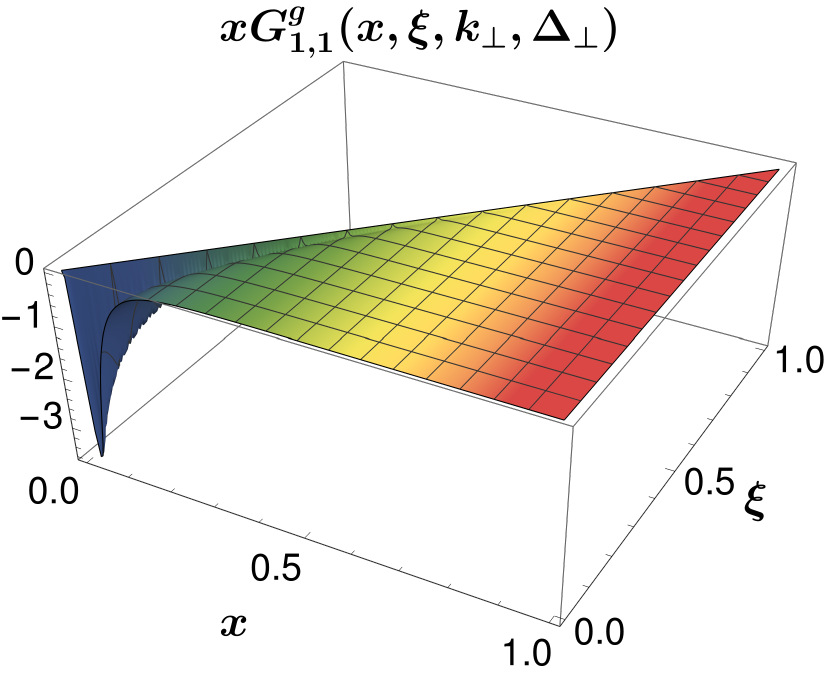}
    \includegraphics[width=0.245\linewidth]{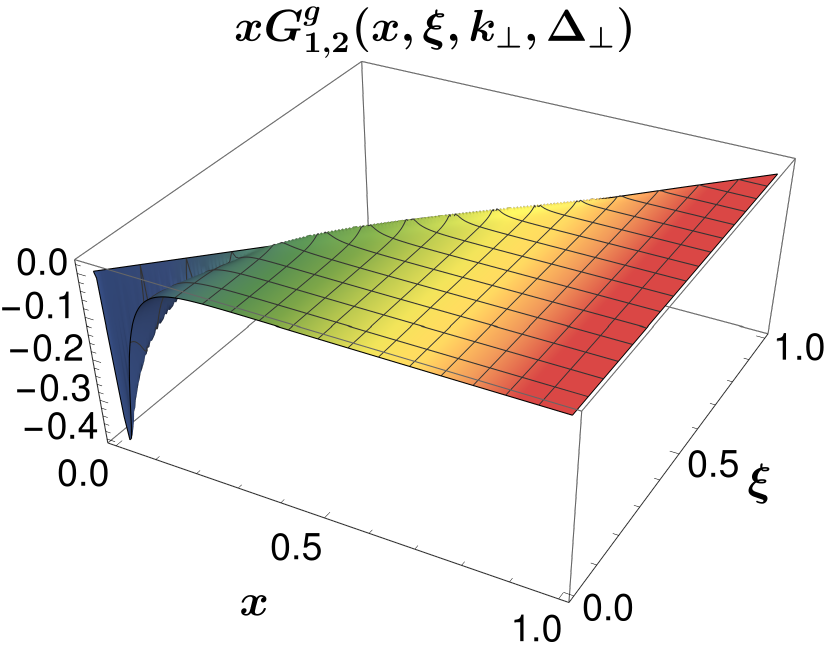}
    \includegraphics[width=0.245\linewidth]{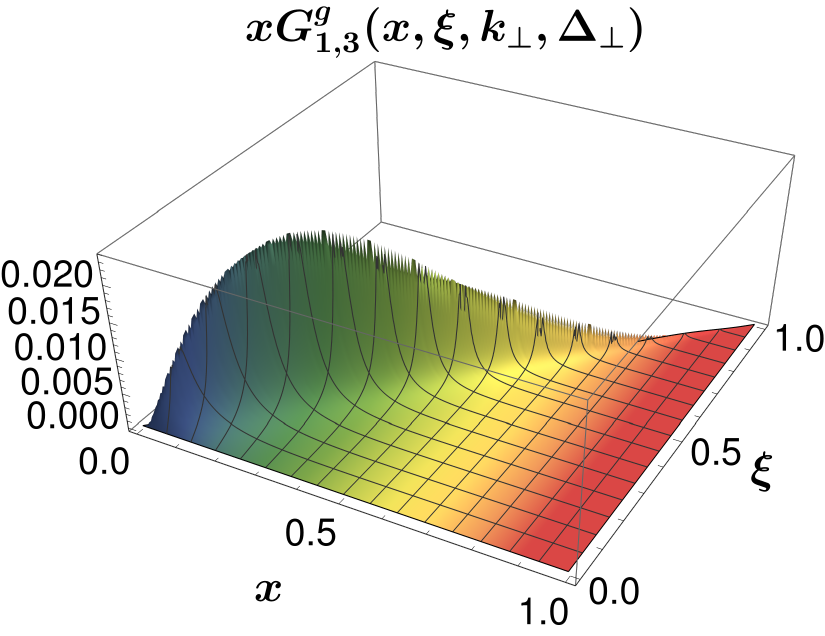}
    \includegraphics[width=0.245\linewidth]{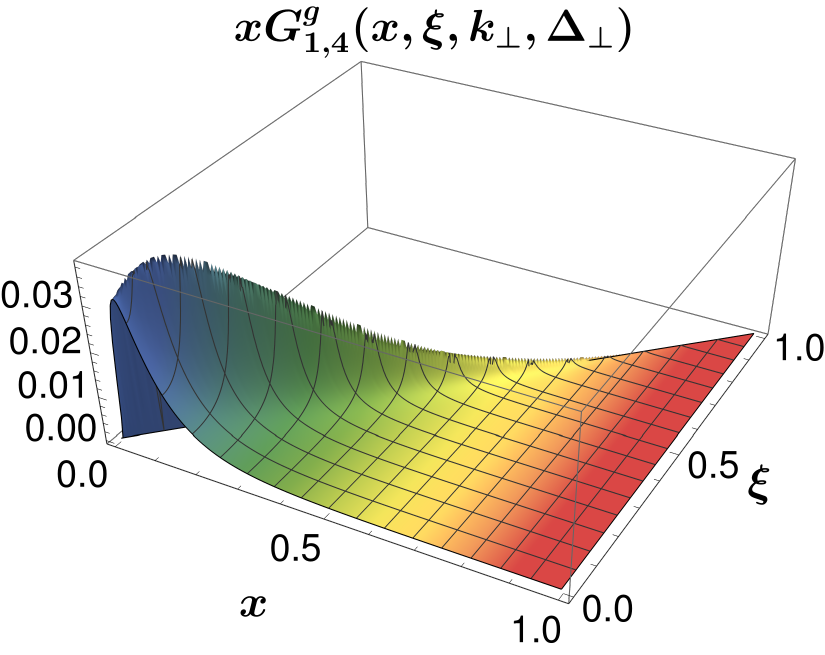}\\
    \includegraphics[width=0.245\linewidth]{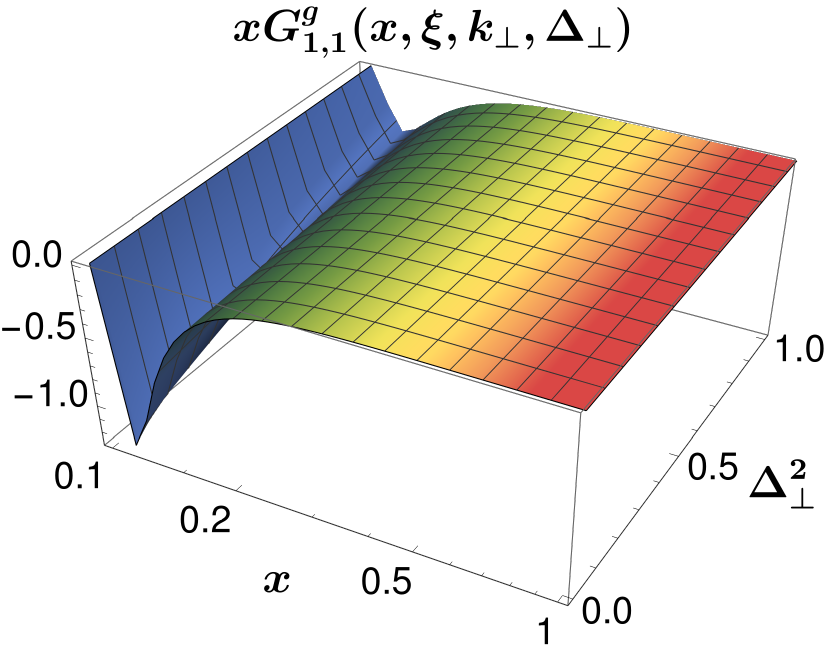}
    \includegraphics[width=0.245\linewidth]{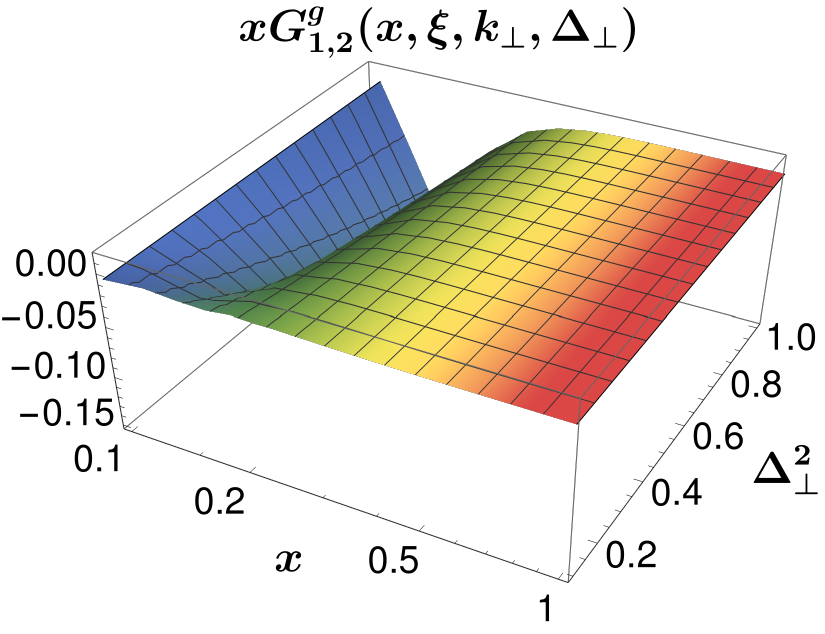}
    \includegraphics[width=0.245\linewidth]{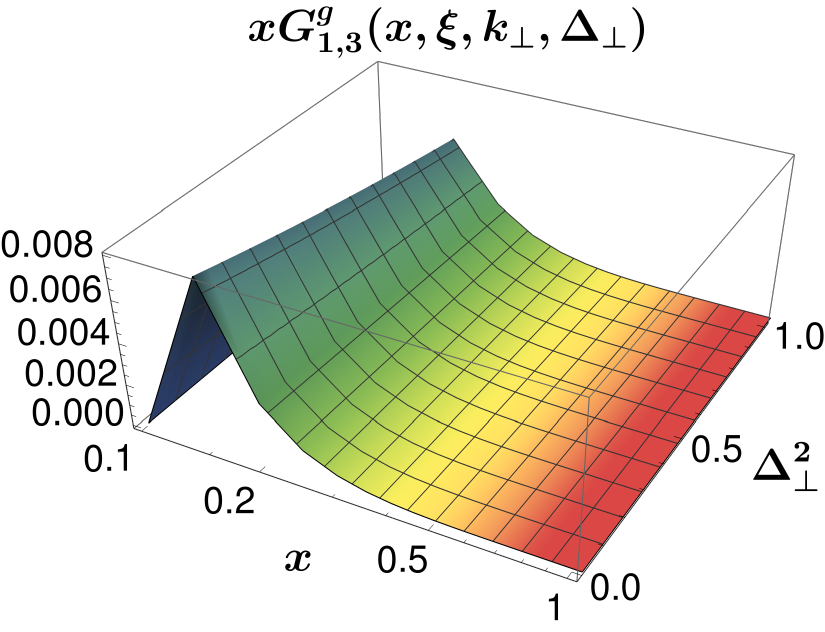}
    \includegraphics[width=0.245\linewidth]{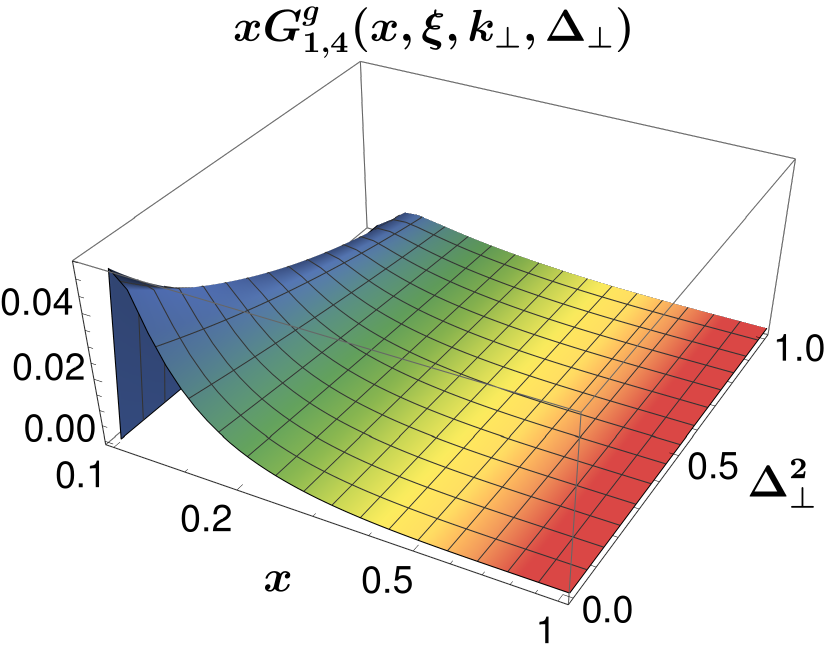}
    \caption{
    \justifying
    Upper panel: Three-dimensional distributions of the four $G$-type gluon GTMDs: $G_{1,1}^{g},~G_{1,2}^{g},~G_{1,3}^{g}$ and $G_{1,4}^{g}$ as functions of the gluon longitudinal momentum fraction $x$ and the skewness parameter $\xi$, evaluated at fixed transverse momentum $k_{\perp}^{2}=0.3~\mathrm{GeV}^2$, transverse momentum transfer $\Delta_{\perp}^{2}=0.2~\mathrm{GeV}^2$, and $k_{\perp}\cdot\Delta_{\perp}=0$.
Lower panel: Same distributions plotted as functions of $x$ and $\Delta_{\perp}^{2}$, with fixed $k_{\perp}^{2}=0.3~\mathrm{GeV}^2$ and $\xi=0.1$.}
    \label{fig:GTMDs_G_type}
\end{figure}
\begin{figure}
    \centering
    \includegraphics[width=0.325\linewidth]{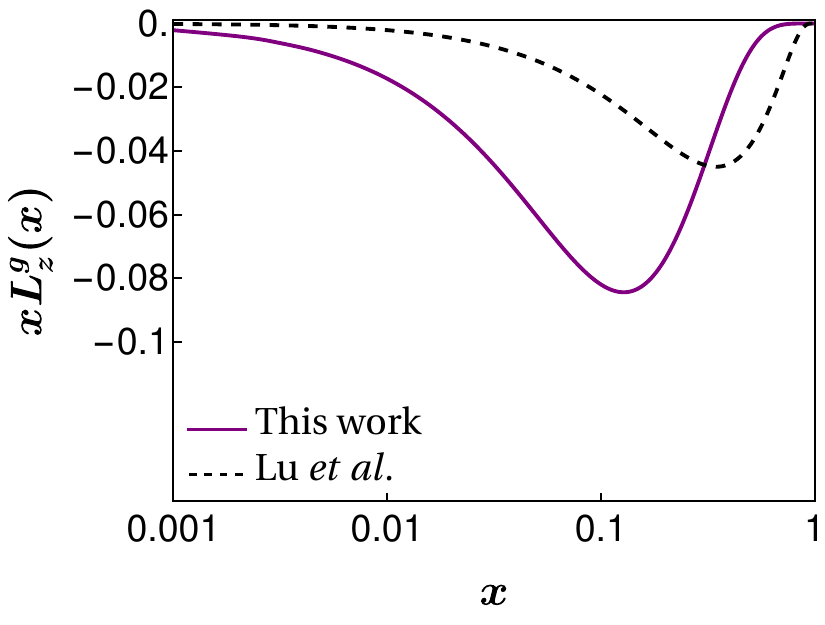}\quad
    \includegraphics[width=0.325\linewidth]{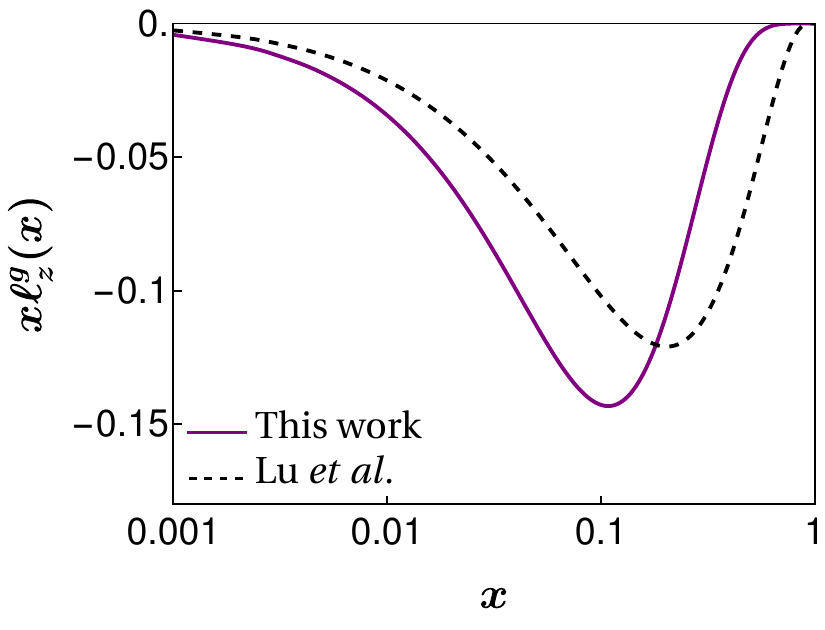}\quad
    \includegraphics[width=0.305\linewidth]{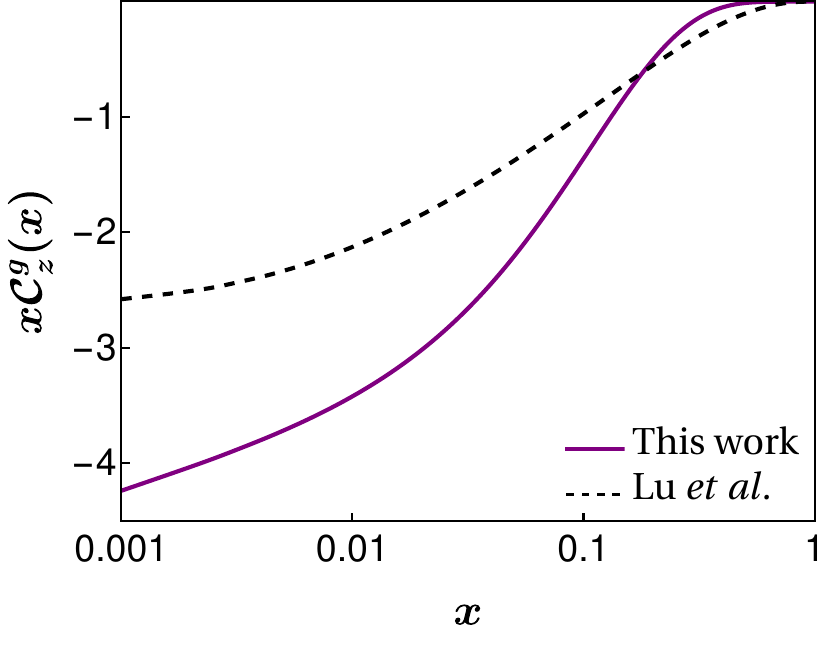}
    \caption{
    \justifying
    Our model predictions for the gluon kinetic OAM $L_{z}^{g}$ (left panel), canonical OAM $\ell_{z}^{g}$ (middle panel) and the gluon spin-orbit correlation factor $\mathcal{C}_{z}^{g}$ (right panel) along with those reported in Refs.~\cite{Tan:2023kbl,Tan:2023vvi} with an another spectator model by Lu \textit{et al.} as a functions of gluon longitudinal momentum fraction $x$.}
    \label{fig:OAM_SO_correlations}
\end{figure}

\subsection[B]{Longitudinally polarized gluon GTMDs}

In Figure~\ref{fig:GTMDs_G_type}, we present the model results for the four $G$-type gluon GTMDs $G_{1,1}^{g},~G_{1,2}^{g},~G_{1,3}^{g}$ and $G_{1,4}^{g}$. The upper panel presents the variation of those four gluon GTMDs with $x$ and $\xi$ for fixed $k_{\perp}^2=0.3~\text{GeV}^{2}$ and $\Delta_{\perp}^{2}=0.2~\text{GeV}^{2}$. Again those GTMDs are also peaked around small $x$ and the peak decreases towards the larger values of $x$. Similarly, the peak also decreases on increasing the skewness variable $\xi$ but with a lesser rate than $x$. The lower panel shows the variation of those four GTMDs with $x$ and $\Delta_{\perp}^{2}$. The peak is around small $x$ and decreases with increasing $x$; similarly, the peak also decreases while we increase the value of $\Delta_{\perp}^{2}$, with a lower rate as compared to when we increase the $x$. We found that the GTMDs $G_{1,1}^{g}$ and $G_{1,2}^{g}$ are negative, while the GTMDs $G_{1,3}^{g}$ and $G_{1,4}^{g}$ are positive in  both $x-\xi$ and $x-\Delta^{2}_{\perp}$ planes, respectively. In another gluon spectator model by Lu \textit{et al.}~\cite{Tan:2024dmz}, the longitudinally polarized gluon GTMDs exhibit the following behavior: $G_{1,1}^{g}$ and $G_{1,2}^{g}$ remain negative, $G_{1,4}^{g}$ is positive, while $G_{1,3}^{g}$ is negative in the small-$x$ region but turns positive at larger values of $x$. Within the light-cone gauge and forward limit at zero skewness the GTMD $G_{1,4}^{g}$ related to the gluon helicity TMD, $g_{1}^{g}(x,\bfk^2)$, while within the GPD limit, it project out to the longitudinally polarized gluon GPD, $\widetilde{H}_{g}(x,t)$. On the other hand the GTMD $G_{1,1}^{g}$ is related to the  gluon spin-orbit correlations factor as follows~\cite{Lorce:2011kd},
\begin{align}\label{eq:SO_correlation}
    \mathcal{C}_{z}^{g}=&\int{\rm d}x{\rm d}^{2}\bfk{\rm d}^{2}\bfb(\bfb\times\bfk)_{z}\mathcal{W}_{UL}^{g}(x,\bfk,\bfb)\nonumber\\
    =&\int{\rm d}x{\rm d}^{2}\bfk\frac{\bfk^{2}}{M^2}G_{1,1}^{g}(x,\xi=0,\bfk,\mathbf{\Delta}_{\perp}=0)
\end{align}
In the left panel of Figure~\ref{fig:OAM_SO_correlations}, we show the variation of the gluon kinetic OAM with the momentum fraction $x$ and compare it with predictions from another gluon spectator model by Lu \textit{et al.}, where the scalar part of LFWFs is modified using the  Brodsky–Huang–Lepage prescription. We observe that the gluon kinetic OAM in our model exhibits a broader distribution with a more pronounced peak compared to the spectator model of Lu \textit{et al.}~\cite{Tan:2023kbl}. Similarly, in the middle panel, we present the model prediction for $x$-dependence of gluon canonical OAM along with the gluon spectator model of Lu \textit{et al.}~\cite{Tan:2023vvi}, and found that it is negatively distributed throughout the entire $x$-region. The $\ell_{z}^{g}$ describes the correlation between proton spin and gluon OAM. From Eq.~\eqref{eq:canonical_OAM}, the positively distributed GTMD $F_{1,4}^{g}$ confirms that the canonical OAM $\ell_{z}^{g}<0$, which indicates that the gluon canonical OAM and the proton spin are anti-aligned to each other. We obtain the numerical value of gluon canonical OAM, $\ell_{z}^{g}=-0.38$, which is consistent with the value reported in~\cite{Tan:2023vvi} as $\ell_{z}^{g}\simeq-0.33$. While, the right panel present the model predictions for $x$-dependence of spin-orbit correlation factor along with the gluon spectator model of Lu \textit{et al.}, and found that $\mathcal{C}_{z}^{g}<0$ for the entire range of $x$, which implies that the gluon spin and OAM are anti-aligned to each other. The numerical value of gluon spin-orbit correlation factor is found $\mathcal{C}_{z}^{g}=-15.6$, which is larger than $\mathcal{C}_{z}^{g}\simeq-10.05$ as reported in Ref.~\cite{Tan:2023vvi}. 
   \begin{figure}
        \centering
        \includegraphics[width=0.325\linewidth]{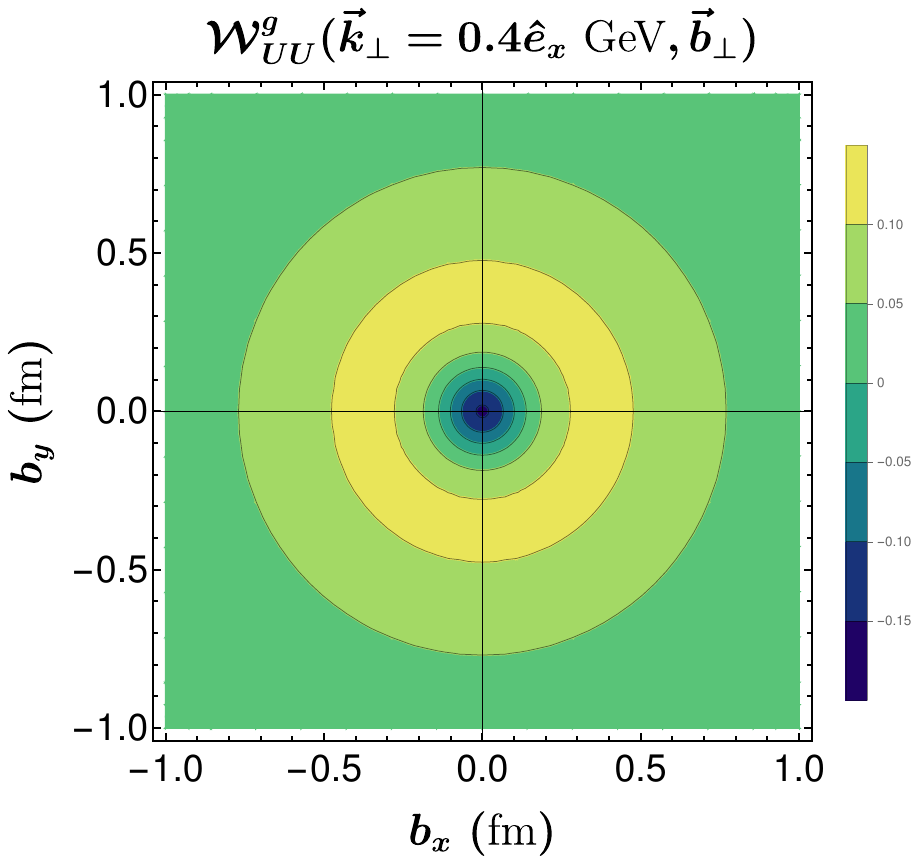}
        \includegraphics[width=0.325\linewidth]{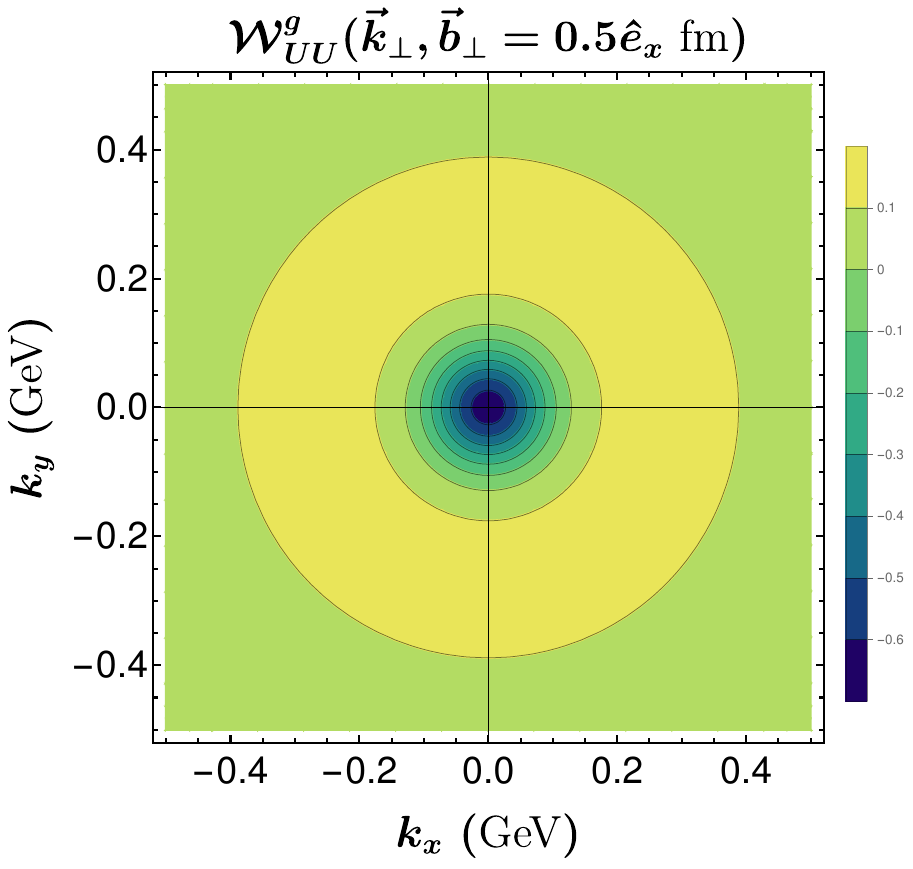}
        \includegraphics[width=0.325\linewidth]{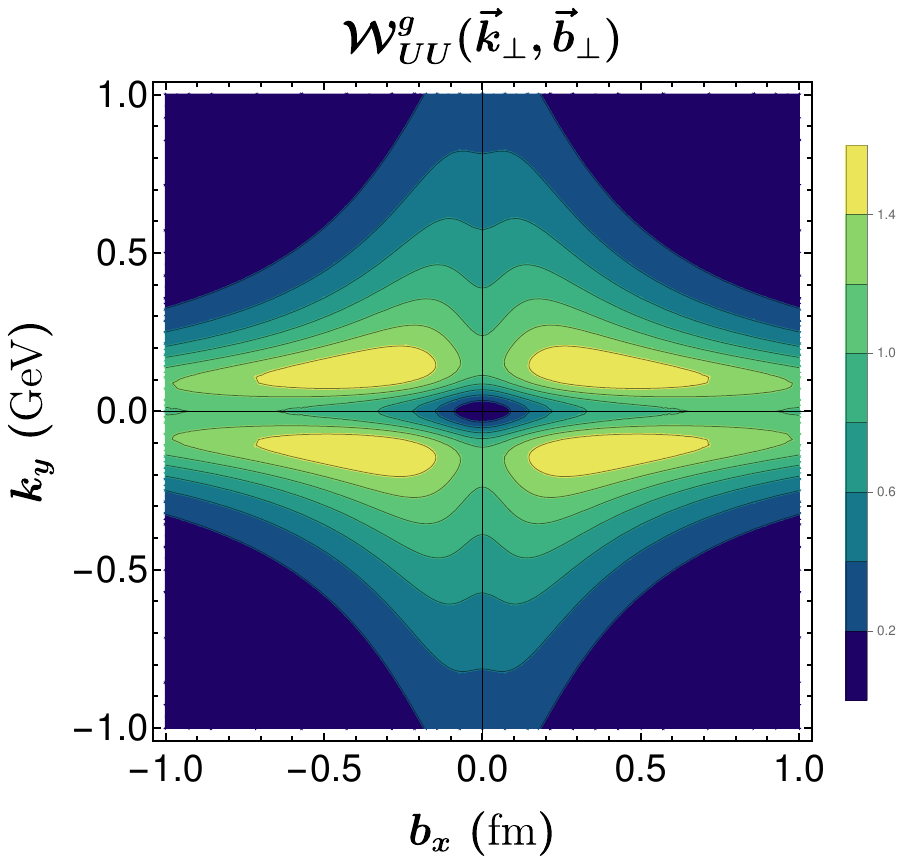}
        \includegraphics[width=0.325\linewidth]{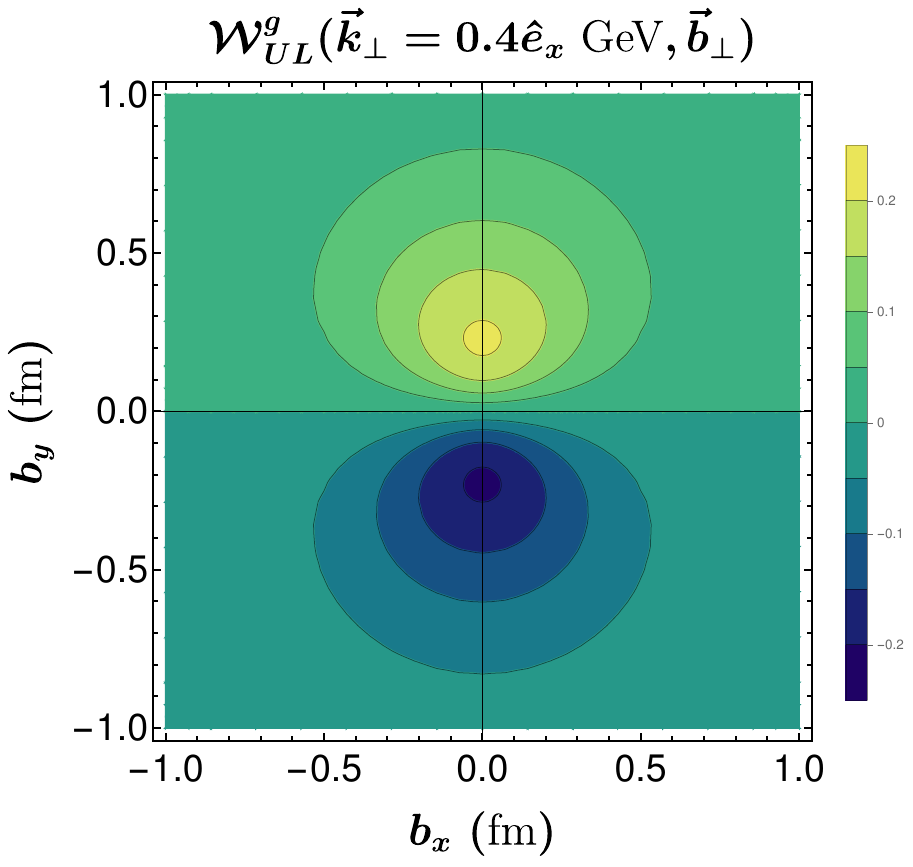}
        \includegraphics[width=0.325\linewidth]{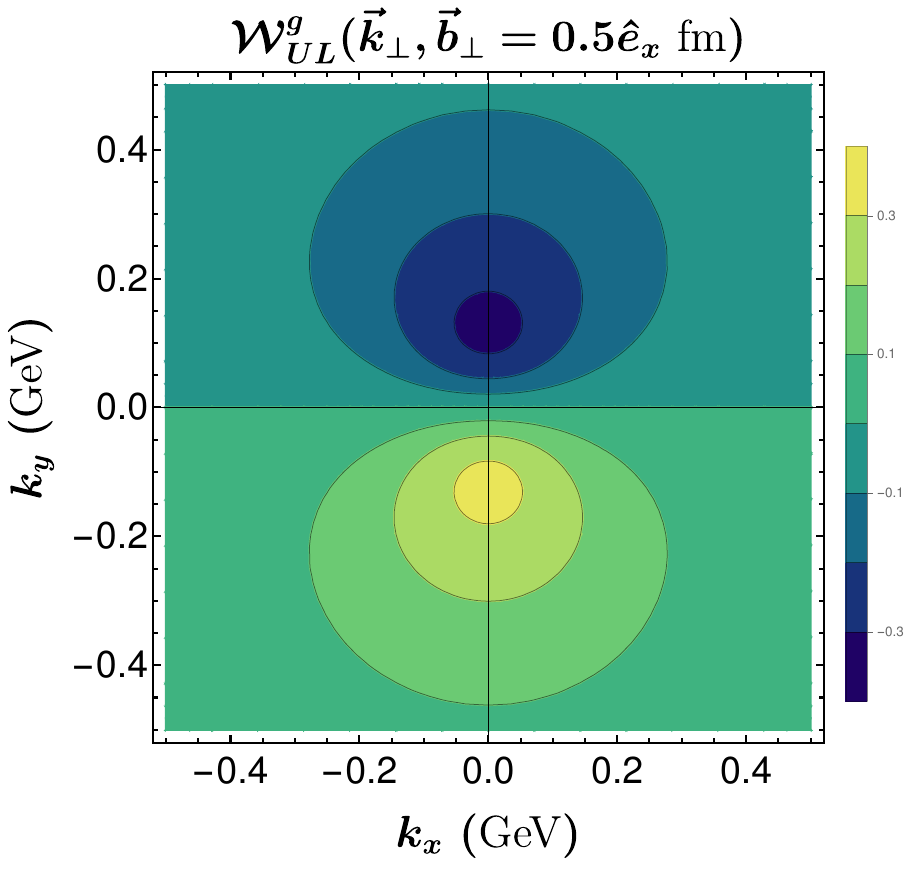}
        \includegraphics[width=0.325\linewidth]{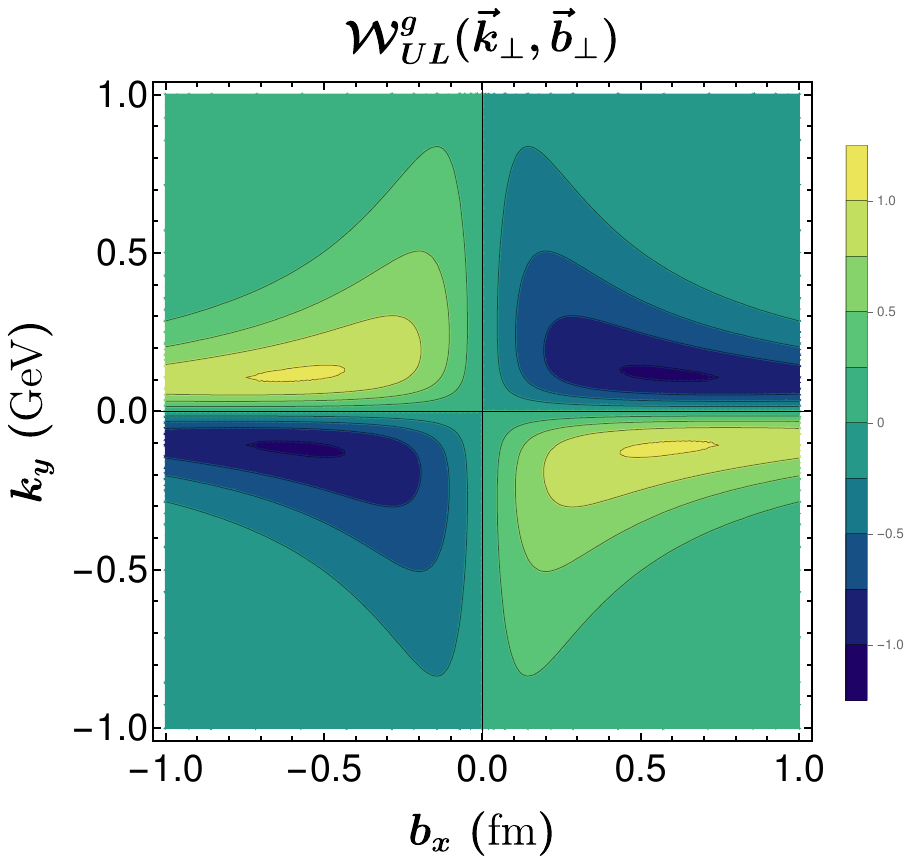}
        \caption{
        \justifying
        Upper panel: The transverse Wigner distributions of unpolarized gluons in an unpolarized proton are shown in three configurations: in the transverse impact parameter plane at fixed transverse momentum $\mathbf{k}_\perp = 0.4\hat{e}_x~\text{GeV}$ (left), in the transverse momentum plane at fixed impact parameter $\mathbf{b}_{\perp} = 0.5\hat{e}_x~\text{fm}$ (middle), and in the mixed transverse plane after integrating over $k_x$ and $b_y$ (right).
        Lower panel: Same as the upper panel, but for longitudinally polarized gluons in an unpolarized proton.}
        \label{fig:Figure4}
    \end{figure}
       \begin{figure}
        \centering
        \includegraphics[width=0.325\linewidth]{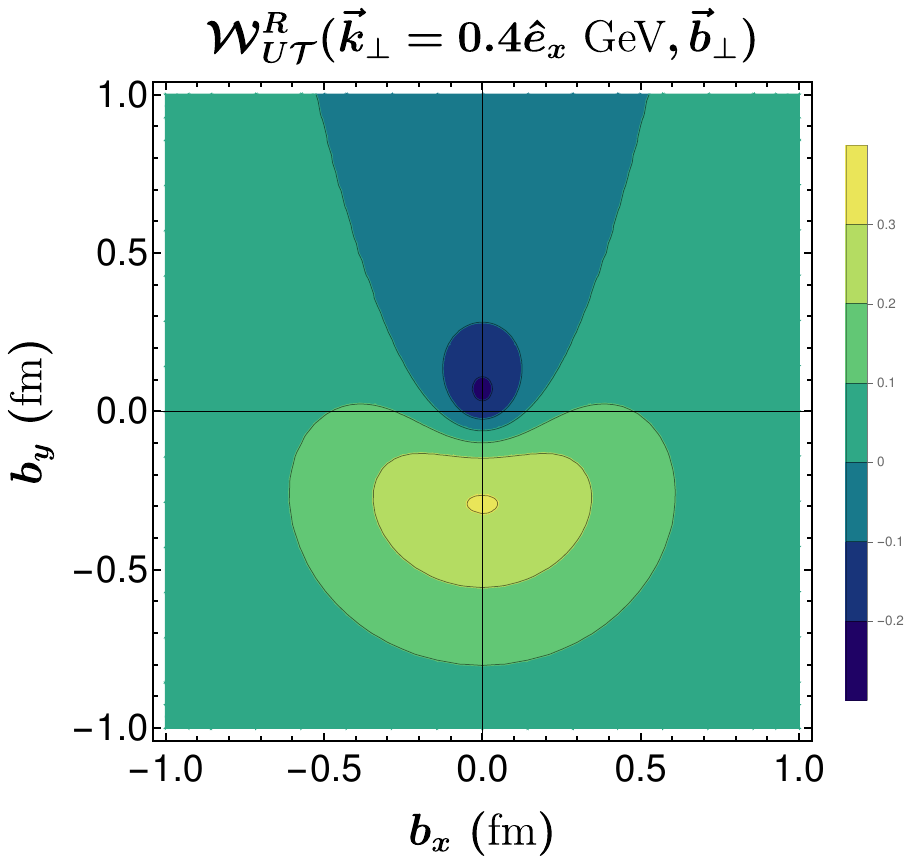}
        \includegraphics[width=0.325\linewidth]{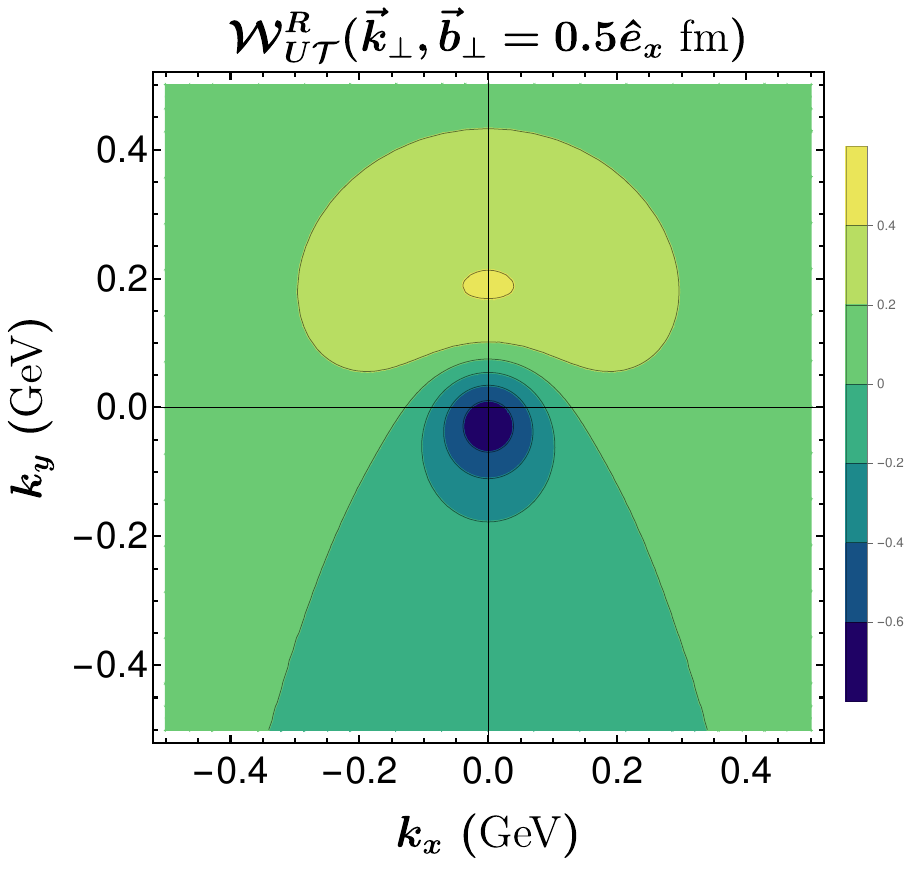}
        \includegraphics[width=0.325\linewidth]{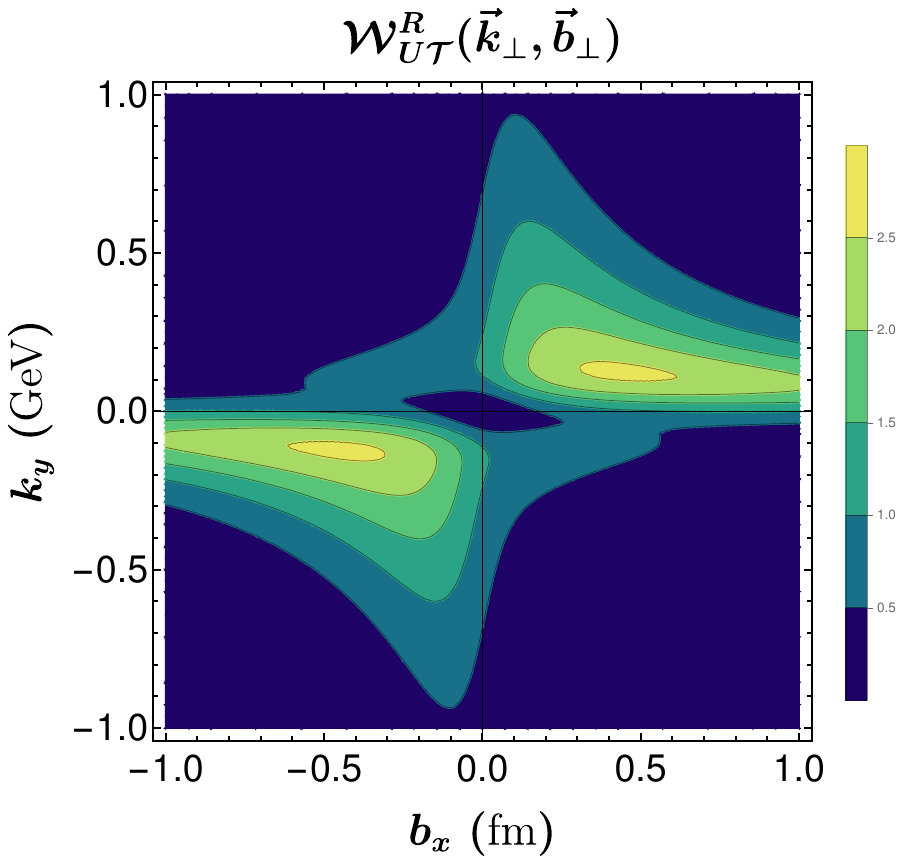}
        \includegraphics[width=0.325\linewidth]{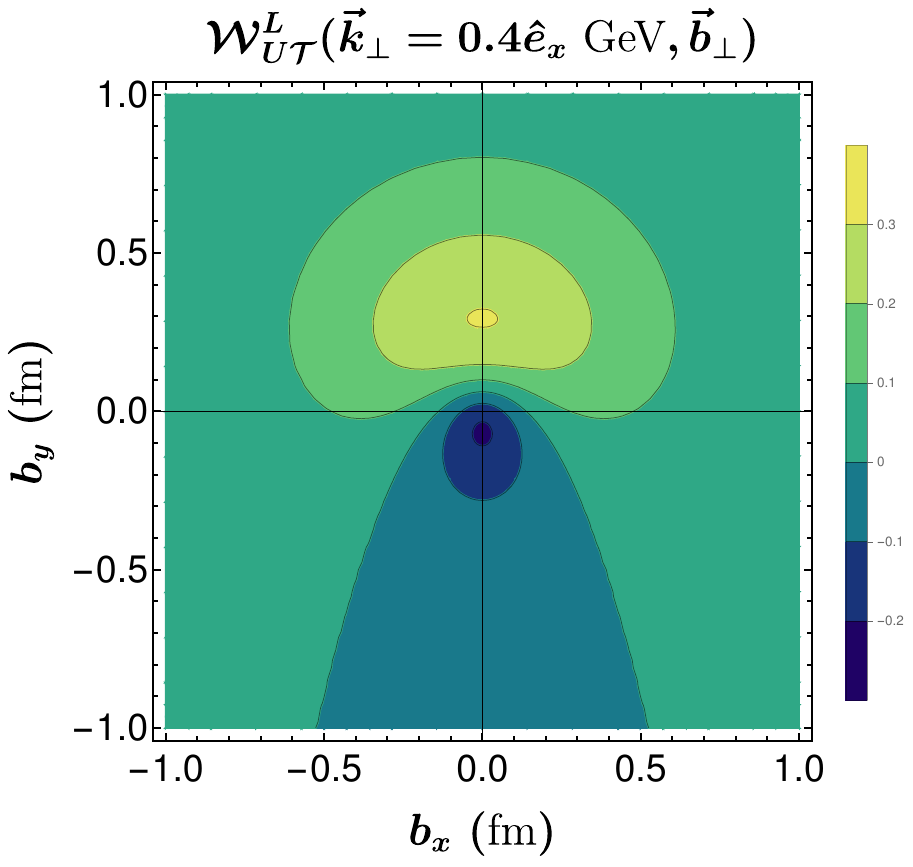}
        \includegraphics[width=0.325\linewidth]{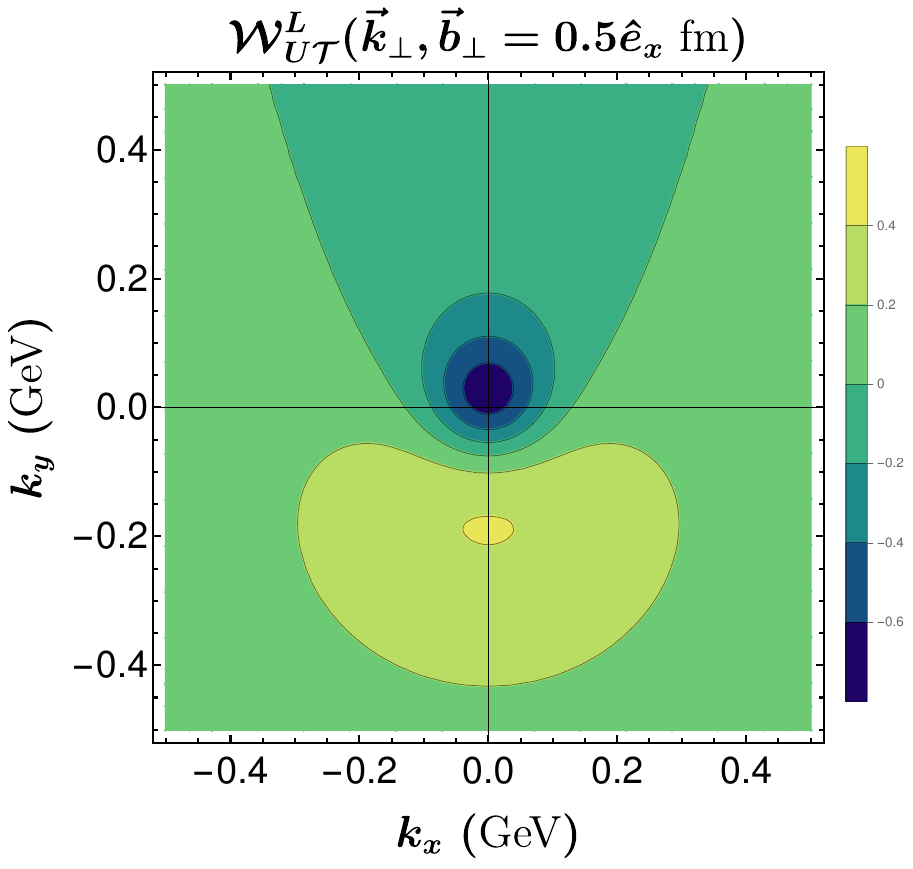}
        \includegraphics[width=0.325\linewidth]{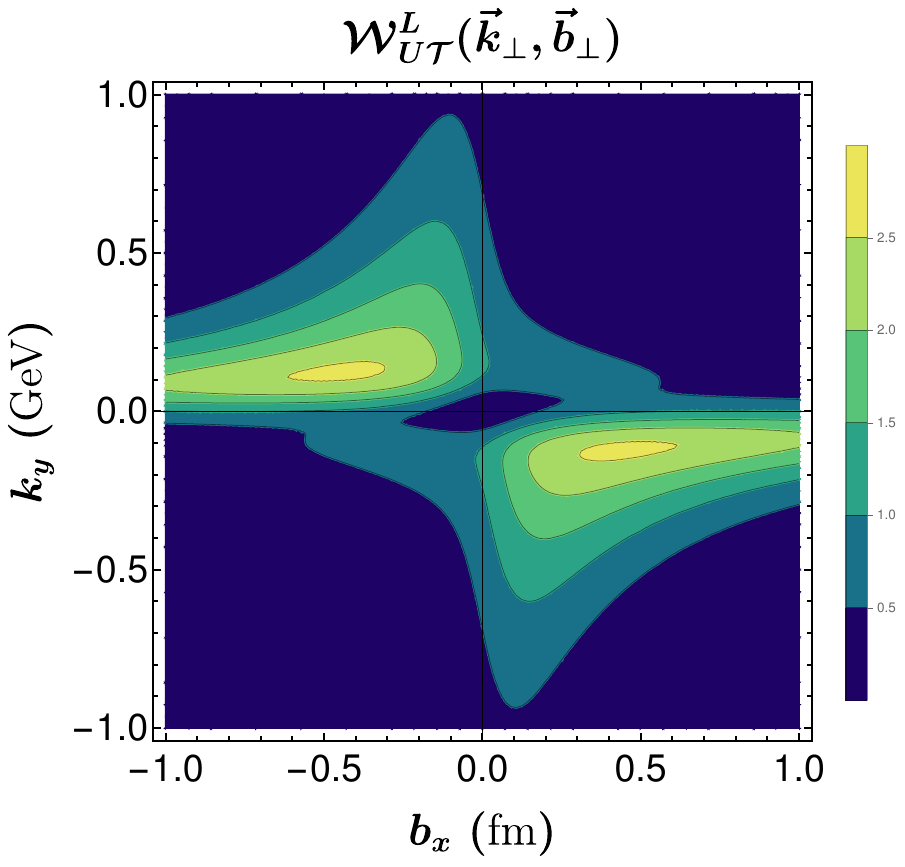}
        \caption{
        \justifying
        Upper panel: The transverse Wigner distributions of right-handed ($R$) linearly polarized gluons in an unpolarized proton are shown in three configurations: in the transverse impact parameter plane at fixed transverse momentum $\mathbf{k}_\perp = 0.4\hat{e}_x~\text{GeV}$ (left), in the transverse momentum plane at fixed impact parameter $\mathbf{b}_{\perp} = 0.5\hat{e}_x~\text{fm}$ (middle), and in the mixed transverse plane after integrating over $k_x$ and $b_y$ (right).
        Lower panel: Same as the upper panel, but for left-handed ($L$) linearly polarized gluons in an unpolarized proton.}
        \label{fig:Figure5}
        \end{figure}
        \begin{figure}
        \centering
        \includegraphics[width=0.325\linewidth]{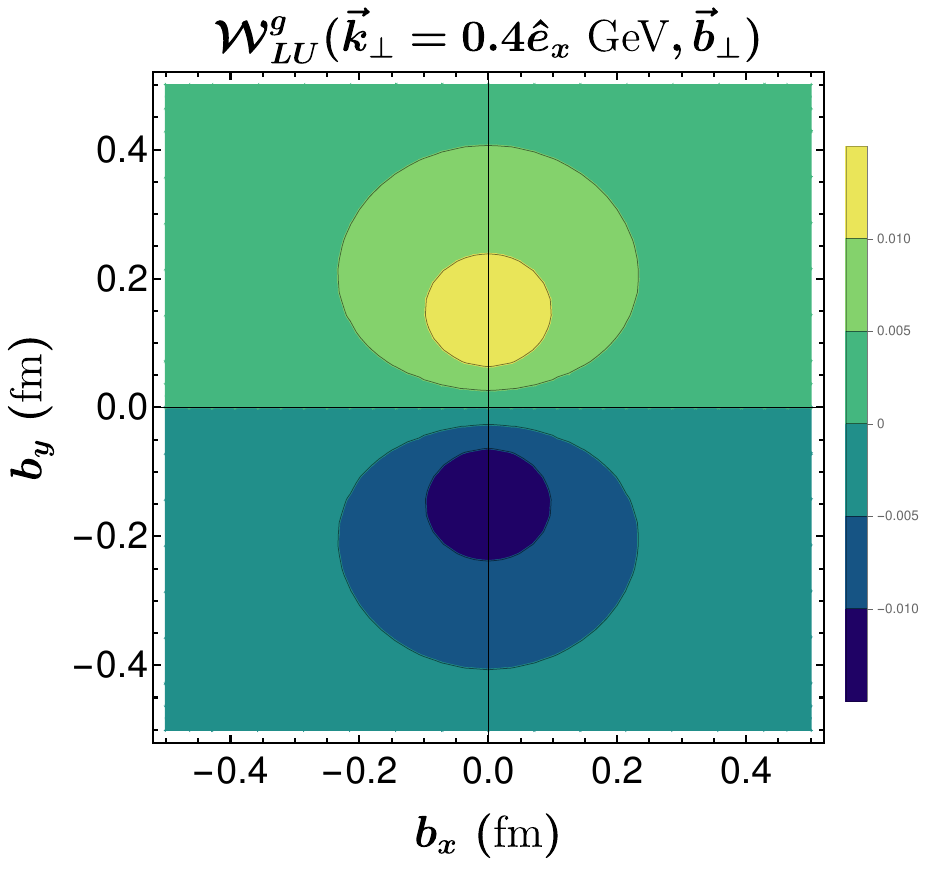}
        \includegraphics[width=0.325\linewidth]{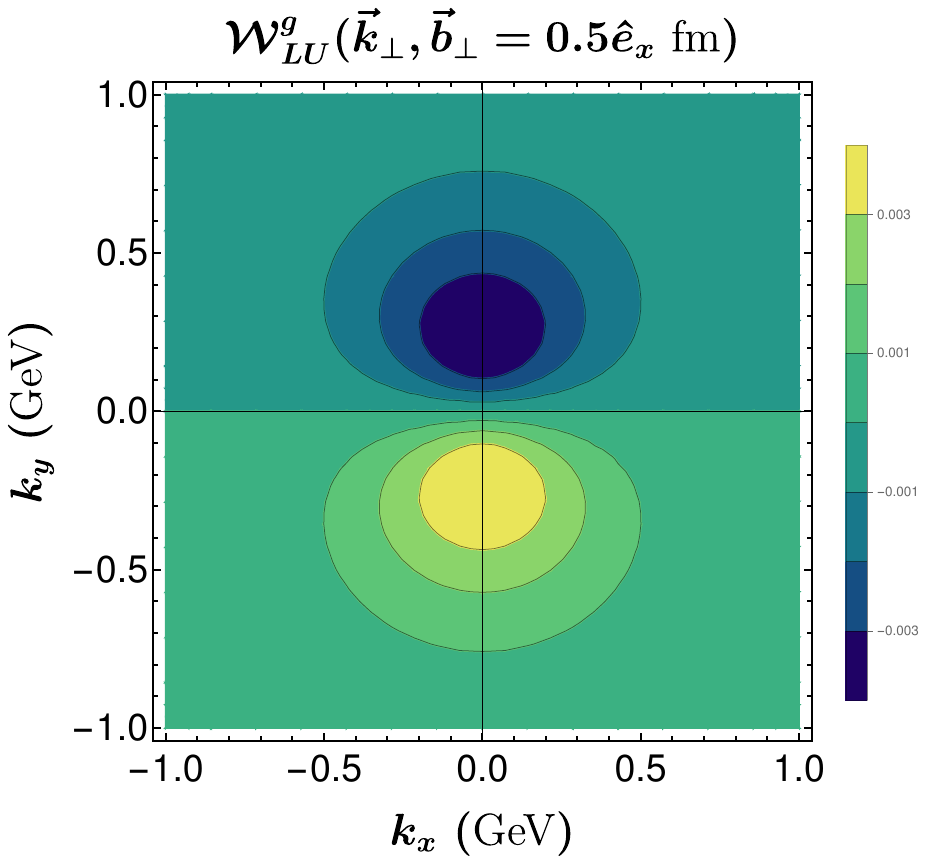}
        \includegraphics[width=0.325\linewidth]{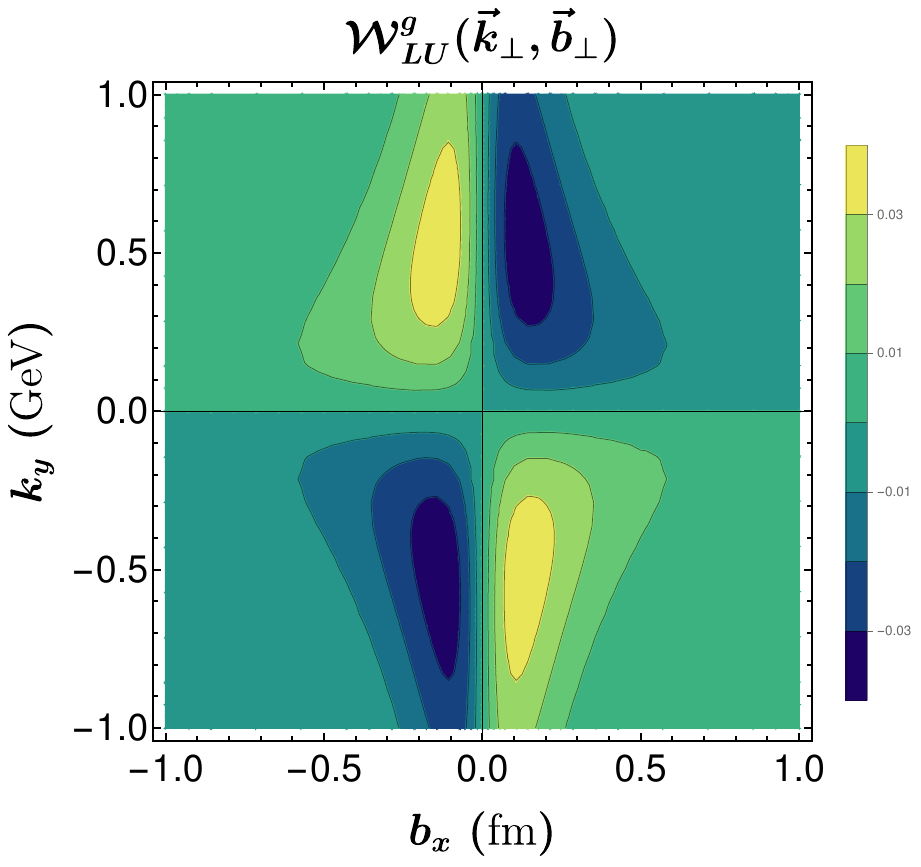}
        \includegraphics[width=0.325\linewidth]{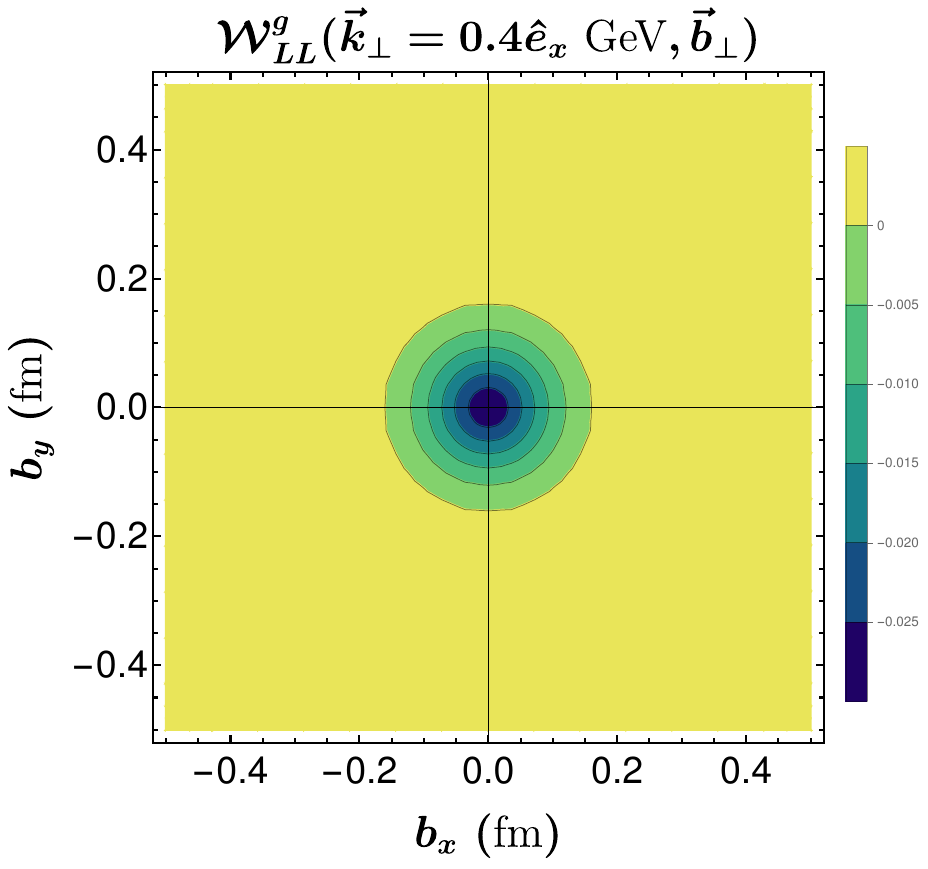}
        \includegraphics[width=0.325\linewidth]{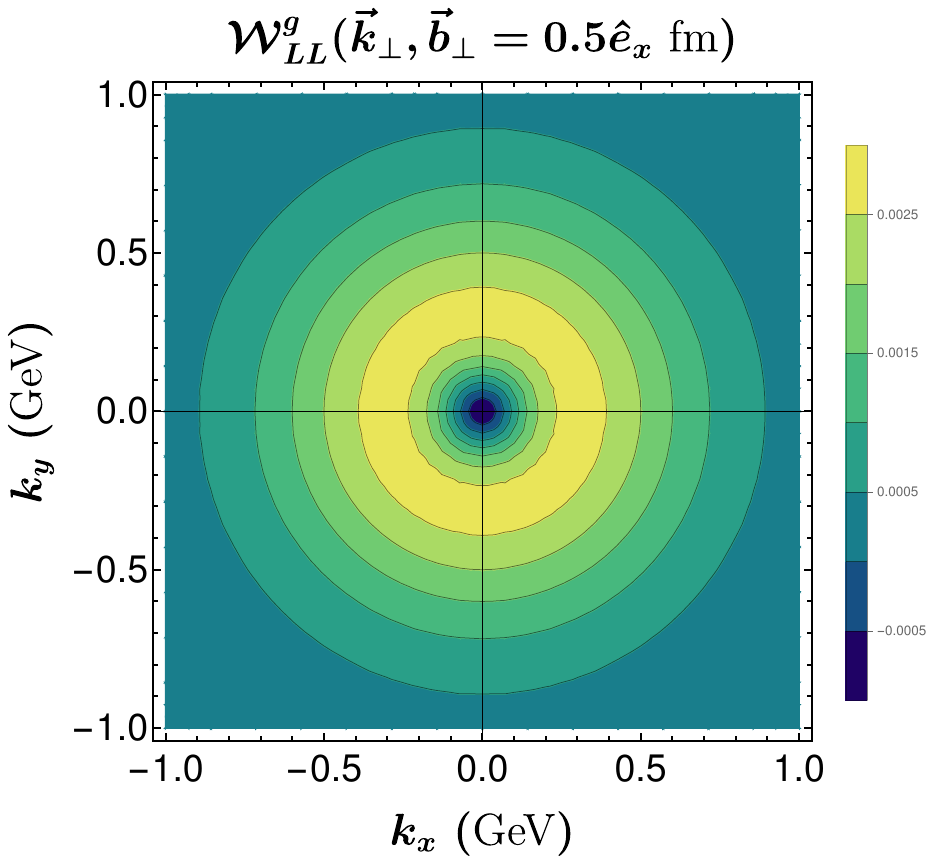}
        \includegraphics[width=0.325\linewidth]{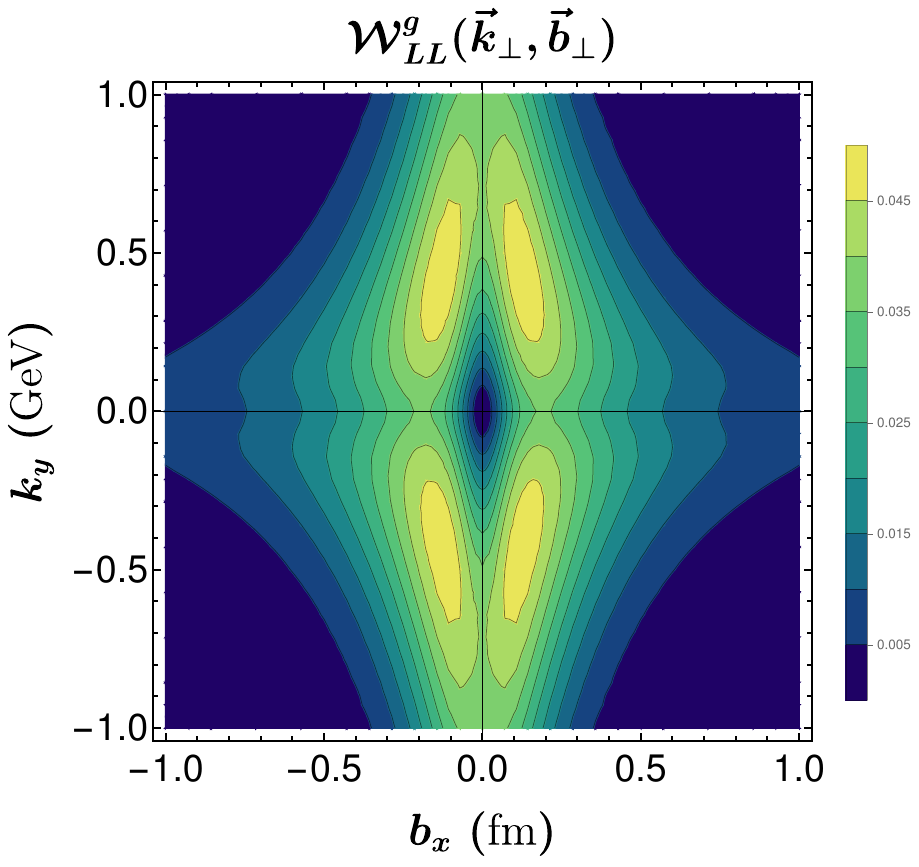}
        \caption{
        \justifying
        Upper panel: The transverse Wigner distributions of unpolarized gluons in a longitudinally polarized proton are shown in three configurations: in the transverse impact parameter plane at fixed transverse momentum $\mathbf{k}_\perp = 0.4\hat{e}_x~\text{GeV}$ (left), in the transverse momentum plane at fixed impact parameter $\mathbf{b}_{\perp} = 0.5\hat{e}_x~\text{fm}$ (middle), and in the mixed transverse plane after integrating over $k_x$ and $b_y$ (right).
        Lower panel: Same as the upper panel, but for longitudinally polarized gluons in a longitudinally polarized proton.}
    \label{fig:Figure6}
    \end{figure}
    \begin{figure}
        \centering
        \includegraphics[width=0.325\linewidth]{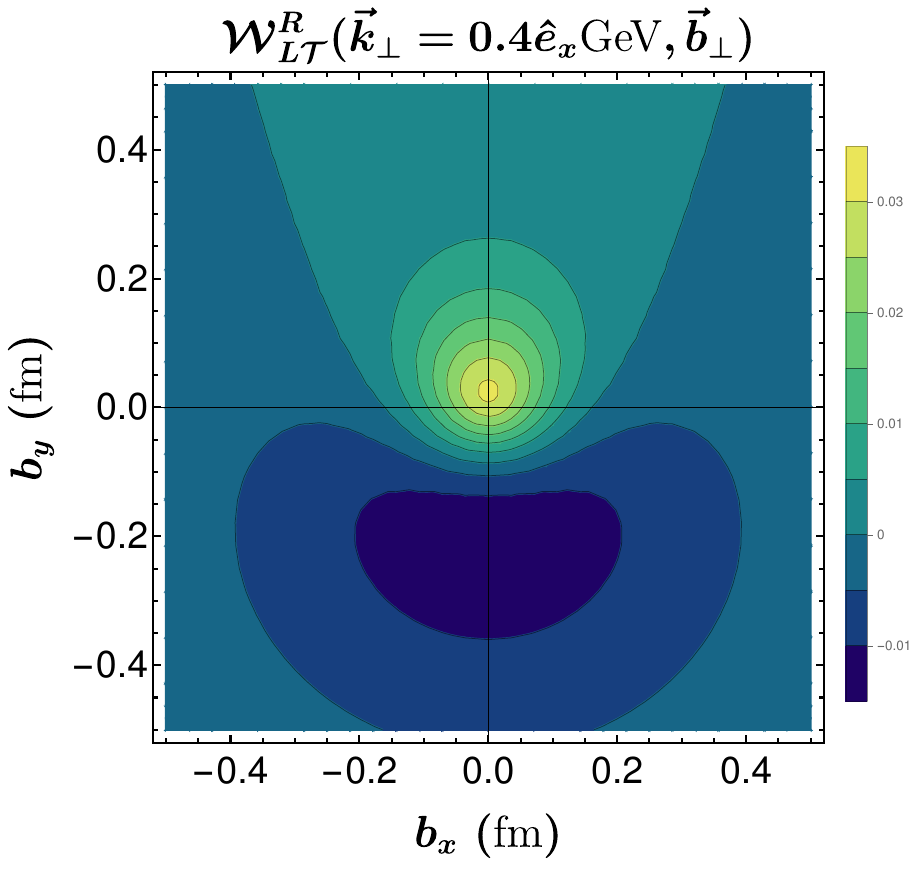}
        \includegraphics[width=0.325\linewidth]{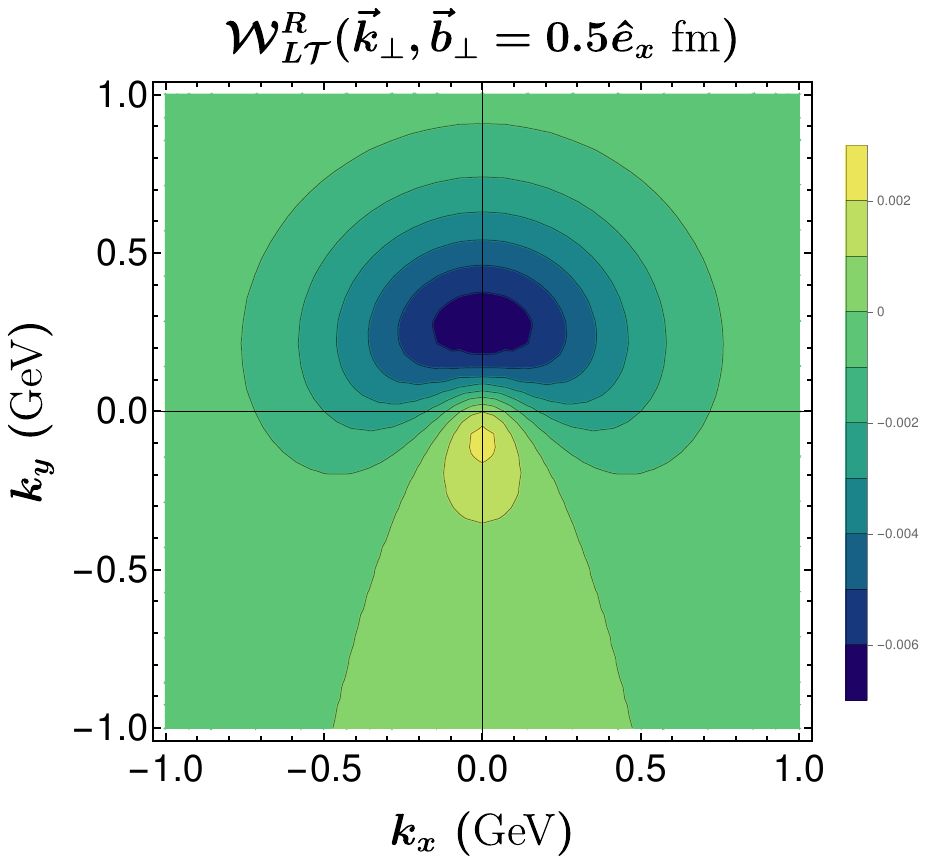}
        \includegraphics[width=0.325\linewidth]{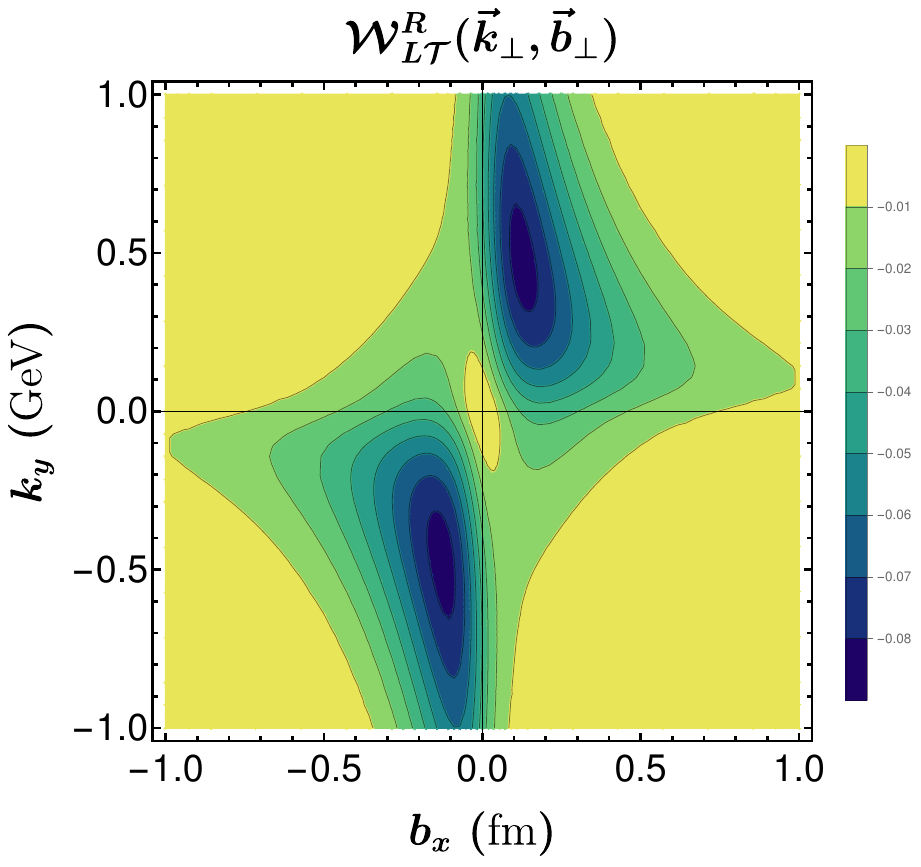}
        \includegraphics[width=0.325\linewidth]{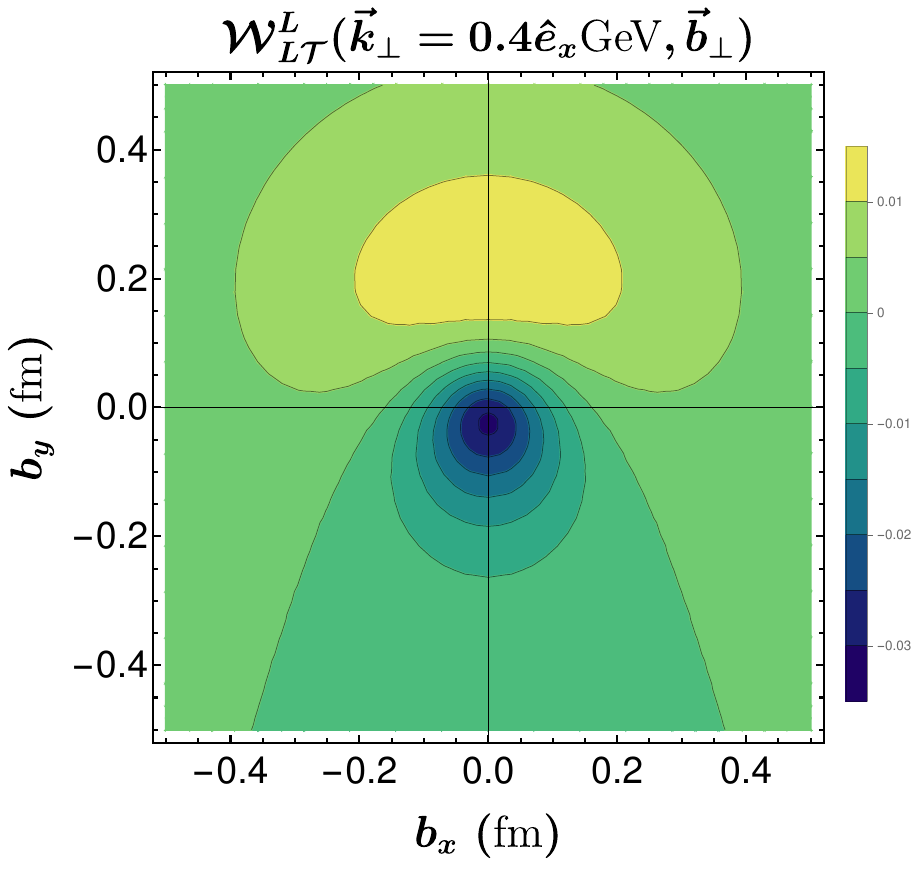}
        \includegraphics[width=0.325\linewidth]{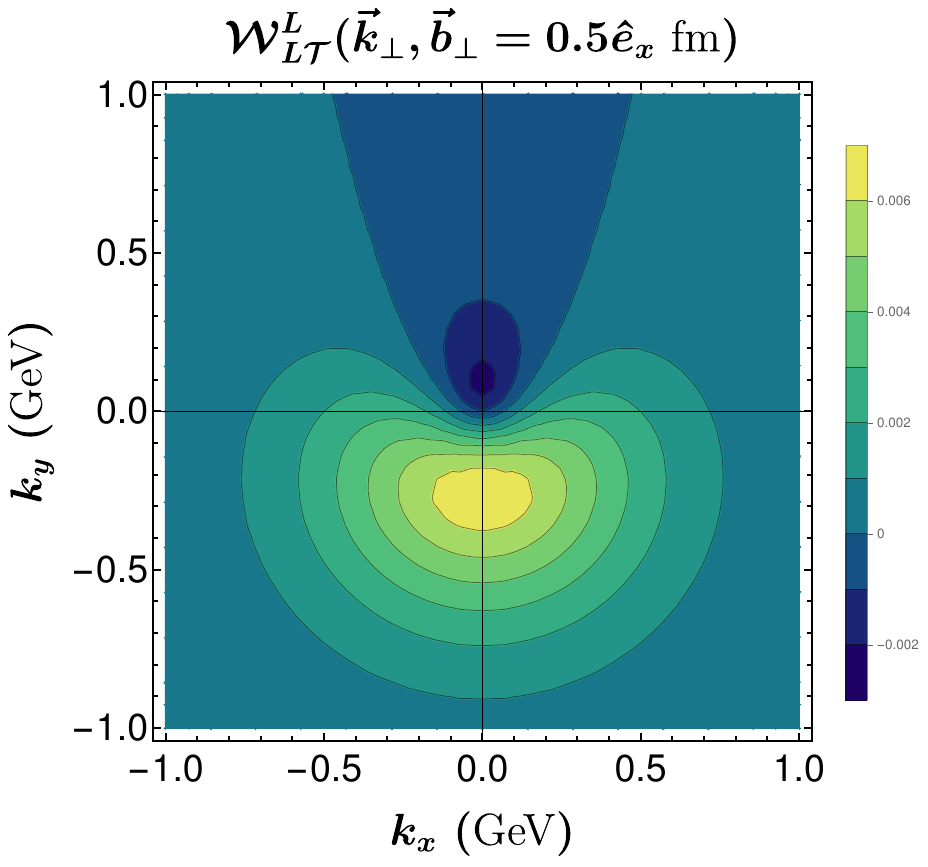}
        \includegraphics[width=0.325\linewidth]{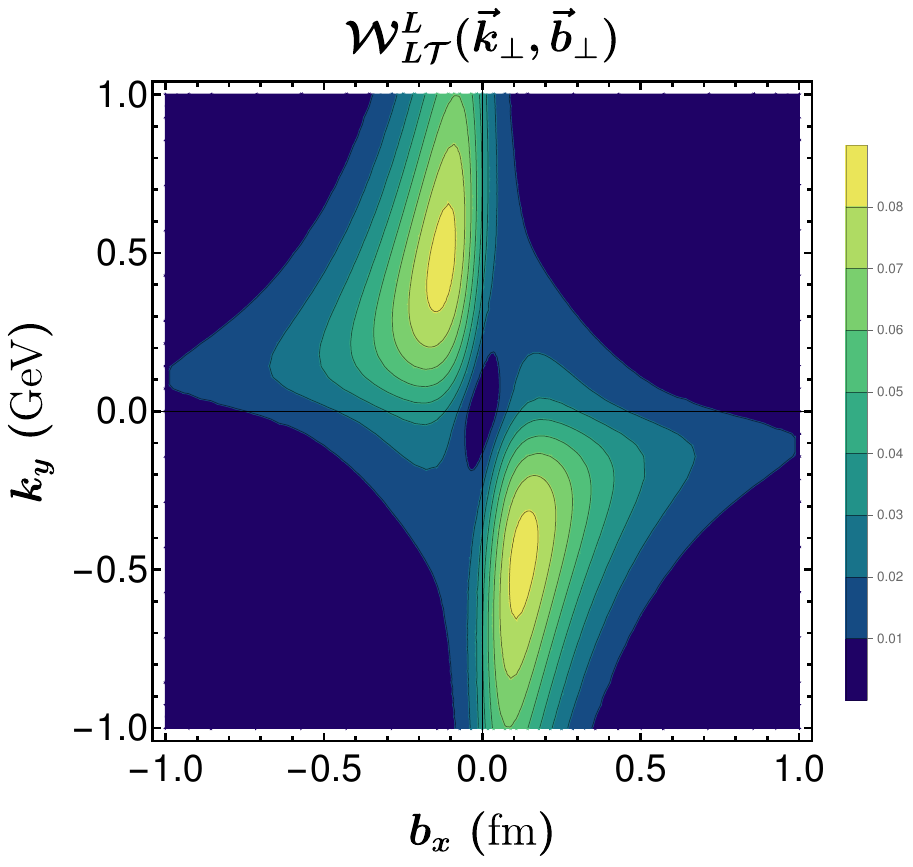}
        \caption{
        \justifying
        Upper panel: The transverse Wigner distributions of right-handed ($R$) linearly polarized gluons in a longitudinally polarized proton are shown in three configurations: in the transverse impact parameter plane at fixed transverse momentum $\mathbf{k}_\perp = 0.4\hat{e}_x~\text{GeV}$ (left), in the transverse momentum plane at fixed impact parameter $\mathbf{b}_{\perp} = 0.5\hat{e}_x~\text{fm}$ (middle), and in the mixed transverse plane after integrating over $k_x$ and $b_y$ (right).
        Lower panel: Same as the upper panel, but for left-handed ($L$) linearly polarized gluons in a longitudinally polarized proton.}
        \label{fig:Figure7}
    \end{figure}
    \begin{figure}
        \centering
        \includegraphics[width=0.325\linewidth]{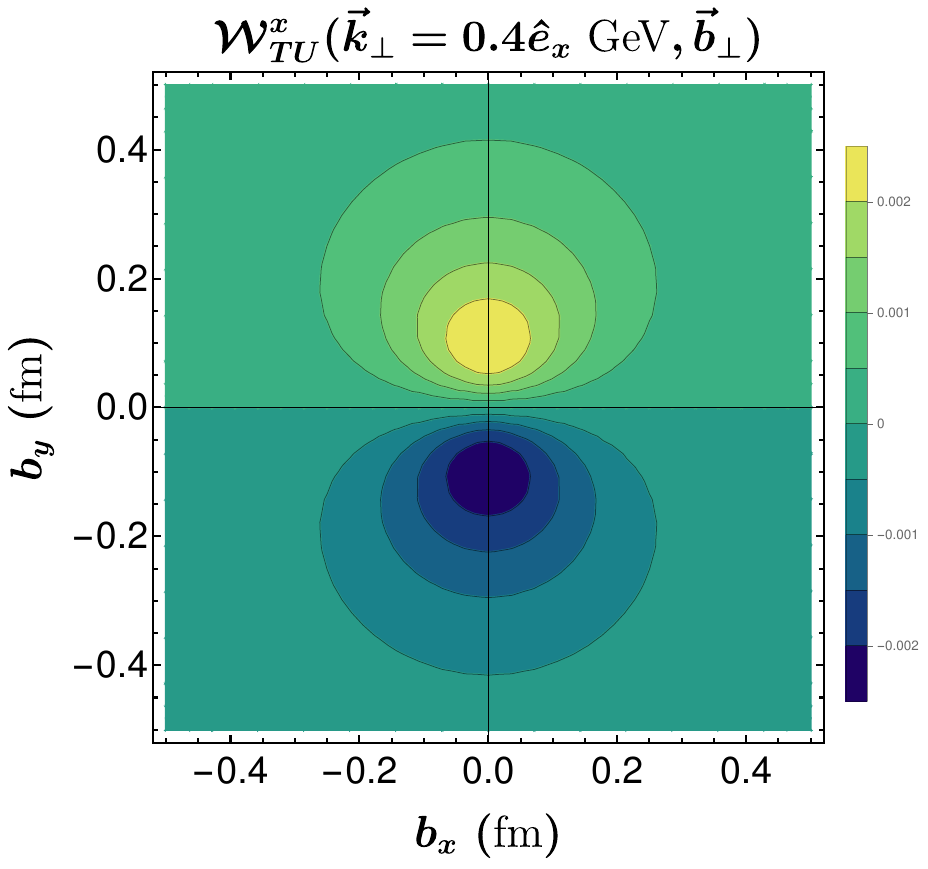}
        \includegraphics[width=0.325\linewidth]{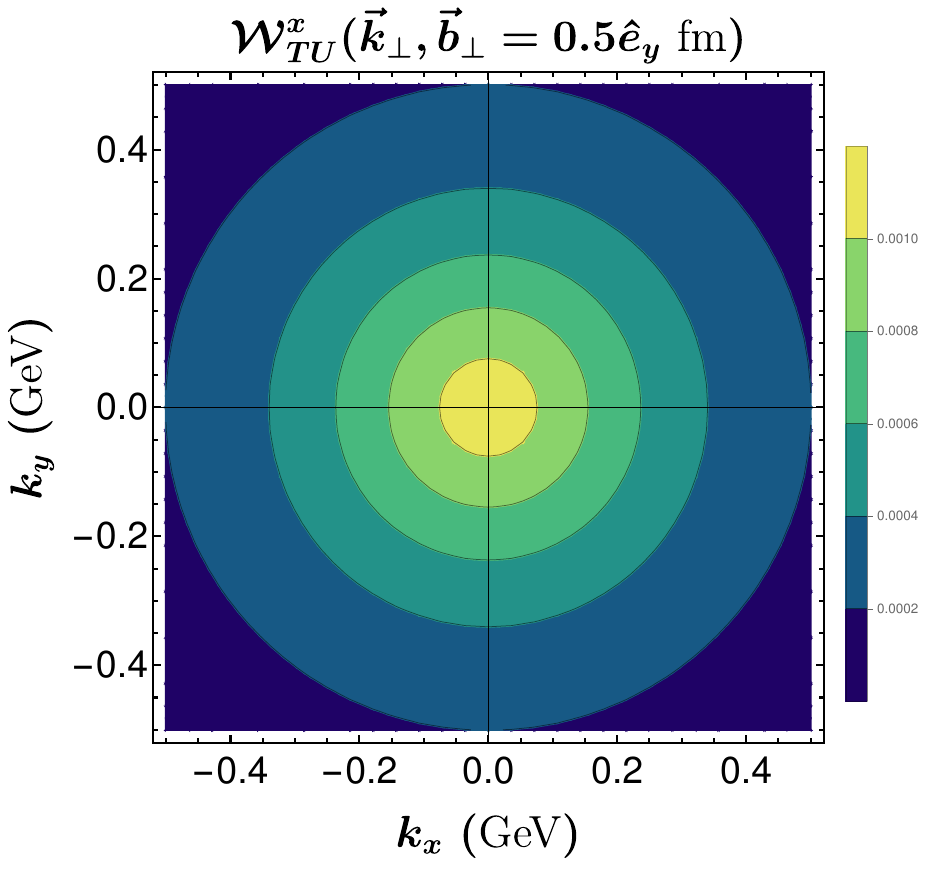}
        \includegraphics[width=0.325\linewidth]{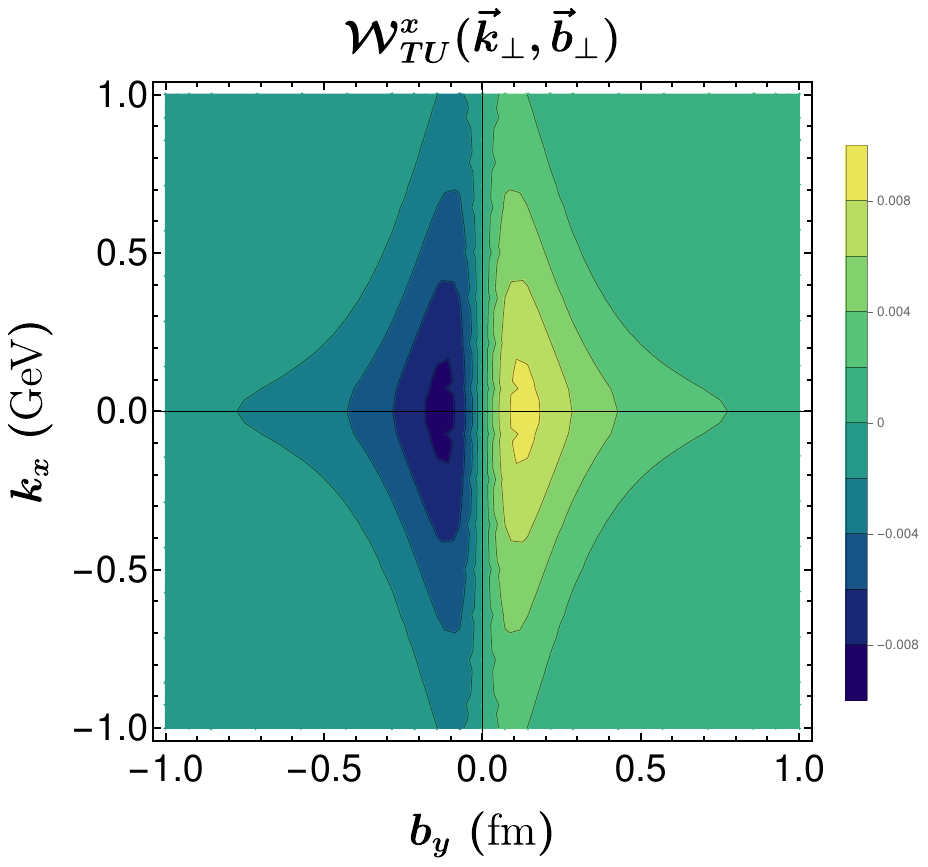}
        \includegraphics[width=0.325\linewidth]{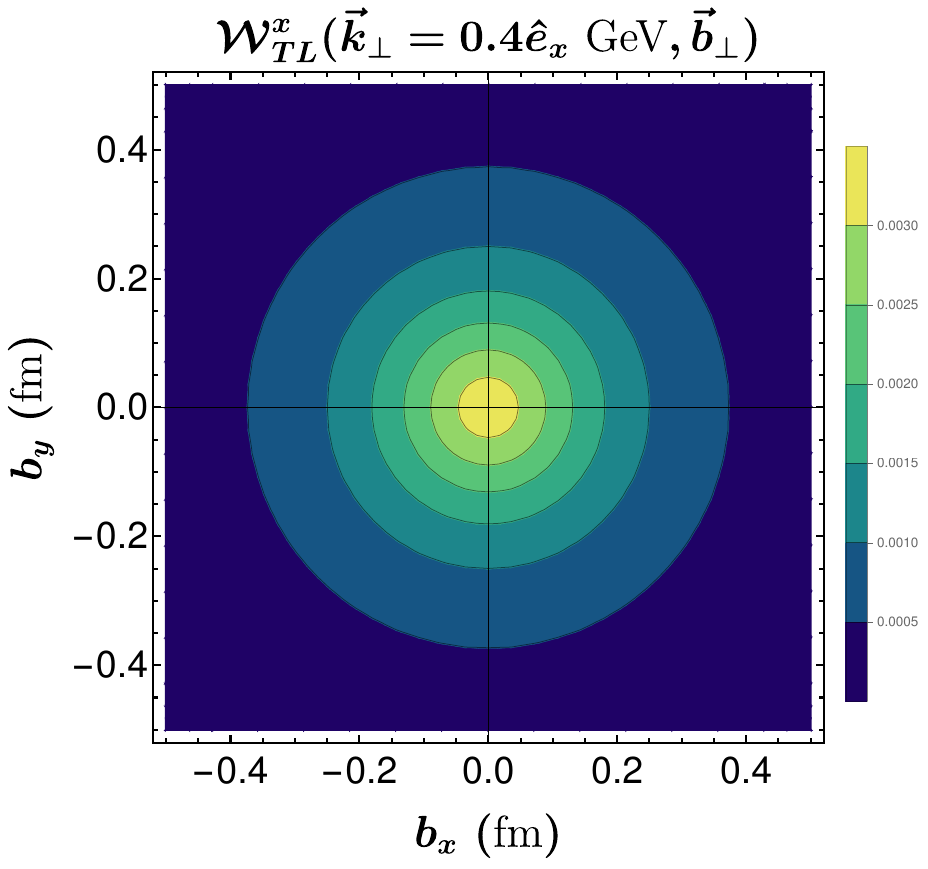}
        \includegraphics[width=0.325\linewidth]{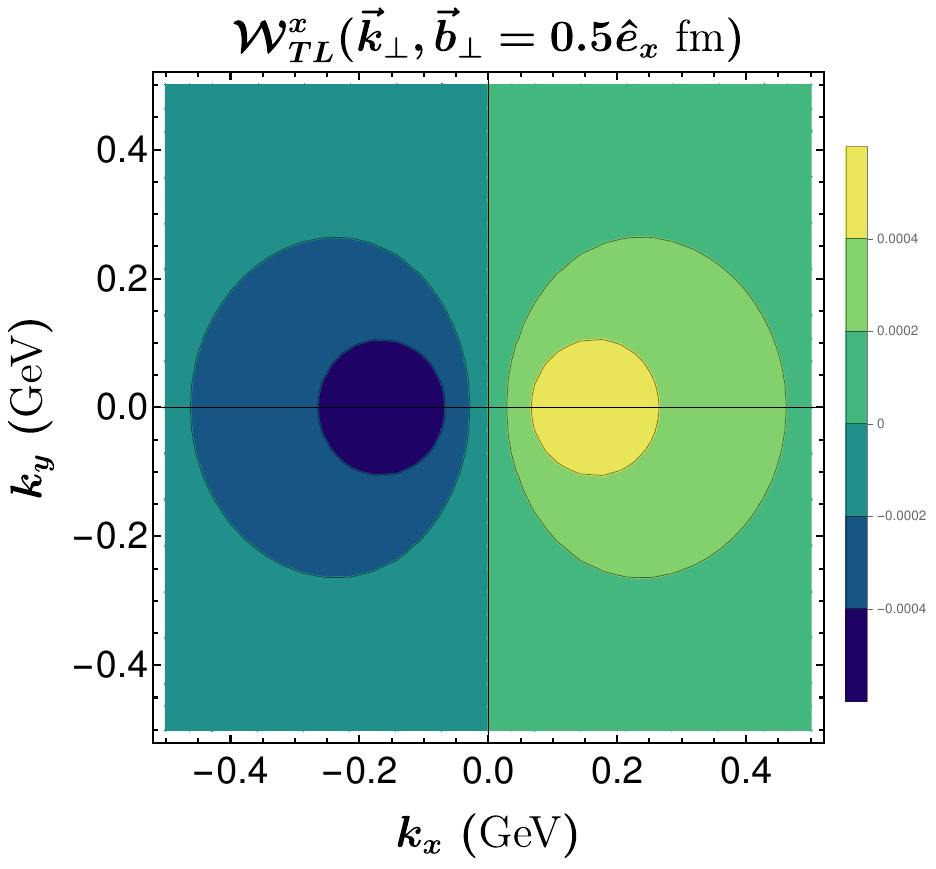}
        \includegraphics[width=0.325\linewidth]{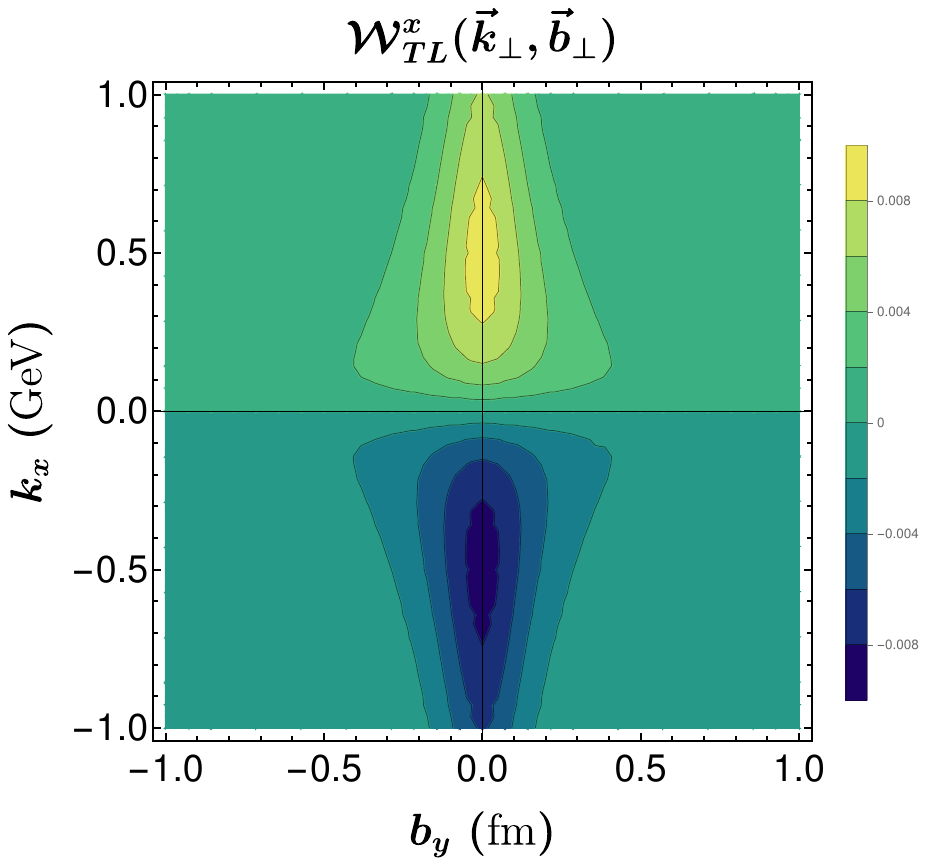}
        \caption{
        \justifying
        Upper panel: The transverse Wigner distributions of unpolarized gluons in a transversely polarized proton are shown in three configurations: in the transverse impact parameter plane at fixed transverse momentum $\mathbf{k}_\perp = 0.4\hat{e}_x~\text{GeV}$ (left), in the transverse momentum plane at fixed impact parameter $\mathbf{b}_{\perp} = 0.5\hat{e}_x~\text{fm}$ (middle), and in the mixed transverse plane after integrating over $k_x$ and $b_y$ (right).
        Lower panel: Same as the upper panel, but for longitudinally polarized gluons in a transversely polarized proton.}
        \label{fig:Figure8}
    \end{figure}     
\subsection{Unpolarized proton Wigner distributions}
In this section, we present the first Mellin moment of the  Wigner distributions, representing the purely transverse four-dimensional phase space—comprising two transverse position and two transverse momentum coordinates. These are hereafter referred to as the transverse Wigner distributions, defined as~\cite{Lorce:2011kd},
\begin{align}
    \mathcal{W}_{\Lambda_{p}\Lambda_{g}}(\bfk,\bfb)=\int{\rm d}x~\mathcal{W}_{\Lambda_{p}\Lambda_{g}}(x,\bfk,\bfb)\,.
\end{align}
The upper panel of Figure~\ref{fig:Figure4} shows the transverse Wigner distribution, $\mathcal{W}_{UU}^{g}(\mathbf{k}_{\perp},\mathbf{b}_{\perp})$ for unpolarized gluons in an unpolarized proton. The distributions are presented in three configurations: in the impact parameter space at fixed transverse momentum $\mathbf{k}_{\perp}$ along $\hat{x}$ and $k_{x} = 0.4~\text{GeV}$ (left panel), in the transverse momentum space at fixed impact parameter $\mathbf{b}_{\perp}$ along $\hat{x}$ with $ b_{x}= 0.5~\text{fm}$ (middle panel), and in the mixed space ($b_x$, $k_y$), i.e., transverse momentum and impact parameter space after integrating over $b_y$ and $k_x$ (right panel), respectively. The distribution attains a negative maximum at the origin, $(b_x = b_y = 0)$ and $(k_x = k_y = 0)$. It then increases smoothly outward, reaching a positive maximum before gradually decreasing and vanishing at larger values of transverse momentum and impact parameter. This behavior is consistently observed in both the transverse momentum and transverse impact parameter planes. It likely originates from the quasiprobabilistic nature of the unpolarized Wigner distribution; however, after integrating over the impact parameter space variables, the distribution becomes strictly positive. Due to the presence of Gaussian terms in the AdS/QCD LFWFs, the unpolarized Wigner distribution, $\mathcal{W}_{UU}^g$, is an even function of both $\bfb$ and $\bfk$. As a result, the average quadrupole distortions~\cite{Lorce:2014mxa} vanish in our model. The unpolarized Wigner distributions $\mathcal{W}_{UU}^g$ are circularly symmetric in both the transverse momentum and transverse impact parameter planes. However, in the mixed plane—defined by transverse momentum and impact parameter, those distributions exhibit axial symmetry. Therefore, there is no preferred configuration between $b_{\perp} \perp k_{\perp}$ and $b_{\perp} \parallel k_{\perp}$ in the mixed space. On the other hand, Ref.~\cite{More:2017zqp} reports that the unpolarized gluon Wigner distribution lacks circular symmetry, exhibiting a positive peak at the center and gradually decreasing away from it, whereas Ref.~\cite{Tan:2023vvi} finds the distribution to be negative at the center and positive in the outer region, also without circular symmetry. The lower panel of Figure~\ref{fig:Figure4} displays the transverse Wigner distributions for a longitudinally polarized gluon inside an unpolarized proton, $\mathcal{W}_{UL}^{g}(\bfk,\bfb)$. We show in the same three configurations as in the upper panel. The transverse Wigner distribution, $\mathcal{W}_{UL}^{g}(\bfk,\bfb)$, is related to the spin-orbit correlation and exhibits a dipolar pattern in both the transverse momentum and impact‑parameter planes, and shows quadrupole structure in the mixed $(b_x,k_y)$ plane. This behavior directly reflects the negative spin–orbit correlation, $\mathcal{C}^g_{z} < 0$, which implies that the gluon OAM is anti-aligned with its spin, as obtained numerically from Eq.~\eqref{eq:SO_correlation}. We also observe that the polarity in momentum space is opposite to that in position space. Within the TMD limit, the $\mathcal{W}^{g}_{UU}$  correspond to the unpolarized gluon TMD $f_{1}^{g}(x,\bfk^2)$, while in the GPD limit, it reduces to the unpolarized gluon GPDs $H^{g}(x,t)$. In contrast, $\mathcal{W}^{g}_{UL}$ does not have a well-defined TMD or GPD limit.
 
Figure~\ref{fig:Figure5} displays the transverse Wigner distributions for linearly polarized gluons inside an unpolarized proton. Specifically, the upper panel corresponds to right-handed polarized gluons, while the lower panel shows left-handed polarized gluons. From left to right, the panels represent the linearly polarized gluon Wigner distributions $\mathcal{W}_{U\mathcal{T}}^{R(L)}$ in the transverse position plane, transverse momentum plane, and mixed plane, respectively. The distributions in the transverse impact parameter plane are shown at a fixed transverse momentum of $k_{\perp} = 0.4$ GeV along the $x$-direction, whereas those in the transverse momentum plane are plotted at a fixed impact parameter of $b_{\perp} = 0.5$ fm along the $x$-direction. The mixed-plane distributions are obtained by integrating out the transverse momentum component $k_{x}$ and the impact parameter component $b_{y}$, and are presented as functions of the remaining transverse momentum $k_{y}$ and transverse impact parameter $b_{x}$, respectively. The qualitative behavior of the left- and right-handed polarized gluon transverse Wigner distributions is similar in shape but opposite in polarity, exhibiting a distorted pattern in the transverse impact-parameter and transverse-momentum planes, as well as in mixed planes, respectively. In particular, the distributions are distorted along the $b_{y}$ and $k_{y}$ axes. In {impact parameter} space, the right-handed linearly polarized gluon distribution exhibits a smaller, narrower peak toward positive $b_{y}$ and a broader, larger peak toward negative $b_{y}$. In contrast, the momentum-space distributions display a broader, smaller peak along the positive $k_{y}$ axis and a narrower, larger peak along the negative $k_{y}$ axis. However, the mixed-space distributions feature two positive peaks. Conversely, the left-handed linearly polarized gluon distributions exhibit the same pattern as the right-handed ones but with opposite orientation. A similar kind of distorted distribution is also presented in a perturbative model where the quark is dressed with a gluon~\cite{More:2017zqp}. In the TMD limit, $\mathcal{W}_{U\mathcal{T}}^{g}$ reduces to the Boer-Mulders gluon TMD $h_{1}^{\perp g}(x,\mathbf{k}_{\perp}^{2})$, while in the GPD limit, it projects to the chiral odd gluon GPDs. 
\subsection{Longitudinally polarized proton Wigner distributions}
In this section, we present the transverse Wigner distributions for unpolarized, longitudinally, and both
right- and left-handed linearly polarized gluons inside a longitudinally polarized proton. 

The upper panel of Figure~\ref{fig:Figure6} shows the transverse Wigner distributions $\mathcal{W}_{LU}^g$ for unpolarized gluons in a longitudinally polarized proton. These distributions are also presented in three configurations:  in impact parameter space with fixed transverse momentum $\mathbf{k_\perp}$ along $\hat{x}$ and $k_x = 0.4\,\text{GeV}$ (left panel), in transverse momentum space at fixed impact parameter $\mathbf{b_\perp}$ along $\hat{x}$ with $b_x = 0.5\,\text{fm}$ (middle panel), and in the mixed space $(b_x, k_y)$, i.e., the transverse momentum and impact parameter space after integration over $b_y$ and $k_x$ (right panel). The distribution is related to the gluon canonical OAM and in both impact parameter space and transverse momentum space exhibits a dipolar structure similar to the transverse Wigner distributions of longitudinally polarized gluons inside an unpolarized proton, $\mathcal{W}_{UL}^g$. In the mixed plane, we also observe a quadrupole pattern, consistent with what is seen for $\mathcal{W}_{UL}^g$, but with an opposite polarity. This behavior directly reflects the negative OAM $l_{z}^g<0$ obtained from Eq.~\eqref{eq:canonical_OAM} numerically. A similar observation has also been reported in Ref.~\cite{Tan:2023vvi} using the gluon spectator model and in Ref.~\cite{More:2017zqp} using the perturbative model. The lower panel of Figure~\ref{fig:Figure6} displays the transverse Wigner distribution for a longitudinally polarized gluon in a longitudinally polarized proton, $\mathcal{W}_{LL}^{g}$, shown in the same three configurations as in the upper panel. Here, the distributions are circularly symmetric in both impact parameter space and transverse momentum space, similar to $\mathcal{W}_{UU}^g$. However, in the mixed plane, the distribution exhibits axial symmetry, consistent with the behavior observed for $\mathcal{W}_{UU}^g$. The distribution $\mathcal{W}_{LL}^{g}$ exhibits a negative peak at the center and positive values in the outer regions in both impact parameter and transverse momentum spaces. While, the mixed-plane distribution shows a positive minimum at the origin, $b_{x}=0$ and $k_{y}=0$. Similar to the unpolarized Wigner distribution, $\mathcal{W}_{UU}^{g}$, the longitudinally polarized Wigner distribution, $\mathcal{W}_{LL}^g$, is also an even function of both $\bfb$ and $\bfk$. As a result, the average quadrupole distortions vanish in this model. In the TMD limit, $\mathcal{W}_{LL}^{g}$ reduces to the gluon helicity TMD $g_{1L}^{g}(x,\mathbf{k}_{\perp}^{2})$, while in the GPD limit, it projects to the gluon helicity GPD $\widetilde{H}^{g}(x,t)$. Similar to $\mathcal{W}^{g}_{UL}$, the transverse Wigner distribution, $\mathcal{W}^{g}_{LU}$, also lacks a well-defined TMD or GPD limit.
\begin{figure}
        \centering
        \includegraphics[width=0.325\linewidth]{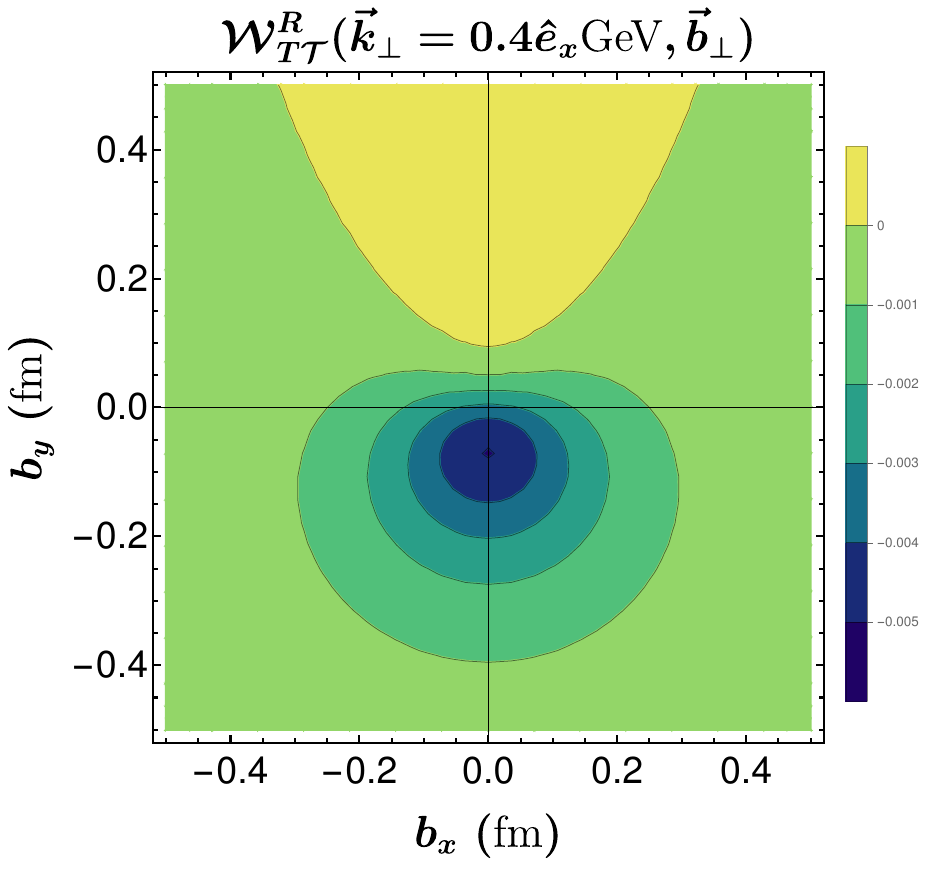}
        \includegraphics[width=0.325\linewidth]{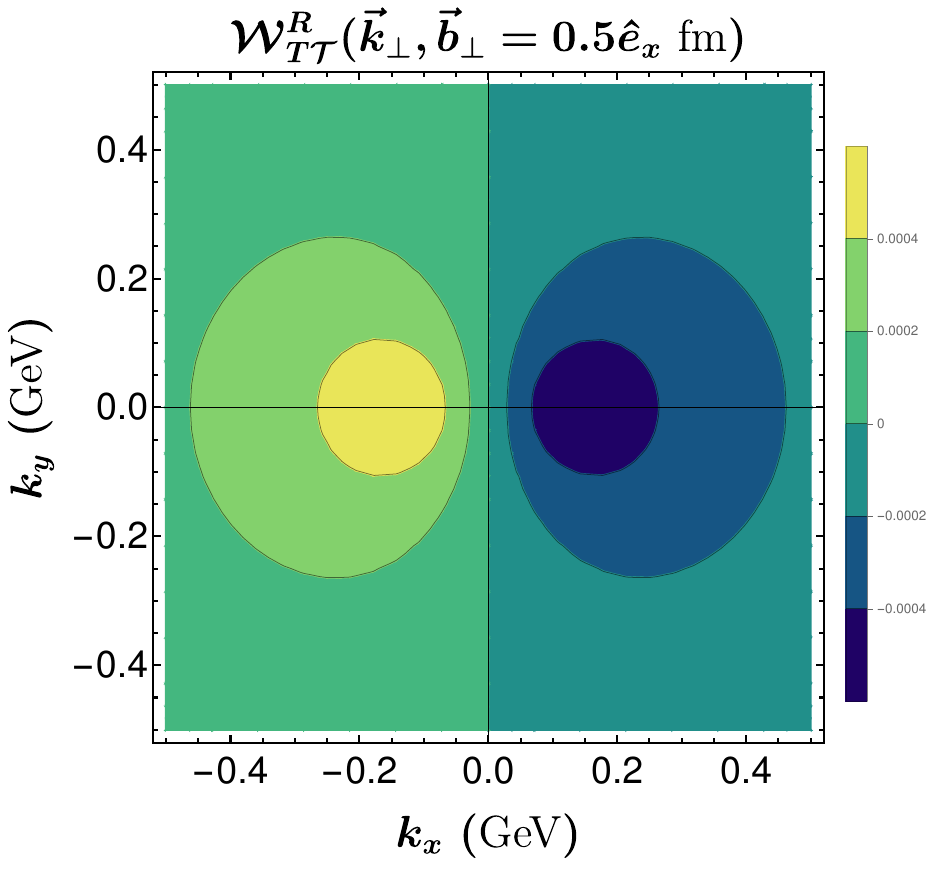}
        \includegraphics[width=0.325\linewidth]{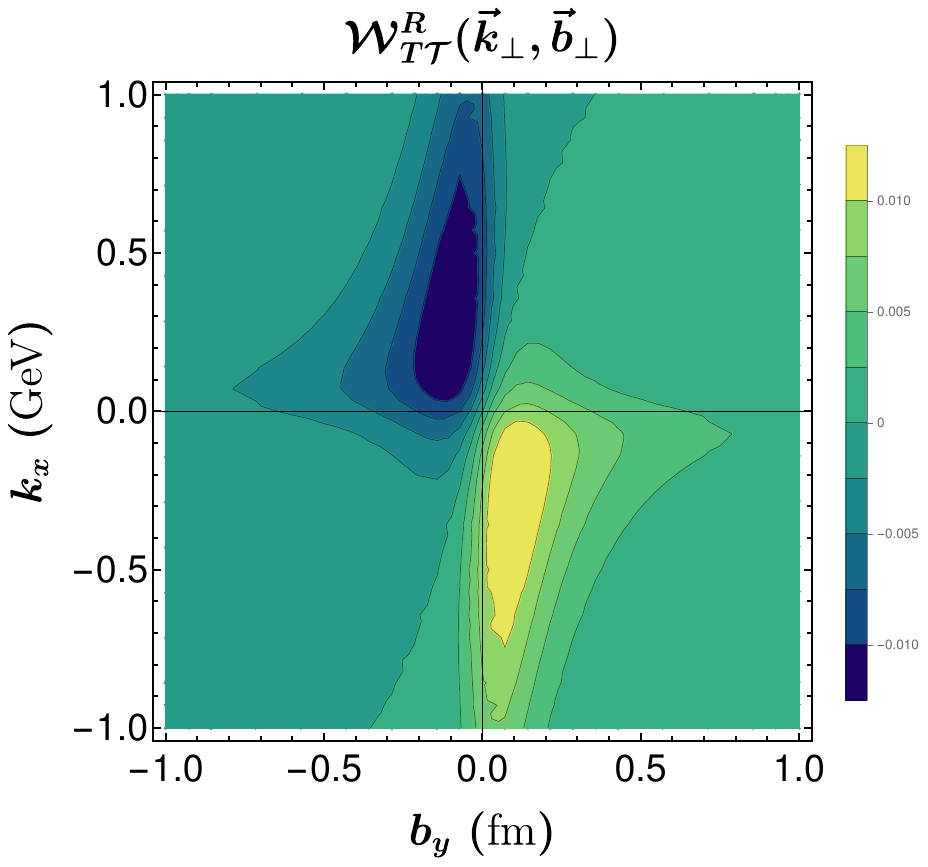}
        \includegraphics[width=0.325\linewidth]{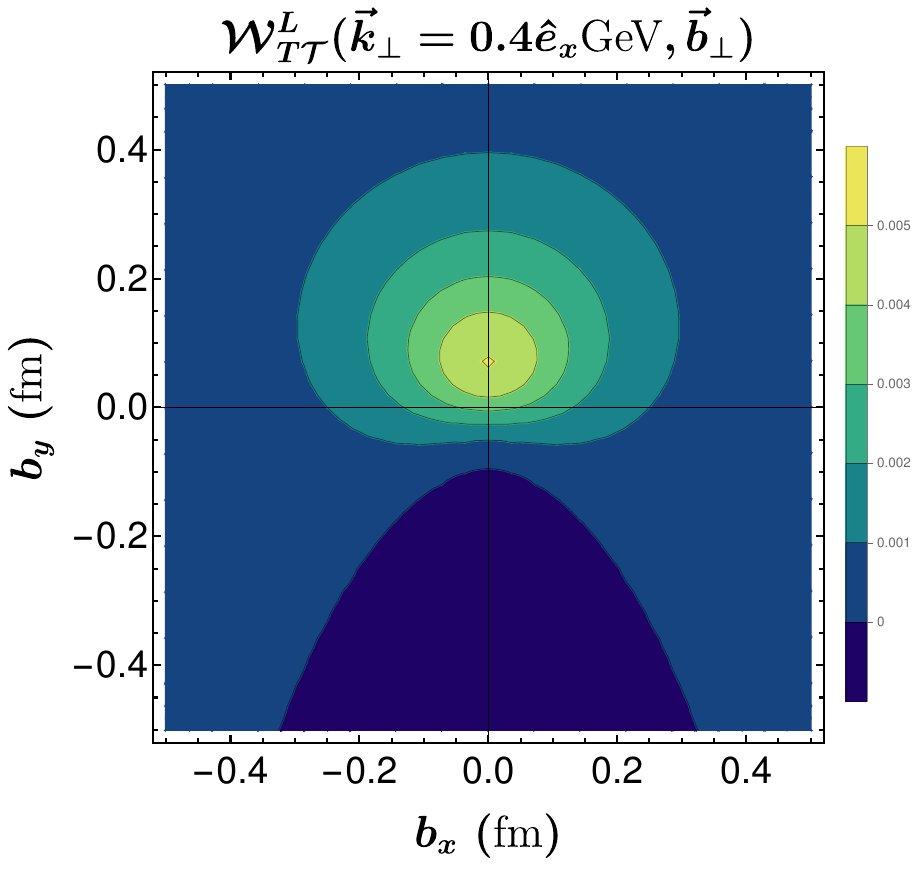}
        \includegraphics[width=0.325\linewidth]{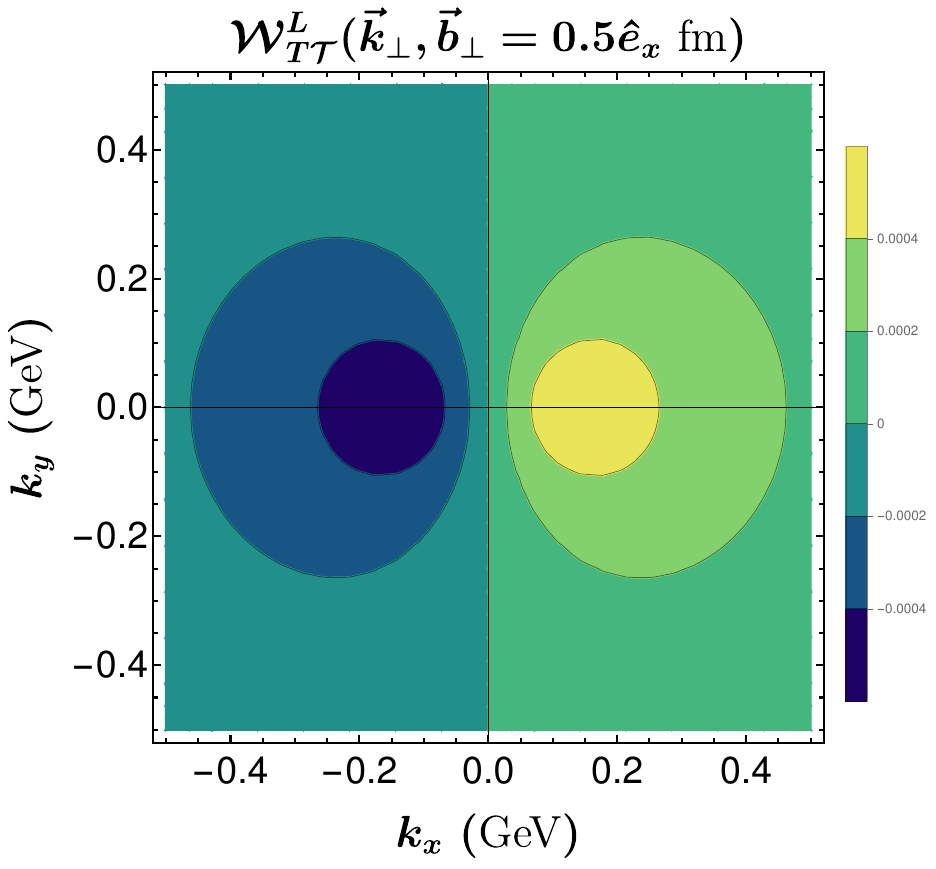}
        \includegraphics[width=0.325\linewidth]{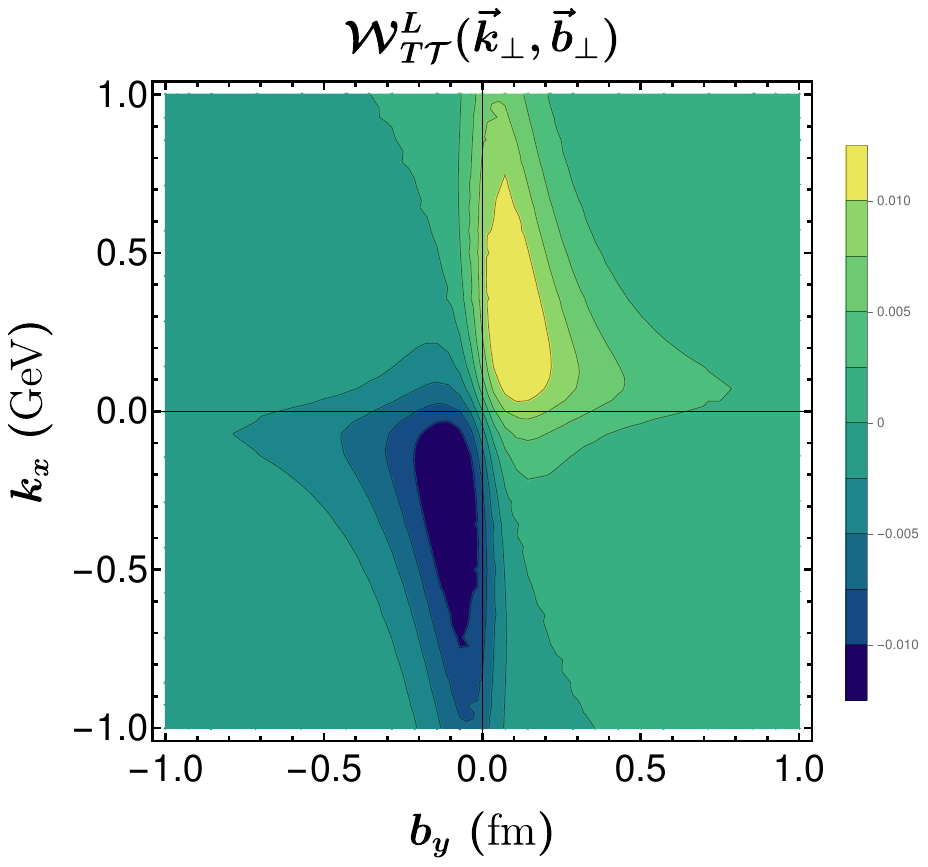}
        \caption{
        \justifying
        Upper panel: The transverse Wigner distributions of right handed ($R$) linearly polarized gluons in a transversely polarized proton are shown in three configurations: in the transverse impact parameter plane at fixed transverse momentum $\mathbf{k}_\perp = 0.4\hat{e}_x~\text{GeV}$ (left), in the transverse momentum plane at fixed impact parameter $\mathbf{b}_{\perp} = 0.5\hat{e}_x~\text{fm}$ (middle), and in the mixed transverse plane after integrating over $k_x$ and $b_y$ (right).
        Lower panel: Same as the upper panel, but for left-handed ($L$) linearly polarized gluons in a transversely polarized proton.}
        \label{fig:Figure9}
\end{figure}     
    
Figure~\ref{fig:Figure7} presents the transverse Wigner distributions of linearly polarized gluons inside a longitudinally polarized proton. Specifically, the upper panel illustrates the distributions of right-handed linearly polarized gluons in three different configurations. The left panel shows the $\mathcal{W}_{L\mathcal{T}}^{R}$ distribution in the transverse impact-parameter plane at a fixed transverse momentum $\bfk$ oriented along the $x$-axis, with $k_{x}=0.4$ GeV. The middle panel displays the distribution in the transverse momentum plane at a fixed transverse impact parameter $\bfb$ along the $x$-axis, with $b_{x}=0.5$ fm. Finally, the right panel depicts the transverse Wigner distribution of right-handed linearly polarized gluons in the mixed $(b_{x}, k_{y})$ plane. On the other hand, the lower panel depicts the distributions of left-handed linearly polarized gluons, $\mathcal{W}_{L\mathcal{T}}^{L}$ in three different configurations same as the upper panel. We observe that the qualitative behavior of both left-handed and right-handed linearly polarized gluon distributions in impact parameter space is distorted dipolar (the distortion is along the $b_{y}$ axis) in nature. Specifically, there is a narrow positive peak along the positive  $b_{y}$ direction for right-handed polarized gluons, while left-handed polarized gluons exhibit the opposite polarity. Similarly, in the transverse momentum plane, the distributions also display a similar distorted dipolar structure, with a narrow positive peak along $k_{y}$ for right-handed polarized gluons and an opposite polarity for left-handed gluons. The mixed-space distributions exhibit two negative peaks for right-handed linearly polarized gluons, whereas for left-handed linearly polarized gluons, the peaks are positive but with a smaller magnitude compared to the right-handed case. Similar observations have also been reported for these distributions in a perturbative model~\cite{More:2017zqp}.
\subsection{Transversely polarized proton Wigner distributions}
In this subsection, we present the transverse Wigner distributions of unpolarized, longitudinally polarized, and both right- and left-handed linearly polarized gluons inside a transversely polarized proton. The upper panel (from left to right) of Figure~\ref{fig:Figure8} displays the transverse Wigner distributions of unpolarized gluons inside a transversely polarized proton, $\mathcal{W}_{TU}^{x}$ in the transverse impact parameter plane, transverse momentum plane, and mixed plane, respectively. The superscript index indicates that the proton is polarized along the $x$-direction. These distributions in the impact parameter space are shown at a fixed transverse momentum of $k_{x} = 0.4$ GeV along the $x$-axis and vanish if the
gluon impact parameter coordinate is parallel to the target polarization. Similarly, the distributions in the transverse momentum plane are presented at a fixed impact parameter of $b_{y} = 0.5$ fm along the $y$-axis. The mixed plane distributions correspond to the $b_{y}$ and $k_{x}$ components, obtained by integrating over the $b_{x}$ and $k_{y}$ variables, respectively. The distributions of unpolarized gluons inside a transversely polarized proton exhibit a dipolar structure in the impact parameter plane and spread towards larger $b_{y}$, while in transverse momentum space, they are circularly symmetric with a positive peak at the center. Notably, this distribution remains symmetric in the transverse momentum plane, but displays a dipolar pattern in the mixed plane, in contrast to $\mathcal{W}_{UL}$ and $\mathcal{W}_{LU}$, which show a quadrupole structure in the mixed plane. The impact parameter space integrated transverse Wigner distribution, $\mathcal{W}_{TU}^{x}$, reduces to the gluon Sivers function $f_{1T}^{\perp g}(x,\bfk^2)$. In contrast, the transverse momentum integrated Wigner distribution, $\mathcal{W}_{TU}^{x}$, reduces to the unpolarized gluon GPD $E^{g}(x,t)$ along with some other distribution. 

Meanwhile, the transverse Wigner distribution of longitudinally polarized gluons inside a transversely polarized proton, $\mathcal{W}_{TL}^{x}$, is shown in the lower panel of Figure~\ref{fig:Figure8}. The panels from left to right correspond to the impact parameter space distributions at fixed transverse momentum $k_{x}=0.4$ GeV. These exhibit circular symmetry in impact parameter space with a positive peak at the center, and they vanish when the gluon transverse momentum is perpendicular to the target polarization, indicating a strong correlation between the gluon transverse momentum and the target polarization. In contrast, the transverse momentum space distributions display a dipolar structure extending towards larger $k_{x}$. A similar dipolar behavior is also observed in the mixed plane. Furthermore, the transverse Wigner distribution $\mathcal{W}_{TL}^{x}$ reduces to the worm-gear gluon TMD, $g_{1T}^{g}(x,\bfk^2)$, upon integrating over the impact parameter space variable, while integration over the transverse-momentum variable yields the longitudinally polarized gluon GPDs $\widetilde{H}^{g}(x,t)$ and $\widetilde{E}^{g}(x,t)$ together with some other distributions.

Figure~\ref{fig:Figure9} presents the transverse Wigner distributions of right- and left-handed linearly polarized gluons inside a transversely polarized proton target along the $x$-axis. 
The upper panel (from left to right) shows the distributions for right-handed linearly polarized gluons, $\mathcal{W}_{T\mathcal{T}}^{R}$, in the impact parameter plane, the transverse momentum plane, and the mixed plane. The lower panel (from left to right) shows the corresponding distributions for left-handed linearly polarized gluons, $\mathcal{W}_{T\mathcal{T}}^{L}$, in the same three configurations. The impact parameter–space distributions are presented at fixed transverse momentum $k_{x}=0.4$ GeV, while the transverse momentum–space distributions are obtained at fixed impact parameter $b_{x}=0.5$ fm. The mixed-space distributions are asymmetric and shown as functions of $b_{y}$ and $k_{x}$, obtained by integrating over $b_{x}$ and $k_{y}$. Both the left- and right-handed linearly polarized gluon distributions exhibit similar overall patterns but with opposite polarity in all three configurations. In the impact parameter plane, $\mathcal{W}_{T\mathcal{T}}^{R}$ is shifted toward negative $b_{y}$, whereas $\mathcal{W}_{T\mathcal{T}}^{L}$ is shifted toward positive $b_{y}$, with the peak inverted in sign. In the transverse momentum plane, both distributions display a dipolar structure with a broad spread, while in mixed space, they extend over a wide region. A similar qualitative behavior has also been predicted for linearly polarized gluons in a transversely polarized dressed quark state~\cite{More:2017zqp}. The transverse Wigner distribution, $\mathcal{W}_{T\mathcal{T}}^{g}$, is related to the  $h_{1T}^{g}(x,\bfk^2)$ gluon TMD, and the GPD limit of this distribution is related to the chiral-odd GPDs  ${H}^{g}_{T}(x,t)$, ${E}^{g}_{T}(x,t)$, $\widetilde{H}^{g}_{T}(x,t)$ and $\widetilde{E}^{g}_{T}$(x,t), along with some other distribution.
\begin{figure}
        \centering
        \includegraphics[width=0.325\linewidth]{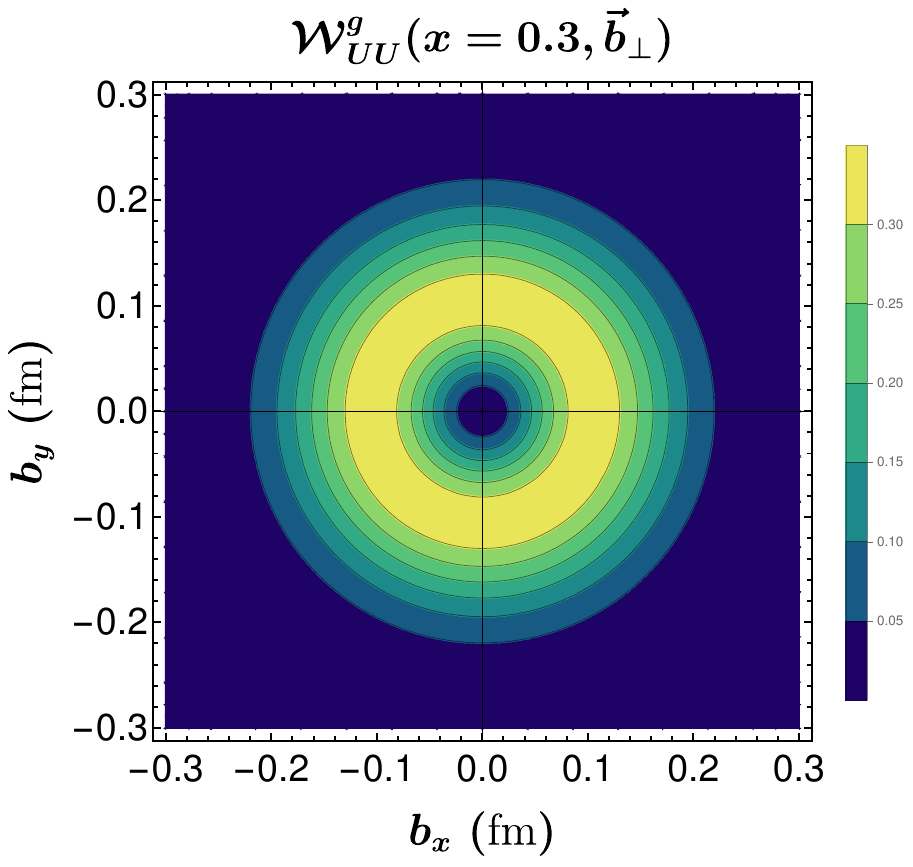}
        \includegraphics[width=0.325\linewidth]{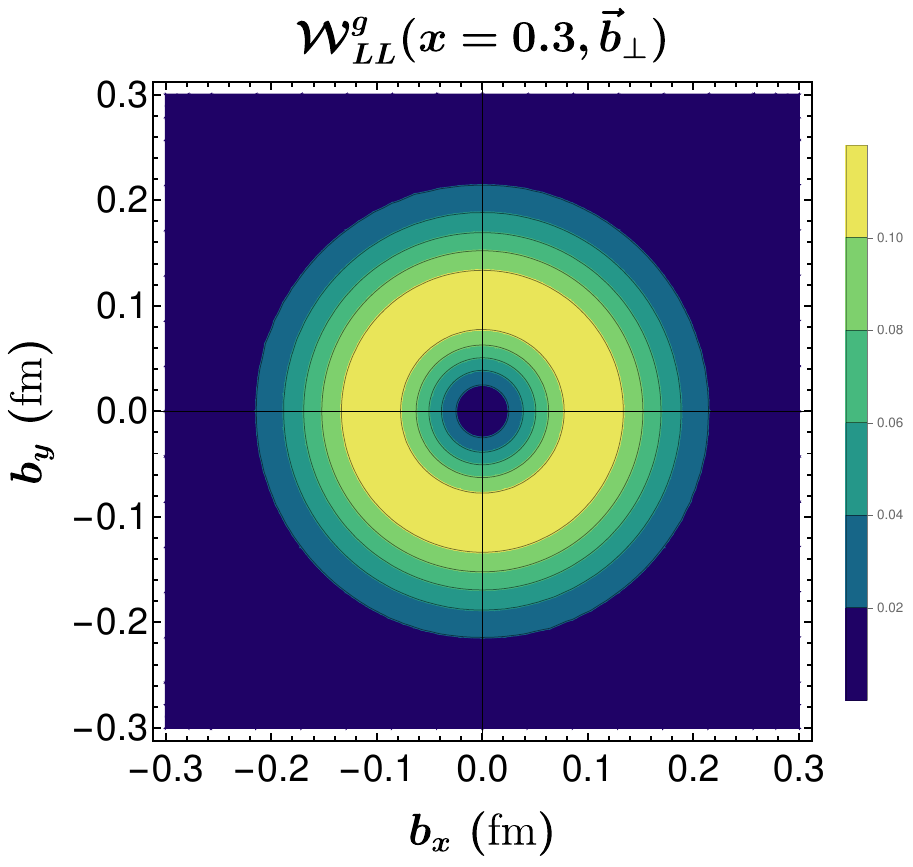}
        \includegraphics[width=0.325\linewidth]{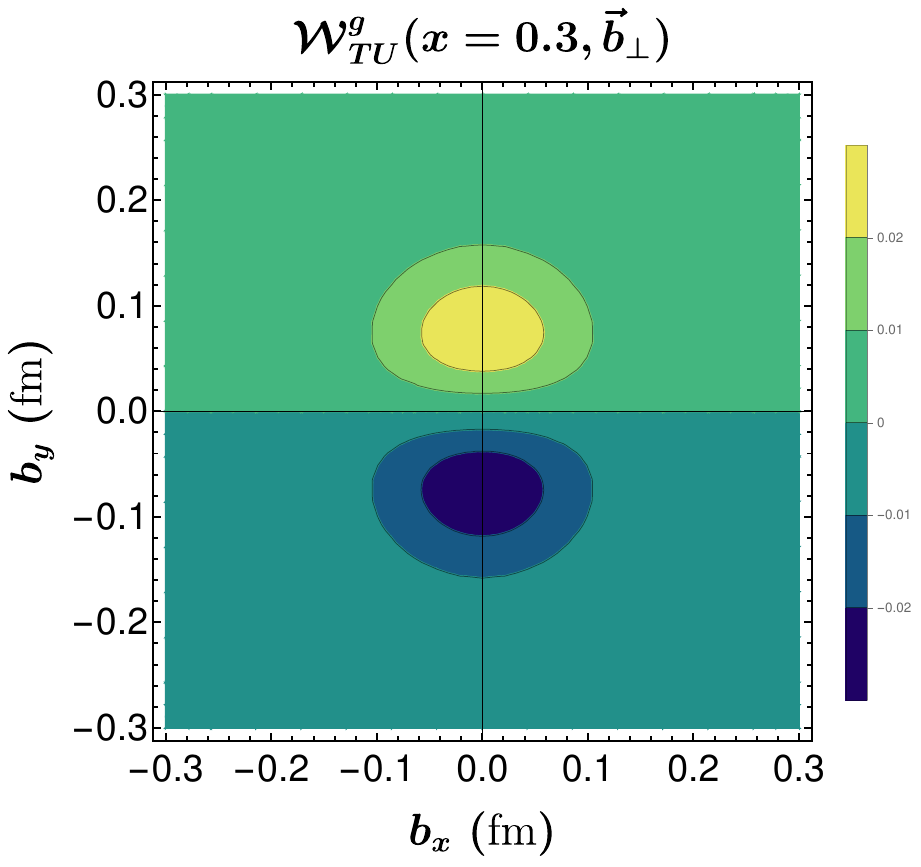}
        \includegraphics[width=0.325\linewidth]{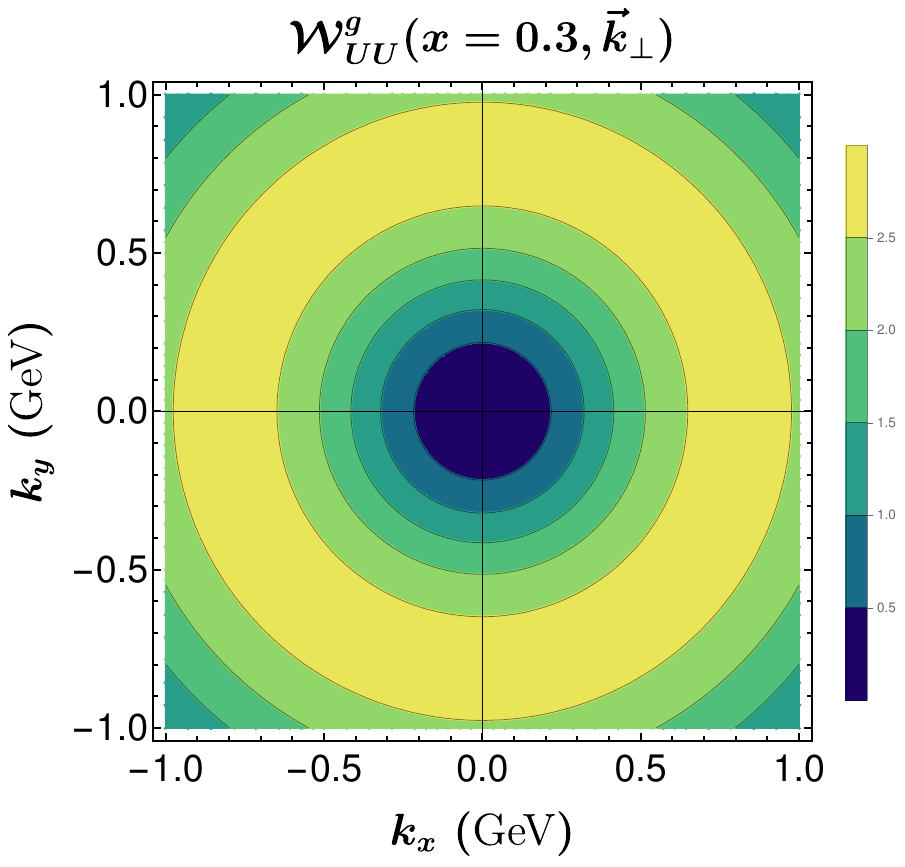}
        \includegraphics[width=0.325\linewidth]{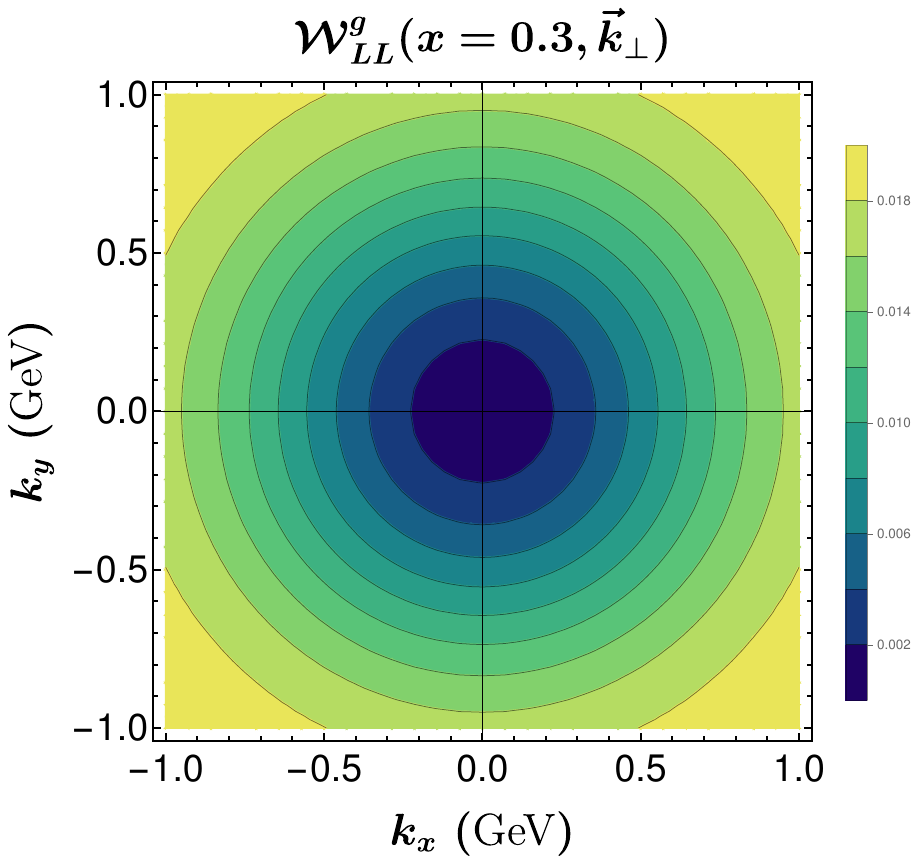}
        \includegraphics[width=0.325\linewidth]{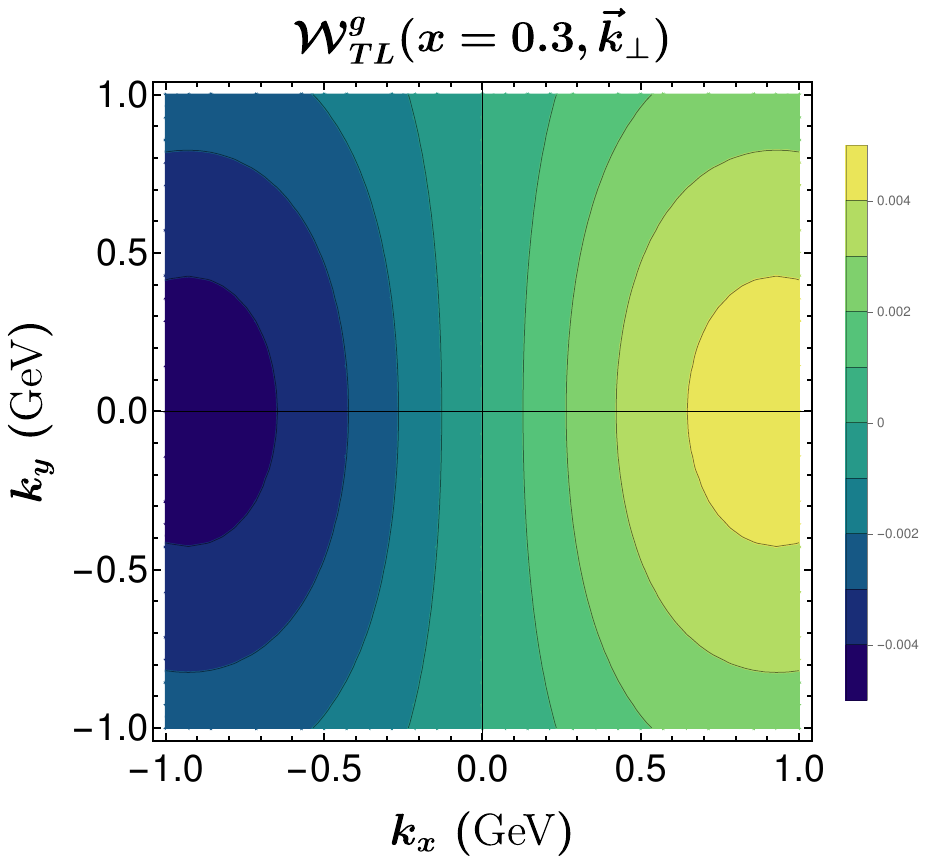}
        \caption{
        \justifying
        Upper panel: The spin densities in impact parameter space for unpolarized gluons in an unpolarized proton (left), longitudinally polarized gluons in a longitudinally polarized proton (middle), and unpolarized gluons in a transversely polarized proton at a fixed value of $x=0.3$. Lower panel: same as upper panel, but in the transverse momentum plane.}
        \label{fig:Figure10}
\end{figure}
%
{{\subsection{Spin densities and spin-spin correlations}
In the previous subsections, we discussed the transverse Wigner distributions in impact-parameter space for fixed transverse momentum, as well as in transverse-momentum space for fixed impact-parameter values. In this subsection, we focus on the spin densities and spin-spin correlations in both the impact-parameter and transverse-momentum planes, evaluated at a fixed gluon longitudinal momentum fraction $x=0.3$, by integrating the Wigner distributions over the transverse momentum and impact parameter variables, respectively. In the upper panel of Figure~\ref{fig:Figure10}, from left to right, we present the spin densities of unpolarized gluons inside an unpolarized target, longitudinally polarized gluons inside a longitudinally polarized target, and unpolarized gluons inside a transversely polarized target. The lower panel shows the corresponding distributions in the transverse-momentum plane. We find that the spin densities of unpolarized and longitudinally polarized gluons inside unpolarized and longitudinally polarized targets are circularly symmetric, with a positive minimum at the center in both impact-parameter space and the transverse-momentum plane. In impact-parameter space, the distributions exhibit a central minimum that increases to a maximum before decreasing again, whereas in the transverse-momentum plane, the distributions increase monotonically. In contrast, the spin densities of unpolarized gluons inside a transversely polarized target exhibit a dipolar pattern in the impact-parameter plane but vanish in the transverse-momentum plane. Similarly, the spin densities of longitudinally polarized gluons inside a transversely polarized targe vanish in the impact-parameter plane, while being distributed widely across the transverse-momentum plane. This indicates a clear correlation between the target’s transverse polarization and the gluon’s transverse momentum and position variables. A similar type of behavior was observed for the gluon spin densities in a perturbative model where the gluons are dressed with a quark target state~\cite{More:2017zqp}. 

By combining the Wigner distribution of unpolarized gluons in an unpolarized proton, $\mathcal{W}_{UU}$, with the polarization-induced distortions $\mathcal{W}_{UL}$, $\mathcal{W}_{LU}$, and $\mathcal{W}_{LL}$ as defined in Eq.~\eqref{eq:spin-spin correlation}, we obtain the spin–spin correlation function $\rho_{\Lambda_{p}\Lambda_{g}}^{g}(x, \bfk, \bfb)$, shown in Figure~\ref{fig:Figure11}. This function characterizes the correlation between the spins of the gluon and proton, whether aligned in the same or opposite directions, in various spatial and momentum configurations. Specifically, the distributions are presented in impact-parameter space at fixed transverse momentum along $\hat{x}$ with $k_{x} = 0.4$ GeV;  transverse-momentum space at fixed impact parameter along $\hat{x}$ with $b_{x} = 0.5$ fm; and a mixed representation, where $b_{y}$ and $k_{x}$ are integrated out. The deformations resulting from gluon and proton polarizations are evident in the shifts of the distributions away from the center, as seen in the left and middle panels. In particular, when the gluon and proton polarizations are parallel (antiparallel), the distribution shifts in the negative (positive) $\hat{b}_{y}$ direction, as illustrated in the upper (lower) panels. From Figure~\ref{fig:Figure4} and Figure~\ref{fig:Figure6}, it is evident that the Wigner distributions $\mathcal{W}_{UU}$ and $\mathcal{W}_{LL}$ exhibit circular symmetry in both the impact parameter and transverse momentum planes. The distortions observed in the spin-spin correlations arise from the contributions of the Wigner distributions $\mathcal{W}_{UL}$ and $\mathcal{W}_{LU}$ in the spin-spin correlation function. From the negative values of both the spin–orbit correlation function, $\mathcal{C}_{z}^{g}<0$, and the canonical OAM, $\ell_{z}^{g}<0$, we infer that the gluon and proton spins are antialigned with the gluon OAM. When the gluon polarization is parallel to the proton spin, i.e., $\rho_{\uparrow\uparrow}^{g}$ or $\rho_{\downarrow\downarrow}^{g}$, the contributions of $\mathcal{W}_{UL}$ and $\mathcal{W}_{LU}$ interfere constructively, leading to a shift in the negative $\hat{b}_{y}$ direction. In contrast, when the gluon polarization is antiparallel to the proton spin, i.e., $\rho_{\uparrow\downarrow}^{g}$ or $\rho_{\downarrow\uparrow}^{g}$, the contributions of $\mathcal{W}_{UL}$ and $\mathcal{W}_{LU}$ interfere destructively. As shown in Figures~\ref{fig:Figure4} and \ref{fig:Figure5}, the correlation between the gluon OAM and gluon spin is stronger than that between the gluon OAM and proton spin, resulting in a net shift in the positive $\hat{b}_{y}$ direction, as illustrated in the upper and lower panels of Figure~\ref{fig:Figure11}. The distortions in the mixed plane show a dipolar distribution. 
\begin{figure}
        \centering
        \includegraphics[width=0.325\linewidth]{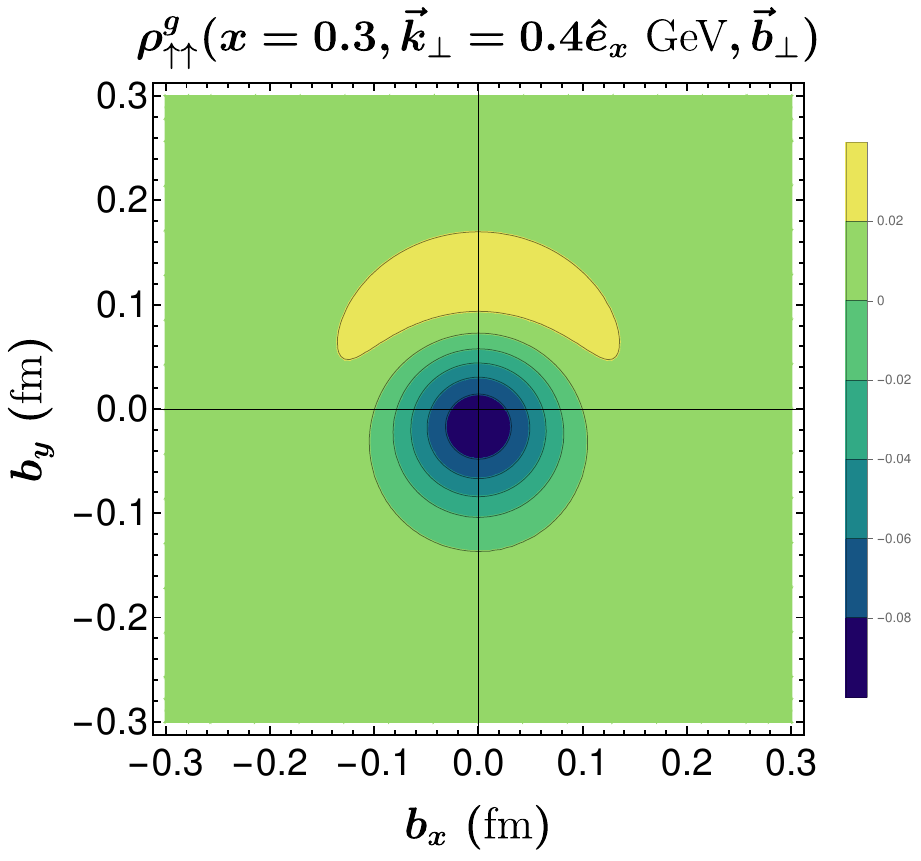}
        \includegraphics[width=0.325\linewidth]{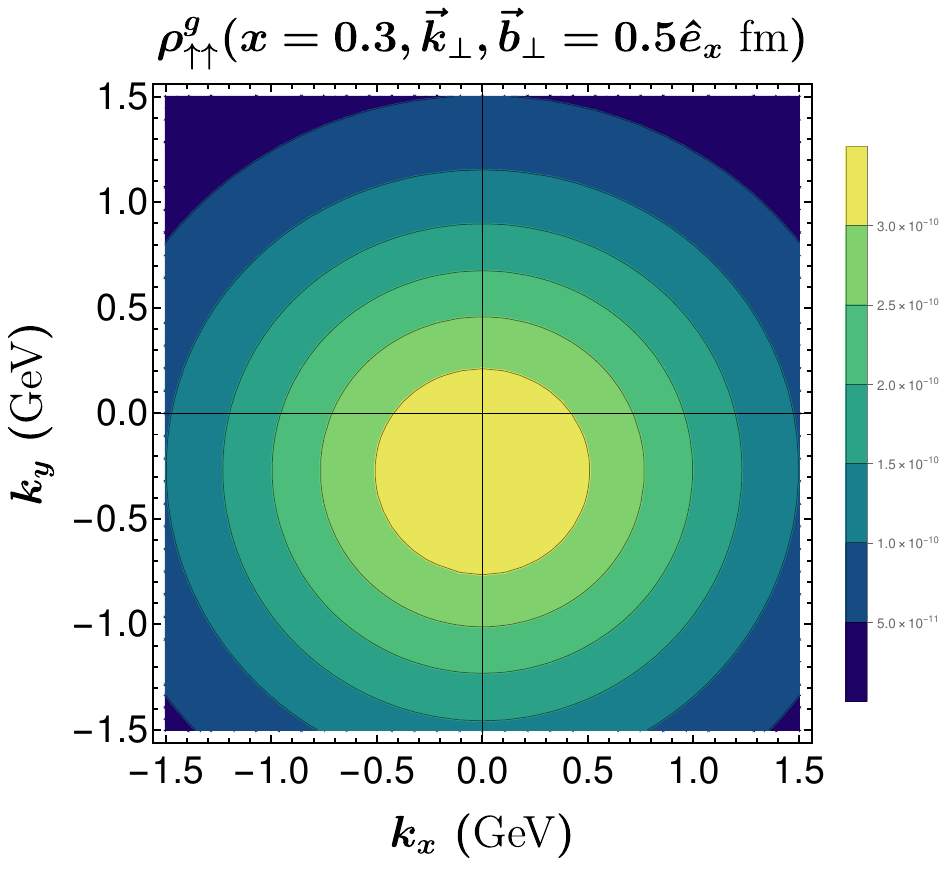}
        \includegraphics[width=0.325\linewidth]{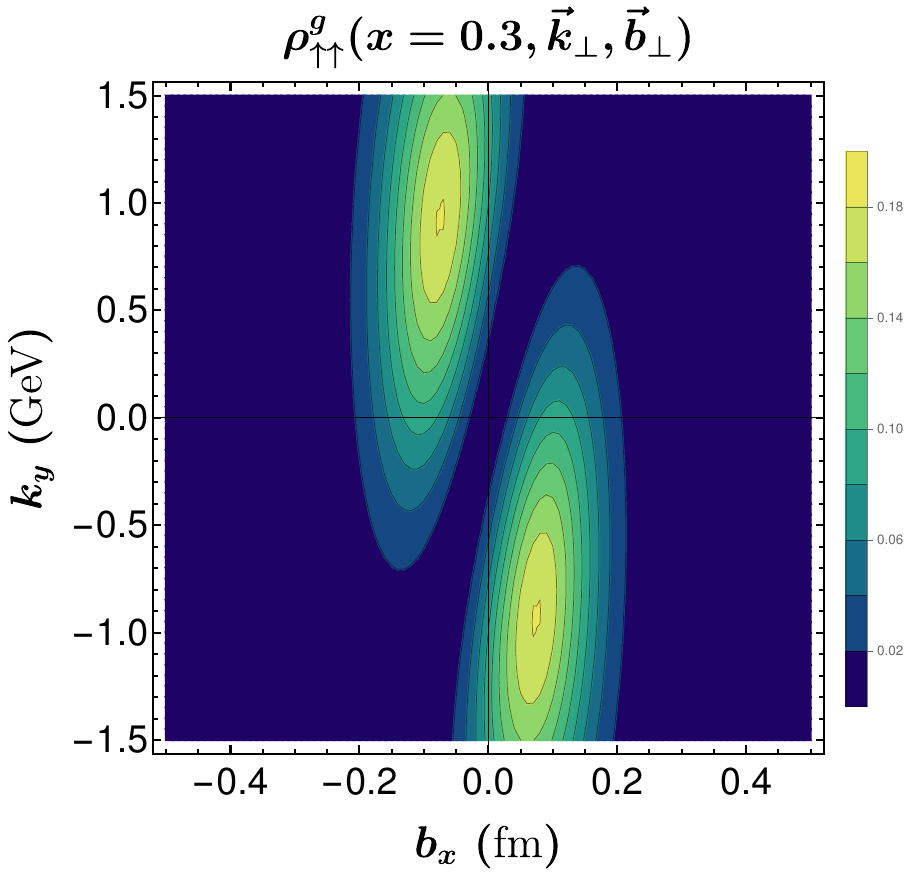}
        \includegraphics[width=0.325\linewidth]{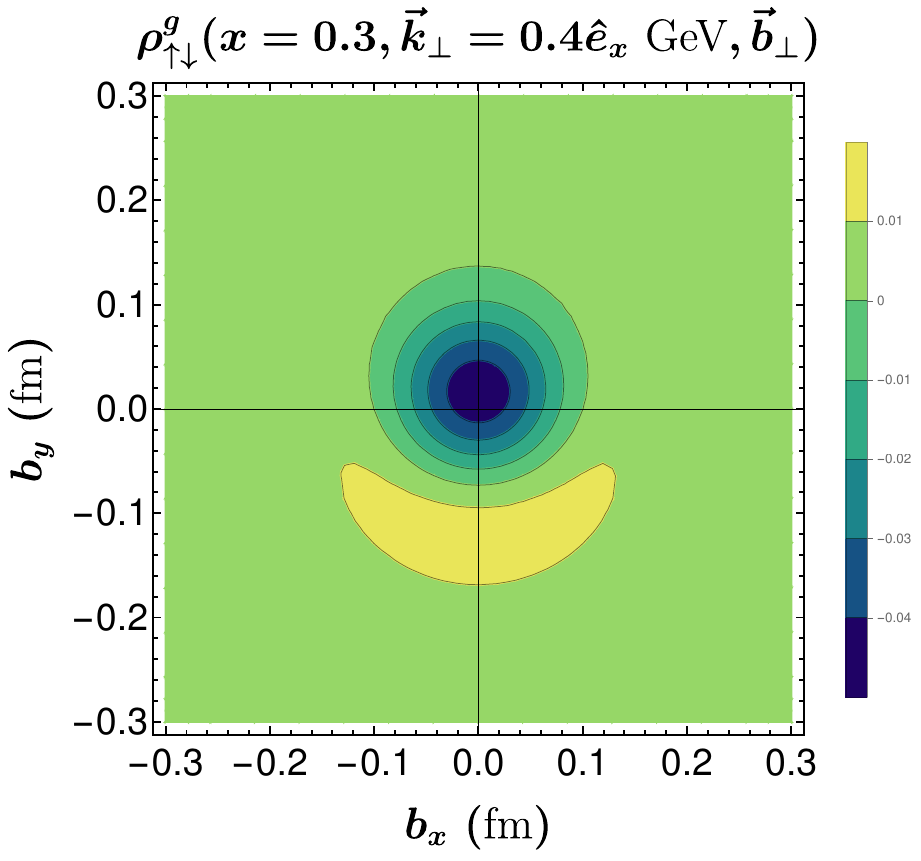}
        \includegraphics[width=0.325\linewidth]{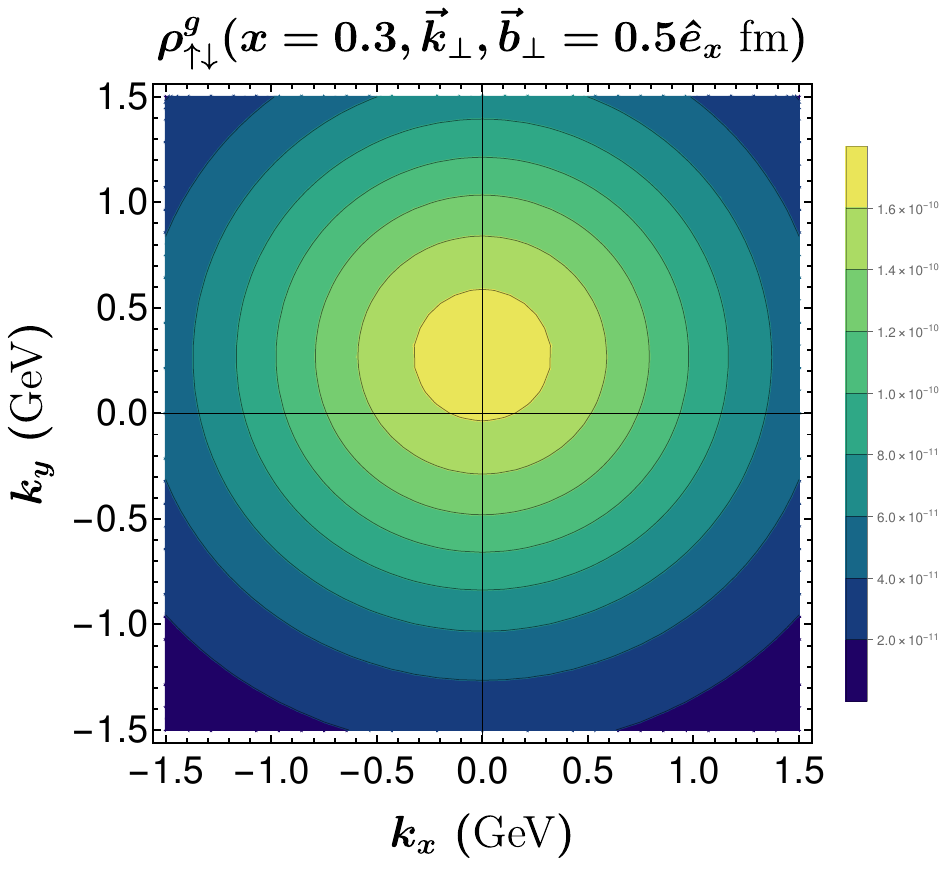}
        \includegraphics[width=0.325\linewidth]{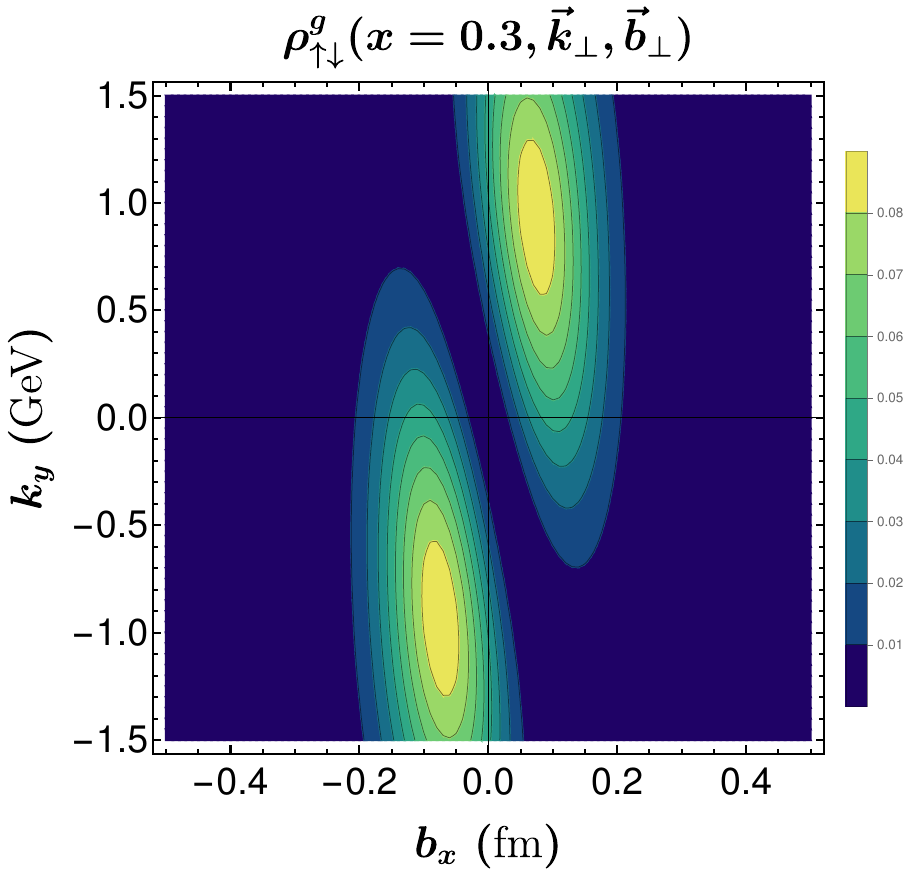}
        \caption{
        \justifying
        Upper panel: Spin–spin correlation function for gluon and proton spins aligned in the same direction, shown in impact-parameter space (left), transverse-momentum plane (middle), and mixed plane (right) at a fixed value of $x=0.3$. Lower panel: Same as the upper panel, but with gluon and proton spins aligned in opposite directions.}
        \label{fig:Figure11}
\end{figure}
}}

\section{conclusion}\label{sec:conclusion}
In this work, we studied the leading-twist gluon GTMDs for unpolarized and longitudinally polarized gluons at nonzero skewness of the proton within {{a recently developed}} light-front gluon spectator model inspired by the soft-wall AdS/QCD framework, {{ where the proton is considered as an active gluon and a spin $1/2$ spectator}}. We obtained the numerical results for the  GTMDs {{with non-zero skewness}} and presented them in the DGLAP region ($x>\xi$). The gluon GTMDs were found to peak near $x \rightarrow \xi$ and gradually shift toward larger values of $x$ with decreasing magnitude, with their dominant contributions concentrated in the small-$x$ region. Furthermore, using the GTMDs $F_{1,4}^{g}$ and $G_{1,1}^{g}$, we evaluated the gluon canonical OAM and the spin–orbit correlation factor, and found that both are distributed negatively with respect to $x$. The negative values of the former indicate a reduction of total gluon angular momentum contribution to the total nucleon spin, while the latter implies that the gluon spin and OAM are anti-aligned. We have also shown that the gluon GPDs and TMDs can be obtained as specific projections of the gluon GTMDs. Using these gluon GPDs, we evaluated the gluon total angular momentum and kinetic OAM, and found our predictions to be consistent with results available in the literature. In particular, the total angular momentum shows good agreement with lattice QCD calculations. Furthermore, we investigated the gluon Wigner distributions for unpolarized, longitudinally polarized, and transversely polarized proton targets, considering unpolarized, longitudinally polarized, and linearly polarized gluons. These Wigner distributions are derived via a two-dimensional Fourier transform of the GTMD correlators {{with respect to}} the transverse momentum transfer at zero skewness. We found that the Wigner distributions are not positive definite in our model. All the Wigner distributions are expressed as the overlap representation of LFWFs and computed analytically. We have shown that the Wigner distributions for linearly polarized gluons can be obtained as a linear combination of unpolarized and longitudinally polarized gluons. We present the numerical results for gluon Wigner distributions {{in impact parameter space}} at fixed $\bfk$, transverse momentum plane at fixed $\bfb$, as well as in the mixed plane. We found that the Wigner distributions of unpolarized gluons in an unpolarized proton, $\mathcal{W}_{UU}^{g}$, as well as those of longitudinally polarized gluons in a longitudinally polarized proton, $\mathcal{W}_{LL}^{g}$, exhibit circular symmetry in both transverse momentum and impact parameter planes. In contrast, the Wigner distributions of longitudinally polarized gluons in an unpolarized proton, $\mathcal{W}_{UL}^{g}$, and those of unpolarized gluons in a longitudinally polarized proton, $\mathcal{W}_{LU}^{g}$, display a dipolar structure in the transverse momentum and impact parameter spaces, while showing a quadrupole pattern in the mixed space. On the other hand, the Wigner distributions of linearly polarized gluons (left- and right-handed) in both unpolarized proton, $\mathcal{W}_{U\mathcal{T}}^{R/L}$, and longitudinally polarized proton, $\mathcal{W}_{L\mathcal{T}}^{R/L}$ reveal distortions in the impact parameter and transverse momentum planes, respectively. Similarly, the Wigner distributions of unpolarized gluons in a transversely polarized proton, $\mathcal{W}_{TU}^{x}$, exhibit a dipolar pattern in impact parameter space and circular symmetry in momentum space, whereas the opposite behavior is observed for the Wigner distributions of longitudinally polarized gluons in a transversely polarized proton, $\mathcal{W}_{TL}^{x}$. Both of these distributions display a dipolar structure in the mixed space. Furthermore, the Wigner distributions of right-handed linearly polarized gluons in a transversely polarized proton, $\mathcal{W}_{T\mathcal{T}}^{R}$, reveal a distortion in impact parameter space and a dipolar pattern in transverse momentum space, while the left-handed linearly polarized gluons, $\mathcal{W}_{T\mathcal{T}}^{L}$, exhibit the opposite behavior. Those distortions provide direct insights into the internal dynamics of the proton. For instance, the unpolarized proton distortions arise due to the intrinsic transverse momentum anisotropy of gluons and describe the gluon quadrupole distributions, for longitudinally polarized proton distortions arise from spin–OAM correlations, and for the transversely polarized proton, they highlight the spin-transversity correlations, by giving dipole-like asymmetries. We also studied the spin densities and spin-spin correlations between the gluon and proton spins in transverse momentum and position space.


\appendix
\section{Bilinear decompositions of gluon-gluon correlator and GTMDs}\label{AppendixA}
The \textit{leading-twist} gluon-gluon GTMD correlator for unpolarized gluons ($\Gamma^{ij}=\delta_{\perp}^{ij}$) can be parametrized in terms of four $F$-type gluon GTMDs~\cite{Lorce:2013pza} as follows,
\begin{align}\label{eq:F-type parametrization}
    		W^g_{\lambda^{\prime\prime}\lambda^{\prime}}=&\delta_\perp^{ij} W^{g[ij]}_{\lambda^{\prime\prime}\lambda^{\prime}}\nonumber\\
	=&\frac{1}{2M}\bar{U}(P^{\prime\prime},\lambda^{\prime\prime})\bigg[F_{1,1}^g+\frac{i\sigma^{i+}k_\perp^i}{P^+}F^g_{1,2} +\frac{i\sigma^{i+}\Delta_\perp^i}{P^+}F^g_{1,3}+\frac{i\sigma^{ij}k_\perp^i\Delta^j_\perp}{M^2}F^g_{1,4}\bigg]U(P^{\prime},\lambda^{\prime}) \nonumber\\
	=&\frac{1}{M\sqrt{1-\xi^2}} \bigg\{\bigg[M\delta_{\lambda^{\prime\prime}\lambda^{\prime}}-\frac{1}{2}(\lambda^{\prime}\Delta_\perp^{(1)}+i\Delta_\perp^{(2)})\delta_{-\lambda^{\prime\prime},\lambda^{\prime}}\bigg]F^g_{1,1} +(1-\xi^2)(\lambda^{\prime} k_\perp^{(1)}+ik_\perp^{(2)})\delta_{-\lambda^{\prime\prime},\lambda^{\prime}}F_{1,2}^g \nonumber \\
	&+(1-\xi^2)(\lambda\Delta_\perp^{(1)}+i\Delta_\perp^{(2)})\delta_{-\lambda^{\prime\prime},\lambda^{\prime}}F^g_{1,3} +\frac{i\epsilon_\perp^{ij}k_\perp^i\Delta_\perp^j}{M^2} \bigg[\lambda^{\prime} M \delta_{\lambda^{\prime\prime},\lambda^{\prime}}-\frac{\xi}{2}(\Delta_\perp^{(1)}+i\lambda^{\prime}\Delta_\perp^{(2)})\delta_{-\lambda^{\prime\prime},\lambda^{\prime}}\bigg]F_{1,4}^g \bigg\}\,.
\end{align}
Similarly, for longitudinally polarized gluons ($\Gamma^{ij}=-i\epsilon_{\perp}^{ij}$), the gluon-gluon GTMD correlator can parametrized in terms of four $G$-type gluon GTMDs~\cite{Lorce:2013pza} as,
\begin{align}\label{eq:G-type parametrization}
    	\widetilde{W}^g_{\lambda^{\prime\prime}\lambda^{\prime}}=&-i\epsilon_\perp^{ij} W^{g[ij]}_{\lambda^{\prime\prime}\lambda^{\prime}} \nonumber\\
	=&\frac{1}{2M}\bar{U}(P^{\prime\prime},\lambda^{\prime\prime})\bigg[-\frac{i\epsilon_\perp^{ij}k_\perp^i\Delta_\perp^j}{M^2} G_{1,1}^g+\frac{i\sigma^{i+}\gamma_5 k_\perp^i}{P^+}G^g_{1,2} +\frac{i\sigma^{i+}\gamma_5 \Delta_\perp^i}{P^+}G^g_{1,3}+i\sigma^{+-}\gamma_5 G^g_{1,4}\bigg]U(P^{\prime},\lambda^{\prime}) \nonumber\\
	=&\frac{1}{M\sqrt{1-\xi^2}} \bigg\{-\frac{i\epsilon^{ij}_\perp k_\perp^i\Delta_\perp^j}{M^2} \bigg[M\delta_{\lambda^{\prime\prime},\lambda^{\prime}}-\frac{1}{2}(\lambda^{\prime}\Delta_\perp^{(1)}+i\Delta_\perp^{(2)})\delta_{-\lambda^{\prime\prime},\lambda^{\prime}}\bigg]G^g_{1,1} +(1-\xi^2)(k_\perp^{(1)}+i\lambda^{\prime} k_\perp^{(2)})\delta_{-\lambda^{\prime\prime},\lambda^{\prime}}G_{1,2}^g \nonumber \\
	&+(1-\xi^2)(\Delta_\perp^{(1)}+i\lambda^{\prime}\Delta_\perp^{(2)})\delta_{-\lambda^{\prime\prime},\lambda^{\prime}}G^g_{1,3} +\bigg[\lambda^{\prime} M \delta_{\lambda^{\prime\prime},\lambda^{\prime}}-\frac{\xi}{2}(\Delta^{(1)}_\perp+i\lambda\Delta^{(2)}_\perp)\delta_{-\lambda^{\prime\prime},\lambda^{\prime}}\bigg]G_{1,4}^g \bigg\}\,.
\end{align}
where, $\bar{U}(p^{\prime\prime},\lambda^{\prime\prime})$ and ${U}(p^{\prime},\lambda^{\prime})$ are the outgoing and incoming Dirac spinors, and $M$ represent the proton mass. The two tensors $\delta_{\perp}^{ij}$ and $\epsilon_{\perp}^{ij}$ can be written in terms of metric tensor $\delta_{\perp}^{ij}=-g_{\perp}^{ij}$ and antisymmetric tensor $\epsilon_{\perp}^{ij}=\epsilon^{+-ij}$ with $\epsilon^{+-12}=1$, respectively.
Further the two $F$-type gluon GTMDs $F_{1,1}^{g}$, and $F_{1,4}^{g}$ can be extract with proton helicity conserving unpolarized gluon GTMD correlators, $W^{g}_{\uparrow\uparrow}$ and $W^{g}_{\downarrow\downarrow}$ as,
{\begin{align}\label{eq:F11}
  W_{\uparrow\uparrow}^{g}(x,\xi,\mathbf{k}_{\perp},\mathbf{\Delta}_{\perp})+W_{\downarrow\downarrow}^{g}(x,\xi,\mathbf{k}_{\perp},\mathbf{\Delta}_{\perp})=&\frac{2}{\sqrt{1-\xi^2}}F^{g}_{1,1}\,,\\
    W_{\uparrow\uparrow}^{g}(x,\xi,\mathbf{k}_{\perp},\mathbf{\Delta}_{\perp})-W_{\downarrow\downarrow}^{g}(x,\xi,\mathbf{k}_{\perp},\mathbf{\Delta}_{\perp})=&\frac{2}{\sqrt{1-\xi^2}}\frac{i\varepsilon^{ij}_{\perp}\mathbf{k}_{\perp}^{i}\mathbf{\Delta}_{\perp}^{j}}{M^2}F^{g}_{1,4}\,.
    \label{eq:F14}
\end{align}
while the remaining two F-type GTMDs, $F_{1,2}^{g}$ and $F_{1,3}^{g}$ can be extracted from proton helicity flip unpolarized gluon GTMD correlators, $W^{g}_{\uparrow\downarrow}$ and $W^{g}_{\downarrow\uparrow}$ as,
  {\begin{align}\label{eq:GTMDsparametrization12}
         W_{\uparrow\downarrow}^{g}(x,\xi,\mathbf{k}_{\perp},\mathbf{\Delta}_{\perp})+W_{\downarrow\uparrow}^{g}(x,\xi,\mathbf{k}_{\perp},\mathbf{\Delta}_{\perp})=&\frac{1}{M\sqrt{1-\xi^2}}\Big[-i\Delta_{\perp}^{(2)}F^{g}_{1,1}+2i(1-\xi^2)\Big{\{}\Delta_{\perp}^{(2)}F^{g}_{1,3}+k_{\perp}^{(2)}F^{g}_{1,2}\Big{\}}\nonumber\\
         &-i\Delta_{\perp}^{(1)}\frac{\xi}{M^2}\varepsilon^{ij}_{\perp}{k}_{\perp}^{i}{\Delta}_{\perp}^{j}F^{g}_{1,4}\Big]\,,\\
          W_{\uparrow\downarrow}^{g}(x,\xi,\mathbf{k}_{\perp},\mathbf{\Delta}_{\perp})-W_{\downarrow\uparrow}^{g}(x,\xi,\mathbf{k}_{\perp},\mathbf{\Delta}_{\perp})=&\frac{1}{M\sqrt{1-\xi^2}}\Big[\Delta_{\perp}^{(1)}F^{g}_{1,1}-2(1-\xi^2)\Big{\{}\Delta_{\perp}^{(1)}F^{g}_{1,3}+k_{\perp}^{(1)}F^{g}_{1,2}\Big{\}}\nonumber\\
          &-\Delta_{\perp}^{(2)}\frac{\xi}{M^2}\varepsilon^{ij}_{\perp}{k}_{\perp}^{i}{\Delta}_{\perp}^{j}F^{g}_{1,4}\Big]\,.
          \label{eq:GTMDsparametrization13}
    \end{align}}
Similarly, the G-type gluon GTMDs $G_{1,1}^{g}$, and $G_{1,4}^{g}$ can be extract from proton helicity conserving longitudinally polarized gluon GTMD correlators $\widetilde{W}^g_{\uparrow\uparrow}$ and $\widetilde{W}^g_{\downarrow\downarrow}$ as,
{\begin{align}\label{eq:G14}
     \widetilde{W}^g_{\uparrow\uparrow}(x,\xi,\mathbf{k}_{\perp},\mathbf{\Delta}_{\perp})-\widetilde{W}^g_{\downarrow\downarrow}(x,\xi,\mathbf{k}_{\perp},\mathbf{\Delta}_{\perp})=&\frac{2}{\sqrt{1-\xi^2}}G^{g}_{1,4}\,,\\
    \widetilde{W}^g_{\uparrow\uparrow}(x,\xi,\mathbf{k}_{\perp},\mathbf{\Delta}_{\perp})+\widetilde{W}^g_{\downarrow\downarrow}(x,\xi,\mathbf{k}_{\perp},\mathbf{\Delta}_{\perp})=&-\frac{2}{M^2{\sqrt{1-\xi^2}}}i\varepsilon^{ij}_{\perp}\mathbf{k}_{\perp}^{i}\mathbf{\Delta}_{\perp}^{j}G^{g}_{1,1}\,.
\end{align}}
while the GTMDs, $G_{1,2}^{g}$ and $G_{1,3}^{g}$ can be extracted from proton helicity flip longitudinally polarized gluon GTMD correlators $\widetilde{W}^g_{\uparrow\downarrow}$ and $\widetilde{W}^g_{\downarrow\uparrow}$ as,
  \begin{align}
\widetilde{W}^g_{\uparrow,\downarrow}(x,\xi,\mathbf{k}_{\perp},\mathbf{\Delta}_{\perp})+\widetilde{W}^g_{\downarrow,\uparrow}(x,\xi,\mathbf{k}_{\perp},\mathbf{\Delta}_{\perp})=&\frac{1}{M\sqrt{1-\xi^2}}\Big[-\frac{\varepsilon_{\perp}^{ij}k_{\perp}^{i}\Delta_{\perp}^{j}}{M^2}\Delta_{\perp}^{(2)}G_{1,1}^{g}+2(1-\xi^2)\Big{\{}k_{\perp}^{(1)}G_{1,2}^{g}+\Delta_{\perp}^{(1)}G_{1,3}^{g}\Big{\}}\nonumber\\
&-{\xi}\Delta_{\perp}^{(1)}G_{1,4}^{g}\Big]\,,\\
\widetilde{W}^g_{\uparrow,\downarrow}(x,\xi,\mathbf{k}_{\perp},\mathbf{\Delta}_{\perp})-\widetilde{W}^g_{\downarrow,\uparrow}(x,\xi,\mathbf{k}_{\perp},\mathbf{\Delta}_{\perp})=&\frac{1}{M\sqrt{1-\xi^2}}\Big[-\frac{i\varepsilon_{\perp}^{ij}k_{\perp}^{i}\Delta_{\perp}^{j}}{M^2}\Delta_{\perp}^{(1)}G_{1,1}^{g}-2i(1-\xi^2)\Big{\{}k_{\perp}^{(2)}G_{1,2}^{g}+\Delta_{\perp}^{(2)}G_{1,3}^{g}\Big{\}}\nonumber\\
&+i{\xi}\Delta_{\perp}^{(2)}G_{1,4}^{g}\Big]\,.
\label{eq:parametrizationG13}
\end{align}
 After employing the proton LFWFs from Eqs.~\eqref{LFWFsuparrow} and~\eqref{LFWFsdownarrow} into Eq.~\eqref{eq:correlator1}, one can obtain the analytical expression for unpolarized gluon GTMD correlators $W^{g}_{\uparrow\uparrow}$, $W^{g}_{\downarrow\downarrow}$, $W^{g}_{\uparrow\downarrow}$, and $W^{g}_{\downarrow\uparrow}$ as follows,
\begin{align}\label{eq:unpolcorrelatorupup}
    W^{g}_{\uparrow\uparrow}(x,\xi,\mathbf{k}_{\perp},\mathbf{\Delta}_{\perp})=&\frac{2N_{g}^{2}}{\pi\kappa^2}\Big[\mathcal{F}_{a}(x^\prime,x^{\prime\prime})\Big(\mathbf{k}_{\perp}^{2}-(1-x^{\prime})(1-x^{\prime\prime})\frac{\mathbf{\Delta}_{\perp}^{2}}{4}\Big)+\mathcal{F}_{b}(x^\prime,x^{\prime\prime})+\Big{\{}\frac{\mathbf{\Delta}_{\perp}^{r}.\mathbf{k}_{\perp}^{l}}{2}\mathcal{F}_{c}(x^{\prime},x^{\prime\prime})-\frac{\mathbf{\Delta}_{\perp}^{l}.\mathbf{k}_{\perp}^{r}}{2}\nonumber\\
    &\times\mathcal{F}_{d}(x^{\prime},x^{\prime\prime})\Big{\}}\Big]\exp\Big[-\mathcal{A}(x^{\prime},x^{\prime\prime})\mathbf{k}_{\perp}^{2}- \mathcal{B}(x^{\prime},x^{\prime\prime})\frac{\mathbf{\Delta}_{\perp}^{2}}{4}- \mathcal{C} (x^{\prime},x^{\prime\prime})(\mathbf{k}_{\perp}.\mathbf{\Delta}_{\perp})\Big]\,,\\
    W^{g}_{\downarrow\downarrow}(x,\xi,\mathbf{k}_{\perp},\mathbf{\Delta}_{\perp})=&\frac{2N_{g}^{2}}{\pi\kappa^2}\Big[\mathcal{F}_{a}(x^\prime,x^{\prime\prime})\Big(\mathbf{k}_{\perp}^{2}-(1-x^{\prime})(1-x^{\prime\prime})\frac{\mathbf{\Delta}_{\perp}^{2}}{4}\Big)+\mathcal{F}_{b}(x^\prime,x^{\prime\prime})+\Big{\{}\frac{\mathbf{\Delta}_{\perp}^{l}.\mathbf{k}_{\perp}^{r}}{2}\mathcal{F}_{c}(x^{\prime},x^{\prime\prime})-\frac{\mathbf{\Delta}_{\perp}^{r}.\mathbf{k}_{\perp}^{l}}{2}\nonumber\\
    &\times \mathcal{F}_{d}(x^{\prime},x^{\prime\prime})\Big{\}}\Big]\times\exp\Big[-\mathcal{A}(x^{\prime},x^{\prime\prime})\mathbf{k}_{\perp}^{2}- \mathcal{B}(x^{\prime},x^{\prime\prime})\frac{\mathbf{\Delta}_{\perp}^{2}}{4}-\mathcal{C} (x^{\prime},x^{\prime\prime})(\mathbf{k}_{\perp}.\mathbf{\Delta}_{\perp})\Big]\,,
    \end{align}    
 \begin{align} 
    W^{g}_{\uparrow\downarrow}(x,\xi,\mathbf{k}_{\perp},\mathbf{\Delta}_{\perp})=&\frac{2N_{g}^{2}}{\pi\kappa^{2}}\Big[-\mathcal{F}_{e}(x^{\prime},x^{\prime\prime})\mathbf{k}_{\perp}^{l}+\mathcal{F}_{f}(x^{\prime},x^{\prime\prime})\frac{\mathbf{\Delta}_{\perp}^{l}}{2}\Big]\exp\Big[-\mathcal{A}(x^{\prime},x^{\prime\prime})\mathbf{k}_{\perp}^{2}- \mathcal{B}(x^{\prime},x^{\prime\prime})\frac{\mathbf{\Delta}_{\perp}^{2}}{4}\nonumber\\
    & -\mathcal{C} (x^{\prime},x^{\prime\prime})(\mathbf{k}_{\perp}.\mathbf{\Delta}_{\perp})\Big]\,,\\
    W^{g}_{\downarrow\uparrow}(x,\xi,\mathbf{k}_{\perp},\mathbf{\Delta}_{\perp}) =&\frac{2N_{g}^{2}}{\pi\kappa^{2}}\Big[\mathcal{F}_{e}(x^{\prime},x^{\prime\prime})\mathbf{k}_{\perp}^{r}-\mathcal{F}_{f}(x^{\prime},x^{\prime\prime})\frac{\mathbf{\Delta}_{\perp}^{r}}{2}\Big]\exp\Big[-\mathcal{A}(x^{\prime},x^{\prime\prime})\mathbf{k}_{\perp}^{2}- \mathcal{B}(x^{\prime},x^{\prime\prime})\frac{\mathbf{\Delta}_{\perp}^{2}}{4}\nonumber\\
    &-\mathcal{C} (x^{\prime},x^{\prime\prime})(\mathbf{k}_{\perp}.\mathbf{\Delta}_{\perp})\Big]\,.
    \label{eq:unpolcorrelatordownup}
\end{align}
Where, $\mathbf{k}^{r(l)}_{\perp}=k_{\perp}^{(1)}\pm ik_{\perp}^{(2)}$ and $\mathbf{\Delta}^{r(l)}_{\perp}=\Delta_{\perp}^{(1)}\pm i\Delta_{\perp}^{(2)}$, respectively. Similarly, the analytical expressions for longitudinally polarized gluon GTMD correlators $\widetilde{W}^{g}_{\uparrow\uparrow}$, $\widetilde{W}^{g}_{\downarrow\downarrow}$, $\widetilde{W}^{g}_{\uparrow\downarrow}$, and $\widetilde{W}^{g}_{\downarrow\uparrow}$ can be obtain by using the proton LFWFs from Eqs.~\eqref{LFWFsuparrow} and~\eqref{LFWFsdownarrow} into Eq.~\eqref{eq:correlator2} as follows,
\begin{align}\label{eq:longpolcorrelatorupup}
 \widetilde{W}^{g}_{\uparrow\uparrow}(x,\xi,\mathbf{k}_{\perp},\mathbf{\Delta}_{\perp})=&\frac{2N_{g}^{2}}{\pi\kappa^2}\Big[\mathcal{F}_{i}(x^\prime,x^{\prime\prime})\Big(\mathbf{k}_{\perp}^{2}-(1-x^{\prime})(1-x^{\prime\prime})\frac{\mathbf{\Delta}_{\perp}^{2}}{4}\Big)+\mathcal{F}_{b}(x^\prime,x^{\prime\prime})+\Big{\{}\frac{\mathbf{\Delta}_{\perp}^{r}.\mathbf{k}_{\perp}^{l}}{2}\mathcal{F}_{g}(x^{\prime},x^{\prime\prime})-\frac{\mathbf{\Delta}_{\perp}^{l}.\mathbf{k}_{\perp}^{r}}{2}\nonumber\\
    &\times\mathcal{F}_{h}(x^{\prime},x^{\prime\prime})\Big{\}}\Big]\exp\Big[-\mathcal{A}(x^{\prime},x^{\prime\prime})\mathbf{k}_{\perp}^{2}- \mathcal{B}(x^{\prime},x^{\prime\prime})\frac{\mathbf{\Delta}_{\perp}^{2}}{4}- \mathcal{C} (x^{\prime},x^{\prime\prime})(\mathbf{k}_{\perp}.\mathbf{\Delta}_{\perp})\Big]\,,\\
    \widetilde{W}_{\downarrow\downarrow}(x,\xi,\mathbf{k}_{\perp},\mathbf{\Delta}_{\perp})=&-\frac{2N_{g}^{2}}{\pi\kappa^2}\Big[\mathcal{F}_{i}(x^\prime,x^{\prime\prime})\Big(\mathbf{k}_{\perp}^{2}-(1-x^{\prime})(1-x^{\prime\prime})\frac{\mathbf{\Delta}_{\perp}^{2}}{4}\Big)+\mathcal{F}_{b}(x^\prime,x^{\prime\prime})+\Big{\{}\frac{\mathbf{\Delta}_{\perp}^{l}.\mathbf{k}_{\perp}^{r}}{2}\mathcal{F}_{g}(x^{\prime},x^{\prime\prime})-\frac{\mathbf{\Delta}_{\perp}^{r}.\mathbf{k}_{\perp}^{l}}{2}\nonumber\\
    &\times\mathcal{F}_{h}(x^{\prime},x^{\prime\prime})\Big{\}}\Big]\exp\Big[-\mathcal{A}(x^{\prime},x^{\prime\prime})\mathbf{k}_{\perp}^{2}- \mathcal{B}(x^{\prime},x^{\prime\prime})\frac{\mathbf{\Delta}_{\perp}^{2}}{4}- \mathcal{C} (x^{\prime},x^{\prime\prime})(\mathbf{k}_{\perp}.\mathbf{\Delta}_{\perp})\Big]\,,\\
   \widetilde{W}^{g}_{\uparrow\downarrow}(x,\xi,\mathbf{k}_{\perp},\mathbf{\Delta}_{\perp})=&\frac{2N_{g}^{2}}{\pi\kappa^{2}}\Big[-\mathcal{F}_{k}(x^\prime,x^{\prime\prime})\mathbf{k}_{\perp}^{l}+\mathcal{F}_{j}(x^\prime,x^{\prime\prime})\frac{\mathbf{\Delta}_{\perp}^{l}}{2}\Big]\exp\Big[-\mathcal{A}(x^{\prime},x^{\prime\prime})\mathbf{k}_{\perp}^{2}- \mathcal{B}(x^{\prime},x^{\prime\prime})\frac{\mathbf{\Delta}_{\perp}^{2}}{4}\nonumber\\
   &-\mathcal{C} (x^{\prime},x^{\prime\prime})(\mathbf{k}_{\perp}.\mathbf{\Delta}_{\perp})\Big]\,,\\
   \widetilde{W}^{g}_{\downarrow\uparrow}(x,\xi,\mathbf{k}_{\perp},\mathbf{\Delta}_{\perp})=&\frac{2N_{g}^{2}}{\pi\kappa^{2}}\Big[-\mathcal{F}_{k}(x^\prime,x^{\prime\prime})\mathbf{k}_{\perp}^{r}+\mathcal{F}_{j}(x^\prime,x^{\prime\prime})\frac{\mathbf{\Delta}_{\perp}^{r}}{2}\Big]\exp\Big[-\mathcal{A}(x^{\prime},x^{\prime\prime})\mathbf{k}_{\perp}^{2}- \mathcal{B}(x^{\prime},x^{\prime\prime})\frac{\mathbf{\Delta}_{\perp}^{2}}{4}\nonumber\\
   &-\mathcal{C} (x^{\prime},x^{\prime\prime})(\mathbf{k}_{\perp}.\mathbf{\Delta}_{\perp})\Big]\,.
   \label{eq:longpolcorrelatordownup}
\end{align}
\section{Parameterized functions}
\begin{align}\label{eq:AppendixFaxx}
    \mathcal{F}_{a}(x^{\prime},x^{\prime\prime})=&\Big{\{}\frac{1+(1-x^{\prime})(1-x^{\prime\prime})}{x^{\prime}x^{\prime\prime}(1-x^{\prime})(1-x^{\prime\prime})}\Big{\}}\sqrt{\frac{\ln\Big(\frac{1}{1-x^{\prime}}\Big)\ln\Big(\frac{1}{1-x^{\prime\prime}}\Big)}{x^{\prime}x^{\prime\prime}}}(x^{\prime}x^{\prime\prime})^{b}\Big((1-x^{\prime})(1-x^{\prime\prime})\Big)^{a}\,,\\
    \mathcal{F}_{b}(x^{\prime},x^{\prime\prime})=&\Big(M-\frac{M_{X}}{(1-x^{\prime})}\Big)\Big(M-\frac{M_{X}}{(1-x^{\prime\prime})}\Big)\sqrt{\frac{\ln\Big(\frac{1}{1-x^{\prime}}\Big)\ln\Big(\frac{1}{1-x^{\prime\prime}}\Big)}{x^{\prime}x^{\prime\prime}}}(x^{\prime}x^{\prime\prime})^{b}\Big((1-x^{\prime})(1-x^{\prime\prime})\Big)^{a}\,,\\
    \mathcal{F}_{c}(x^{\prime},x^{\prime\prime})=&\Big{\{}\frac{1-(1-x^{\prime\prime})^2}{x^{\prime}x^{\prime\prime}(1-x^{\prime\prime})}\Big{\}}\sqrt{\frac{\ln\Big(\frac{1}{1-x^{\prime}}\Big)\ln\Big(\frac{1}{1-x^{\prime\prime}}\Big)}{x^{\prime}x^{\prime\prime}}}(x^{\prime}x^{\prime\prime})^{b}\Big((1-x^{\prime})(1-x^{\prime\prime})\Big)^{a}\,,\\
    \mathcal{F}_{d}(x^{\prime},x^{\prime\prime})=&\Big{\{}\frac{1-(1-x^{\prime})^2}{x^{\prime}x^{\prime\prime}(1-x^{\prime})}\Big{\}}\sqrt{\frac{\ln\Big(\frac{1}{1-x^{\prime}}\Big)\ln\Big(\frac{1}{1-x^{\prime\prime}}\Big)}{x^{\prime}x^{\prime\prime}}}(x^{\prime}x^{\prime\prime})^{b}\Big((1-x^{\prime})(1-x^{\prime\prime})\Big)^{a}\,,\\
    \mathcal{F}_{e}(x^{\prime},x^{\prime\prime})=&\Big{\{}\frac{1}{x^{\prime\prime}}\Big(M-\frac{M_{X}}{1-x^{\prime}}\Big)-\frac{1}{x^{\prime}}\Big(M-\frac{M_{X}}{1-x^{\prime\prime}}\Big)\Big{\}}\sqrt{\frac{\ln\Big(\frac{1}{1-x^{\prime}}\Big)\ln\Big(\frac{1}{1-x^{\prime\prime}}\Big)}{x^{\prime}x^{\prime\prime}}}(x^{\prime}x^{\prime\prime})^{b}\Big((1-x^{\prime})(1-x^{\prime\prime})\Big)^{a}\label{eq:Fe}\,,\\
    \mathcal{F}_{f}(x^{\prime},x^{\prime\prime})=&\Big{\{}\frac{(1-x^{\prime})}{x^{\prime}}\Big(M-\frac{M_{X}}{1-x^{\prime\prime}}\Big)+\frac{(1-x^{\prime\prime})}{x^{\prime\prime}}\Big(M-\frac{M_{X}}{1-x^{\prime}}\Big)\Big{\}}\sqrt{\frac{\ln\Big(\frac{1}{1-x^{\prime}}\Big)\ln\Big(\frac{1}{1-x^{\prime\prime}}\Big)}{x^{\prime}x^{\prime\prime}}}(x^{\prime}x^{\prime\prime})^{b}\Big((1-x^{\prime})(1-x^{\prime\prime})\Big)^{a}\,,
    \end{align}
    \begin{align}
    \mathcal{F}_{g}(x^{\prime},x^{\prime\prime})=&\Big{\{}\frac{1+(1-x^{\prime\prime})^2}{x^{\prime}x^{\prime\prime}(1-x^{\prime\prime})}\Big{\}}\sqrt{\frac{\ln\Big(\frac{1}{1-x^{\prime}}\Big)\ln\Big(\frac{1}{1-x^{\prime\prime}}\Big)}{x^{\prime}x^{\prime\prime}}}(x^{\prime}x^{\prime\prime})^{b}\Big((1-x^{\prime})(1-x^{\prime\prime})\Big)^{a}\,,\\
    \mathcal{F}_{h}(x^{\prime},x^{\prime\prime})=&\Big{\{}\frac{1+(1-x^{\prime})^2}{x^{\prime}x^{\prime\prime}(1-x^{\prime})}\Big{\}}\sqrt{\frac{\ln\Big(\frac{1}{1-x^{\prime}}\Big)\ln\Big(\frac{1}{1-x^{\prime\prime}}\Big)}{x^{\prime}x^{\prime\prime}}}(x^{\prime}x^{\prime\prime})^{b}\Big((1-x^{\prime})(1-x^{\prime\prime})\Big)^{a}\,,\\
    \mathcal{F}_{i}(x^{\prime},x^{\prime\prime})=&\Big{\{}\frac{1-(1-x^{\prime})(1-x^{\prime\prime})}{x^{\prime}x^{\prime\prime}(1-x^{\prime})(1-x^{\prime\prime})}\Big{\}}\sqrt{\frac{\ln\Big(\frac{1}{1-x^{\prime}}\Big)\ln\Big(\frac{1}{1-x^{\prime\prime}}\Big)}{x^{\prime}x^{\prime\prime}}}(x^{\prime}x^{\prime\prime})^{b}\Big((1-x^{\prime})(1-x^{\prime\prime})\Big)^{a}\,,\\
\mathcal{F}_{j}(x^{\prime},x^{\prime\prime})=&\Big{\{}\frac{(1-x^{\prime\prime})}{x^{\prime\prime}}\Big(M-\frac{M_{X}}{1-x^{\prime}}\Big)-\frac{(1-x^{\prime})}{x^{\prime}}\Big(M-\frac{M_{X}}{1-x^{\prime\prime}}\Big)\Big{\}}\sqrt{\frac{\ln\Big(\frac{1}{1-x^{\prime}}\Big)\ln\Big(\frac{1}{1-x^{\prime\prime}}\Big)}{x^{\prime}x^{\prime\prime}}}(x^{\prime}x^{\prime\prime})^{b}\Big((1-x^{\prime})(1-x^{\prime\prime})\Big)^{a}\label{eq:Fj}\,,\\
    \mathcal{F}_{k}(x^{\prime},x^{\prime\prime})=&\Big{\{}\frac{1}{x^{\prime\prime}}\Big(M-\frac{M_{X}}{1-x^{\prime}}\Big)+\frac{1}{x^{\prime}}\Big(M-\frac{M_{X}}{1-x^{\prime\prime}}\Big)\Big{\}}\sqrt{\frac{\ln\Big(\frac{1}{1-x^{\prime}}\Big)\ln\Big(\frac{1}{1-x^{\prime\prime}}\Big)}{x^{\prime}x^{\prime\prime}}}(x^{\prime}x^{\prime\prime})^{b}\Big((1-x^{\prime})(1-x^{\prime\prime})\Big)^{a}\,,\\
    \mathcal{A}(x^{\prime},x^{\prime\prime})=&\Bigg{\{}\frac{\ln\Big(\frac{1}{1-x^{\prime}}\Big)}{2\kappa^{2}x^{\prime2}}+\frac{\ln\Big(\frac{1}{1-x^{\prime\prime}}\Big)}{2\kappa^{2}x^{\prime\prime2}}\Bigg{\}}\,,\\
    \mathcal{B}(x^{\prime},x^{\prime\prime})=&\Bigg{\{}\frac{(1-x^{\prime})^{2}\ln\Big(\frac{1}{1-x^{\prime}}\Big)}{2\kappa^{2}x^{\prime2}}+\frac{(1-x^{\prime\prime})^{2}\ln\Big(\frac{1}{1-x^{\prime\prime}}\Big)}{2\kappa^{2}x^{\prime\prime2}}\Bigg{\}}\,,\\
    \mathcal{C}(x^{\prime},x^{\prime\prime})=&\Bigg{\{}\frac{(1-x^{\prime})\ln\Big(\frac{1}{1-x^{\prime}}\Big)}{2\kappa^{2}x^{\prime2}}-\frac{(1-x^{\prime\prime})\ln\Big(\frac{1}{1-x^{\prime\prime}}\Big)}{2\kappa^{2}x^{\prime\prime2}}\Bigg{\}}\,,\\
    \tilde{a}(x)=&\frac{1}{2\kappa^2x^2}\ln\Big(\frac{1}{1-x}\Big)\,.
    \label{eq:AppendixCxx}
\end{align}

\bibliography{ref.bib}	
	
\end{document}